%% file: main.tex
\documentclass[11pt]{article}
\usepackage{graphicx} 

\usepackage{fancyhdr}
\usepackage{isomath}
\usepackage{mathtools} 
\usepackage{amsbsy}
\usepackage{amssymb,amsthm}
\usepackage{amscd,amsfonts}
\usepackage{multirow}
\usepackage{enumitem}
\usepackage{bigints}
\usepackage{tikz}
\usetikzlibrary{arrows.meta, patterns, decorations.markings}
\usepackage{graphicx}
\usepackage{verbatim}
\usepackage{euscript}
\usepackage{overpic}
\usepackage{alltt}
\usepackage{stmaryrd}
\usepackage{booktabs}
\usepackage{relsize}
\usepackage{enumerate}
\usepackage{url}
\usepackage[stable]{footmisc}
\usepackage{breakurl}
\usepackage{hyperref}
\usepackage{comment}
\usepackage[numbers]{natbib}
\usepackage[font=small,labelfont=bf]{caption}
\usepackage[font=small,labelfont=bf]{subcaption}
\usepackage{float}
\usepackage{mwe}
\usepackage{algorithm}
\usepackage{algpseudocode}
\usepackage{adjustbox}
\usepackage{xcolor}
\usepackage[super]{nth}
\usepackage{soul}
\usepackage{centernot}
\usepackage{bm}

\DeclareGraphicsExtensions{.eps,.pdf}
\input{temp-definitions}

\usepackage[margin=1in]{geometry}

\theoremstyle{definition}

\newcommand{\bs}{\boldsymbol}

\date{}
\begin{document}
\title{Finite-width adiabatic shear banding and dislocation patterning in mesoscale polycrystalline aggregates}

\author{Siddharth Singh\thanks{Department of Civil \& Environmental Engineering, Carnegie Mellon University, Pittsburgh, PA 15213.} \and Rajat Arora\thanks{Apple, Inc., 333 Dexter Ave., N. Seattle, WA 98109.} \and Janith Wanni\thanks{Department of Materials Science and Engineering, University of Wisconsin-Madison, Madison, WI, 53706.} \and Charles Adkins\thanks{Department of Materials Science and Engineering, University of Wisconsin-Madison, Madison, WI, 53706.} \and Raymond Rasmussen\thanks{Department of Mechanical Engineering, University of Wisconsin, Madison, WI 53706.} \and Noah J. Schmelzer\thanks{Department of Civil and Systems Engineering, Johns Hopkins University, Baltimore, MD 21218.} \and Dan J. Thoma\thanks{Department of Materials Science and Engineering, University of Wisconsin-Madison, Madison, WI, 53706.} \and Curt A. Bronkhorst\thanks{Department of Mechanical Engineering, University of Wisconsin, Madison, WI 53706.} \and Amit Acharya\thanks{Department of Civil \& Environmental Engineering, and Center for Nonlinear Analysis, Carnegie Mellon University, Pittsburgh, PA 15213, email: acharyaamit@cmu.edu. Corresponding author.}}

\maketitle
\begin{abstract}
\noindent Dynamic shear banding under adiabatic conditions in a mesoscale polycrystalline aggregate is studied using a model of mesoscale dislocation mechanics and experiments. The model involves a length scale related to hardening induced by excess/polar/geometrically necessary dislocation (GND) density, and utilizes a simple classical crystal plasticity model with isotropic Voce law hardening. Simulations of statistically representative volume elements  of a polycrystal determined from experimental samples are conducted. Studies in 2-d (section) and 3-d capture the experimentally observed finite-width shear bands and the formation of low-angle subgrain boundaries even in the absence of heat conduction in the model, as well as size-dependent strengthening for grain sizes from $1$ to $20 \, \mu m$. The 2-d and large-scale 3-d simulations, the latter involving $1$ million finite elements, provide access to the progressive evolution of material strength, stress state, and temperature in the course of large deformations. GND distributions accumulate at grain boundaries and form patterned structures within grain interiors, offering insight into the microstructural changes that precede failure in adiabatic shear bands. Mesh-converged, delocalized and localized plastic flow to very large deformations without softening is observed for a significant range of parameters, reflecting a competition between GND hardening and thermal softening in setting the non-softening steady state in the absence of other ductile damage mechanisms in the model.
\end{abstract}

\section{Introduction}\label{sec:intro}
\noindent 
    Adiabatic shear bands are a critical damage and failure mechanism in microstructural metals subjected to high-strain-rate loading \cite{DoddBai1985,DoddBai2012,DoddBai2015,Wright2002}. In applications involving impact, ballistic penetration, high-speed machining, and crash worthiness, deformation occurs rapidly enough that heat generated by plastic work cannot diffuse away from the deforming region. Under these adiabatic conditions, thermal and recrystallization softening competes with strain and strain-rate hardening, resulting in the localization of deformation into narrow bands that often serve as precursors to fracture. Experiments reveal that these bands have a finite width and are accompanied by significant microstructural changes, including grain refinement and subgrain formation \cite{DoddBai2015,MAGAGNOSC2021102992,Derby1991,Hines1997,Hines1998,Meyers2000,Meyers2001,Meyers2003}. Understanding the formation and microstructural evolution of these adiabatic shear bands is therefore essential for predicting material performance under conditions of dynamic loading.

The Split Hopkinson Pressure Bar (SHPB) has long served as the standard experimental technique for characterizing material response at strain rates of $10^2$ to $10^4 ~ s^{-1}$ \cite{SHPBT}. Stress wave measurements in the input and output bars provide well-characterized loading conditions, and  make SHPB experiments particularly suitable for studying shear band formation in a controlled and repeatable setting. When combined with the top-hat specimen geometry, which concentrates shear deformation within a narrow section of the sample by design, the technique enables localization to occur in a predetermined region. It also produces a nominally shear deformation state that isolates shear band behavior from competing deformation and damage modes \cite{Hartmann1981, meyer1986hat,Xue2005,Xue2008,Bronkhorst2006,Bronkhorst2007,Bronkhorst2010}.

However, post-mortem analysis reveals only the final microstructure, not the path that led to it. The progressive evolution of internal fields, from the onset of plastic instability to the formation of a mature shear band, remains inaccessible by experiment. Physical theory and simulation can address this gap by providing access to the full spatiotemporal evolution of these fields, while boundary conditions informed by experimental measurements enable validation against experimental observables.

Early observations and understanding of shear bands under adiabatic dynamic conditions are due to Zener and Holloman \cite{zener1944effect}. The early theoretical foundations for understanding shear band formation under dynamic loading were established through linear stability and perturbation analyses \cite{Bai1981,Bai1982,BurnsTrucano1982,Clifton1984,Wu1984,Anand1987}, which identified the competition between strain-rate hardening, thermal softening, and inertia as the governing factors. In \cite{Fressengeas1987}, Fressengeas and Molinari developed a modified linear stability approach using relative perturbations (absolute perturbations normalized by homogeneous solutions), demonstrating better agreement with nonlinear results. Fully nonlinear numerical simulations extended these insights beyond the linear regime \cite{wright1985,wright1987,Needleman1989,Shaw_clif_89,fressengeas1989}. Shawki and coworkers \cite{shawki_94_1,shawki1994b} developed energy-based criteria demonstrating that localization onset is controlled primarily by inertial forces. Early analytical work by Dodd and Bai \cite{DoddBai1985} established that the shear band width is governed by the thermal length scale introduced through heat conduction, demonstrated later in computations by \cite{cherukuri1995}. In a complementary line of work primarily in the quasi-static setting, Aifantis and Zbib \cite{aifantis1987physics,ZbibAifantis1988,ZbibAifantis1989,ZbibAifantis1992} introduced strain gradients into the flow stress, with an associated length-scale required on dimensional grounds, and showed, through linear stability and nonlinear analyses, that this modification regularizes the localization problem and sets the shear band width. Comprehensive reviews of these early foundational works can be found in \cite{shawki_94_1}, \cite{Wright2002}, and \cite{DoddBai2012}. More recent work has suggested a potential role of significant structural evolution within the material leading to adiabatic shear band formation and maturation \cite{JIN2019416,MOURAD20171,LIEOU2018107,LIEOU2019171,ClaytonJAM2025,CLAYTON2024105880,CLAYTONActaMech2025}. Substantial questions remain about material structural evolution at high deformation rates related to adiabatic shear banding and experimental observations. 

More recent computational efforts have employed crystal plasticity frameworks to capture microstructural influences on shear band formation \cite{Watanabe1998771,Inal2002983,Curt_06}. These studies have shown that shear bands can initiate near triple junctions and propagate across grain boundaries, and that texture evolution plays a critical role in localization. However, conventional crystal plasticity theories do not incorporate an intrinsic material length scale and thus cannot capture experimentally observed size-dependent strengthening at the micron scale and below. Various gradient plasticity theories \cite{Fleck2001, Gurtin20025,Evers20042379,Kuroda20061789} have been proposed to address this limitation  which is also expected to affect the prediction of experimentally observed finite width of shear bands. However, an assessment of such models reveals that fundamental issues remain \cite{Mu2014,a_arora}, even apart from representing a physically direct association of the gradient effect with the mechanics of dislocation distributions. In the context of shear banding, Kuroda and Needleman \cite{KurodaNeedleman2019} have studied higher-order gradient crystal plasticity in the context of dynamic tensile loading of planar single crystals, showing that the interaction between an inertia-induced length scale and the intrinsic material length scale can produce non-monotonic transitions in the localization mode with specimen size, and that the gradient theory regularizes shear band development. 

Mesoscale Field Dislocation Mechanics (MFDM) \cite{acharya2006size,acharya2011microcanonical} offers an alternative that incorporates an intrinsic length scale through the work-hardening response of the material. That physical description requires the (magnitude of the) Burgers vector to enter into a coefficient that mediates the hardening produced due to geometrically necessary dislocations (GNDs), adapting the works of \cite{mecking1981kinetics,estrin1984unified,acharya2000grain} to the setting of MFDM. The theory/model also accounts for the stress fields of GNDs and their spatiotemporal evolution. MFDM has successfully captured size effects in constrained shearing \cite{roy2006size,PDA11,Arora_acharya,Arora_zhang_acharya}, orientation-dependent behavior in micropillars \cite{a_arora}, and kink band formation in nanometallic laminates \cite{a_arora_2}. Finite-width localization, size-dependent strengthening, and dislocation-driven microstructural evolution are desirable attributes for the study of adiabatic shear band formation and structural evolution at the mesoscale.

In this work, we employ a finite deformation version of Reduced MFDM (RMFDM), first introduced for quasi-static, small deformation analysis in \cite{RPA_den} and implemented using the Fast Fourier Transform in \cite{DBTL,BTL_2,BTL_1} for large polycrystalline assemblies. Here, we extend RMFDM to the finite deformation, dynamic setting, with a reformulation specifically designed for phenomena with finite propagation speeds. A particular feature is the discretization of the convected derivative of the evolution of elastic distortion in a structure-preserving manner. While this implementation maintains theoretical equivalence to the version used in \cite{Arora_zhang_acharya}, it is numerically distinct. We calibrate the model against SHPB experiments on a quenched and tempered low-carbon steel and simulate shear band formation in representative volume elements spanning the shear band region. The simulations capture the experimentally observed finite widths of shear bands. They also display size-dependent strengthening for grain sizes from $1$ to $20 \, \mu m$, and the formation of low-angle subgrain boundaries, the latter also observed in experiments \cite{HUGHES20002985} and the ones performed in this work. 

In the absence of direct experimental evidence of time-dependent evolution at the length and time scales involved for comparison (as well as the controlled boundary conditions facilitated in simulation), the simulations provide insight into the progressive evolution of GND density, material strength, stress state, and temperature within the specific model employed.  This work is complementary to that of \cite{Rasmussen2026}, also concerned with the study of dynamic, adiabatic shear banding modeled at the full macroscopic sample level, without explicit consideration of a polycrystalline aggregate and associated crystal mechanics. 

The rest of this Section contains some notational details and terminology used in this work. 

\noindent\underline{Notation and terminology}: Vectors and tensors are represented by boldface lower and upper-case letters. The action of a second order tensor $\bfA$ on a vector $\bfb$ is denoted by $\bfA \bfb$. The inner product of two vectors is denoted by $\bfa \cdot \bfb$, while the inner product of two-second order tensors is denoted by $\bfA:\bfB$. A rectangular Cartesian coordinate system is invoked for ambient space and all (vector) tensor components are expressed with respect to the basis of this coordinate system. $(\cdot)_{,i}$ denotes the partial derivative of the quantity $(\cdot)$ w.r.t.\ the $x_i$ coordinate of this coordinate system. $\bfe_i$ denotes the unit vector in the $x_i$ direction.  The time derivative of a quantity is denoted by $\dot{(\cdot)}$. Einstein's summation convection is always implied unless mentioned otherwise. The symbols $\grad$, $\divergence$, and $\curl$ denote the gradient, divergence, and curl on the current configuration. For a second order tensor $\bfA$, vectors $\bfv$, $\bfa$, and $\bfc$, a spatially constant vector field $\bfb$, the operations of $\divergence$, $\curl$, and cross-product of a tensor $(\times)$ with a vector are defined as follows: 
\begin{subequations}
\begin{align*}
    \left( \divergence \bfA \right) \cdot \bfb = \divergence \left( \bfA^T \bfb \right),& \qquad \forall \:\: \bfb \\
    \bfb \cdot \left( \curl \bfA \right) \bfc = \left[ \curl\left(\bfA^T \bfb \right)\right] \cdot \bfc,& \qquad \forall \: \: \bfb, \bfc \\
    \bfc \cdot \left( \bfA \times \bfv \right) \bfa = \left[ \left(\bfA^T \bfc \right) \times \bfv \right] \cdot \bfa,& \qquad \forall \: \: \bfa, \bfc.
\end{align*}
\end{subequations}
In rectangular Cartesian coordinates, these are denoted by
\begin{subequations}
\begin{align*}
    \left( \divergence \bfA \right)_i =  A_{ij,j}, \quad
    \left( \curl \bfA \right)_{ri} = \varepsilon_{ijk} A_{rk,j}, \quad
    \left( \bfA \times \bfv \right)_{ri} = \varepsilon_{ijk} A_{rj} v_{k},
\end{align*}
\end{subequations}
where $\varepsilon_{ijk}$ are the components of the third order alternating tensor $\bfX$. 

All strain measures used in discussion of results refer to nominal strain.

\noindent\emph{Localization}: In this work we will identify a region as strain-localized when a context-dependent measure of deformation in that region exceeds the average of the same measure over the complement of the region in the domain by a factor of approximately three or more.

\noindent\emph{Convergence}: The term convergence will be used predominantly in the context of convergence of simulation results w.r.t.~mesh refinement. The usage will not be in a strict mathematical sense utilizing a norm(s), but in an approximate physical sense allowing for deviations deemed reasonable, given the large magnitudes and high rates of deformation involved.

\section{Theory}\label{sec:Theory}

We present the governing equations for finite deformation MFDM \eqref{eq:MFDM} and Reduced MFDM (RMFDM) \eqref{eq:RMFDM}. 
\begin{subequations}
\label{eq:MFDM}
\begin{gather}
     \mathring{\bs{\alpha}} \equiv (\divergence(\bs{v})) \bs{\alpha} + \dot{\bs{\alpha}} - \bs{\alpha} \bfL^T = - \curl (\bs{\alpha} \times \bfV +  \bfL^p) \label{alpha_conservation} \\
    \begin{array}{c} 
      \bfW = \bs{\chi} + \grad \bff, \: \bs{F}^e = \bs{W}^{-1} \\
      \curl \bs{\chi} = - \bs{\alpha}  \\
    \divergence \bs{\chi}  = \bs{0}          
    \end{array} \Bigg \}
    \label{eq:MFDM_b}\\ 
    \divergence \left( \grad \dot{\bs{f}} \right) = \divergence \left( \bs{\alpha} \times \bs{V} +  \bfL^p - \dot{\bs{\chi}} - \bs{\chi} \bs{L} \right) \label{plastic_eq} \\
    \divergence \left[ \bfT (\bfW) \right] =\rho \dot{\bfv}  \label{eq:MFDM_d}
\end{gather}
\end{subequations}
\begin{subequations}
\label{eq:RMFDM}
\begin{gather}
     \mathring{\bs{\alpha}} \equiv (\divergence(\bs{v})) \bs{\alpha} + \dot{\bs{\alpha}} - \bs{\alpha} \bfL^T = - \curl (\bs{\alpha} \times \bfV +  \bfL^p) \label{eq:alpha_conservation_dy} \\
   \bs{F}^e = \bs{W}^{-1} \\
   \dot{\bs{W}} + \bs{W} \bs{L} = \bs{\alpha} \times \bfV + \bfL^p \label{eq:w_evol}\\
    \divergence \left[ \bfT (\bfW) \right] =\rho \dot{\bfv}
    \label{eq:RMFDM_d}
\end{gather}
\end{subequations}
Here, $\bfW$ denotes the inverse of the elastic distortion tensor $\bfF^e$, decomposed in MFDM into an incompatible part $\bs{\chi}$ and a compatible part $\grad \bff$, where $\bff$ is the plastic position vector. The tensor $\bfalpha$ represents the dislocation density, $\bfv$ is the material velocity with velocity gradient $\bfL = \grad \bfv$, and $\bfT$ is the symmetric Cauchy stress. The mass density is denoted by $\rho$, and $\bfV$ represents the dislocation density field. All fields in the MFDM system are understood as space-time running averages of their counterparts in the underlying FDM theory \cite{Arora_acharya_2}. The field $\bfL^p$, the plastic strain rate produced by statistical dislocations (SD), captures the averaged plastic straining rate arising from dislocations not resolved by the GND/Excess dislocation density, as discussed in  \cite[Sec.~3]{Arora_acharya_2}. The plastic strain rate of the GND density is given by $\bfalpha \times \bfV$. 

The systems \eqref{eq:MFDM}-\eqref{eq:RMFDM}, in principle, are two theoretically equivalent forms of MFDM theory as discussed in \cite[App.~B]{acharya2011microcanonical}. The alternate forms result from how the evolution of the inverse elastic distortion $\bfW$ is described. In the context of numerical algorithms, \eqref{eq:MFDM} is better suited for situations where quasi-static balance of linear momentum is a good idealization. Here, the elastic stress field of a dislocation density distribution is subject to `instantaneous propagation' throughout the domain. Under dynamic balance of linear momentum without ignoring inertia, this is not true, and \eqref{eq:RMFDM} is better suited for numerical computation of finite speed of propagation phenomena. On the other hand, unlike MFDM, it does not allow the calculation of the elastic distortion field consistent with a prescribed GND tensor field, see \eqref{eq:MFDM_b} above. It is the form \eqref{eq:RMFDM} of the MFDM that we adopt in this work, which entails the use of a slightly different algorithm than the one described in \cite{Arora_zhang_acharya}. As noted in the introduction, RMFDM was first employed for quasi-static, small deformation analysis in \cite{RPA_den}; here we extend its use to the finite deformation and dynamic settings. Separately, an FFT-based implementation of RMFDM has been applied in \cite{DBTL,BTL_2,BTL_1} to study large polycrystalline assemblies under quasi-static conditions.

MFDM requires constitutive statements for the stress $\bfT$, the plastic distortion rate $\bfL^p$, and the dislocation velocity $\bfV$. The details of the thermodynamically consistent constitutive formulations are presented in \cite{Arora_zhang_acharya}. To complete the systems of equations, boundary conditions are required for $\bfalpha, \bfv$ and initial conditions for $\bfalpha, \bfv, \bfW$, and the motion map. The constitutive equations for the theory are given in Tables \ref{tab:constitutive_relation_T}-\ref{tab:constitutive_relation_g} below.

\subsection{Boundary conditions}\label{sec:boun_con}
\begin{itemize}
    \item The $\bs{\alpha}$ evolution equation needs a convective boundary condition of the form $(\bfalpha \times \bfV + \bfL^p) \times \bfn = \bs{\Phi} $, where $\bs{\Phi}$ is a second order tensor valued function. $\bs{\Phi}$ is a function of time and position on the boundary. It characterizes the flux of dislocations at the surface, defined by outward unit normal field $\bfn$, satisfying the constraint $\bs{\Phi} \bfn = \bfzero$.

    The following boundary conditions are valid for the evolution of $\bfalpha$: 
    \begin{itemize}
    \item \textit{Constrained}: This condition applies, $\bs{\Phi} (\bfx , t) = \bfzero$ at a point $\bfx$ on the boundary, for all times. This restricts movement of the dislocations at the external boundary to just be tangential, along with making sure that no dislocation is exiting the body. This makes the boundary plastically constrained, and hence referred to as no-slip or rigid boundary condition.

    \item \textit{Unconstrained}: Specification of dislocation flux $\bfalpha (\bfV \cdot \bfn)$ on the inflow part of the boundary, along with $\hat{\bfL}^p \times \bfn$ simply evaluated at the boundary, can be used as a less restrictive boundary condition. The quantity ($\curl \bfalpha \times \bfn$) is imposed to be zero for all the calculations in this project.

    \end{itemize}
    \item Based on the problem of interest, the material velocity boundary conditions are applied. 
\end{itemize}

\subsection{Constitutive relations}\label{sec:const}

In the MFDM framework, constitutive relations are required for the stress $\bfT$, the plastic distortion rate $\bfL^p$, and the dislocation velocity $\bfV$. The thermodynamically consistent formulations of these constitutive relations are detailed in Sec.~3.1 of \cite{Arora_acharya}. The constitutive relation for the Cauchy stress and mesoscopic core energy density is summarized in Table \ref{tab:constitutive_relation_T}. Additionally, Tables \ref{tab:constitutive_relation_Lp} and \ref{tab:constitutive_relation_V} outline the constitutive relations for the SD plastic distortion rate, $\bfL^p$, and GND dislocation velocity, $\bfV$, respectively. The evolution equation for material strength is provided in Table \ref{tab:constitutive_relation_g}, building on the work \cite{acharya2000grain}.

The appearance of the thermally-softened strength $g_\theta$ in the expressions for the scalar rates $\hat{\gamma}^\kappa, \hat{\gamma}$ is of note in relation to the representation of higher plastic straining in regions of higher temperature, for otherwise equal driving (resolved shear/deviatoric) stresses.

{\renewcommand{\arraystretch}{1.8}
\begin{table}[H]
    \centering
    \begin{tabular}{l c}
    \hline 
        Saint-Venant-Kirchhoff Material & 
            \(\displaystyle 
                \phi(\bfW) = \frac{1}{2 \rho^*} \bfE^e : \mathbb{C} : \bfE^e,  \quad \bfT = \bfF^e [\mathbb{C} : \bfE^e] {\bfF^e}^T 
              \) \\   
         Core energy density &
          \(\displaystyle
              \Upsilon (\bfalpha) = \frac{1}{2 \rho^*} \epsilon \bfalpha : \bfalpha
          \) \\ 
    \hline
    \end{tabular}
    \caption{Constitutive relations for Cauchy stress and core energy density.}
    \label{tab:constitutive_relation_T}
\end{table}}
\begin{table}[htbp]
    \centering
    \begin{tabular}{l c}
    \hline 
        \multirow{2}{6em}{$J_2$ plasticity}  & 
            $\displaystyle
                \hat{\bfL}^p = \hat{\gamma} \bfW \frac{\bfT^{'}}{|\bfT^{'}|}; \quad 
                \hat{\gamma} = \hat{\gamma}_0 \left( \frac{|\bfT^{'}|}{\sqrt{2} \, g_\theta} \right)^{\frac{1}{m}}
            $ \\ 
            &  $\displaystyle \bfL^p = \hat{\bfL}^p + l^2 \hat{\gamma} \curl \bfalpha $ \\
    \hline
        \multirow{4}{6em}{Crystal plasticity}  & 
            $\displaystyle
                \hat{\bfL}^p = \bfW \left(\sum_{k}^{n_{sl}}  \hat{\gamma}^k \bfm^k \otimes \bfn^k\right)_{sym}
            $ \\ 
            &  $\displaystyle \bfL^p = \hat{\bfL}^p + \left( \frac{l^2}{n_{sl}} \sum_{k}^{n_{sl}} |\hat{\gamma}^k|\right) \operatorname{curl} \bfalpha $ \\ 
            &  $\displaystyle \hat{\gamma}^k = \operatorname{sgn} (\tau^k) \hat{\gamma}_0 \left(\frac{|\tau^k|}{g_{\theta}}\right)^{\frac{1}{m}} $ \\ 
            &  $\displaystyle \tau^k = \bs m^k \cdot \bfT \, \bs n^k; \qquad \bs m^k = \bs F^e \bs m_0^k; \qquad \bs n^k = (\bfF^e)^{-T} \bs n_0^k$ \\ 
    \hline
    \end{tabular}
    \caption{Constitutive relations for plastic strain rate of SDs, $\bfL^p$, for the MFDM variants of $J_2$ and Crystal plasticity.}
    \label{tab:constitutive_relation_Lp}
\end{table}

{\renewcommand{\arraystretch}{1.8}
\begin{table}[H]
    \centering
    \begin{tabular}{c c}
    \hline
    \( \displaystyle \dot{g} = h(\bs{\alpha}, g) \left(|\bs{F}^e \bs{\alpha} \times \bs{V}| + \sum_{k}^{n_{sl}} |\hat{\gamma}^k|\right); \)
    & \( \displaystyle h(\bs{\alpha}, g) = \begin{cases} \dfrac{\mu^2 \eta^2 b}{2 (g-g_0)} k_{0} |\bs{\alpha}| + h_{0} \left(\dfrac{g_s-g}{g_s - g_0} \right), & g_0 \leq g \leq g_s \\[1.2em] \dfrac{\mu^2 \eta^2 b}{2 (g-g_0)} k_{0} |\bs{\alpha}|, & g > g_s \end{cases} \) \\
    \hline
    \end{tabular}
    \caption{Constitutive relations for material strength $g$.}
    \label{tab:constitutive_relation_g}
\end{table}
{\renewcommand{\arraystretch}{1.8}
\begin{table}[H]
    \centering
    \begin{tabular}{c c c}
    \hline 
        \(\displaystyle T^{'}_{ij} = T_{ij} - \frac{T_{mm}}{3} \delta_{ij}\);
        &  \(\displaystyle
            a_i = \frac{1}{3} T_{mm} \varepsilon_{ijk} F^{e}_{jp} \alpha_{pk};
            \)    
            &  \(\displaystyle c_l = \varepsilon_{ijk} T^{'}_{jr} F^{e}_{rp} \alpha_{pk} \)  \\
    \( \displaystyle \bfd = \bfc - \left( \bfc - \frac{\bfa}{|\bfa|} \right) \frac{\bfa}{|\bfa|}; \) 
    &\(\hat{\gamma}_{avg} = \left( \frac{1}{n_{sl}} \sum_{k}^{n_{sl}} |\hat{\gamma}^k|\right)\)&
    \\
     \(\displaystyle \bfV = \zeta \frac{\bfd}{|\bfd|} ; \)  
    & \(\displaystyle \zeta = \left(\frac{\mu}{g}\right)^2 \eta^2 b \hat{\gamma}\)& \\
    \hline
    \end{tabular}
    \caption{Constitutive relations for dislocation velocity $\bfV$. $\bfalpha \times \bfV$ gives the plastic strain rate of GNDs.}
    \label{tab:constitutive_relation_V}
\end{table}}
The various input material parameters and descriptors entering the constitutive equations in Tables \ref{tab:constitutive_relation_T}-\ref{tab:constitutive_relation_g} are as follows: $\rho^*$ is the mass density of the pure, unstretched elastic lattice, $\mathbb{C}$ is the fourth order elastic tensor, $\mu$ is the shear modulus, \( m \) represents the material rate sensitivity, while  $\hat{\gamma}_0$ and $\hat{\gamma}_k$ denote the reference strain rate and the magnitude of statistical dislocations (SD) slipping rate on the \( k^{\text{th}} \) slip system in the crystal plasticity model, respectively. The number of slip systems is given by \( n_{sl} \), with \( \bs m^k, \bs {n}^k \) and \( \bs m^k_0, \bs n^k_0 \) representing the slip direction and slip plane normal in the current and unstretched lattice configurations respectively. The material strength parameters include the initial, \( g_0 \),  and saturation, \( g_s \), while \( h_0 \) characterizes the Stage 2 hardening rate. The constants \( k_0 \) and \( \eta \) govern the GND related hardening contributions \cite{acharya2000grain}. Additionally, \( b \) denotes the Burgers vector magnitude of a full dislocation in the crystalline material. 
The quantity $l$ is a material length related to the gross modeling of mesoscale effects of dislocation core energy. The length $l$ controls the refinement of the GND microstructure and does not influence the length scale effects as shown in \cite{Arora_zhang_acharya}. Both $k_0$ and $l$ are the only two additional parameters to the  well-accepted model of classical crystal plasticity theory \cite{asaro1983micromechanics,KALIDINDI1992537,Bronk1992,ANAND20045359}.

To model thermal softening, we update the material strength law to model the effect of change in the temperature on the material according to \eqref{eqn:g_theta}. The updated material strength, $g_{\theta}$, is defined from the current strength, $g$, as shown in \eqref{eqn:g_theta} through a simple, commonly used power-law softening relationship, where $\theta$ and $\theta_0$ represent the current temperature and a reference (often room) temperature:
\begin{equation}
    \label{eqn:g_theta}
    g_{\theta} = g\left(\frac{\theta}{\theta_0}\right)^{-p}.
\end{equation}
The temperature of the body, $\theta$, is updated using idealized balance of energy under adiabatic conditions according to
\begin{equation}
    \label{eq:T_evolve}
    \dot{\theta} = \beta \, \bfT : \bfW^{-1}(\bfalpha \times \bfV + \bfL^p),
\end{equation}

\begin{equation}
\label{eq:Beta_form}
    \beta = \dfrac{\kappa}{\rho c_p},
\end{equation}
with $\kappa$ the Taylor-Quinney coefficient, $\rho$ the mass density, and $c_p$ the specific heat capacity of the material.

The isotropic elastic moduli are updated as
\begin{equation}
\begin{aligned}
\mu &= \mu_0\left(1 - \frac{D_0}{\exp\left(\frac{\theta_0}{\theta}\right) - 1}\right) \qquad \mbox{ and } \qquad 
\lambda &= \frac{2\mu\nu}{1 - 2\nu},
\end{aligned}
\label{eq:ES_evolve}
\end{equation}
where $\mu_0$ is the shear modulus at reference temperature $\theta_0$, and $D_0$ is a parameter accounting for the softening in the elastic moduli.

\noindent \emph{Classical $J_2$ and Crystal plasticity formulations from MFDM}: To produce results corresponding to the  classical plasticity versions of these constitutive models, we simply set $l = 0, k_0 = 0, \bfV = \bfzero$. This effectively also implies that the $\bfalpha$ evolution equation \eqref{eq:alpha_conservation_dy} does not have to be solved in such cases.

\section{Algorithm and verification}\label{sec:AV}
The numerical algorithm used to solve the system \eqref{eq:RMFDM} is briefly mentioned here, with more details to be found in \cite{Arora_zhang_acharya}.

The weak forms of the governing equations of MFDM at finite deformation used in this paper for dynamic cases are presented. This is followed by the algorithm summarizing the implementation.

The FEM formulation and computational algorithm outlined here remain general and do not depend on specific constitutive choices for $\bfL^p$, $\bfV$, or $\bfT$. To solve the governing equations, we employ a numerical scheme based on discrete time increments. Considering a typical step from $t^n$ to $t^{n+1}$, any quantity $(\cdot)$ evaluated at these time instances is denoted as $(\cdot)^n$ and $(\cdot)^{n+1}$, respectively. The time increment $\Delta t^n$ is given by $\Delta t^n = t^{n+1} - t^n$.

\begin{itemize}
    \item \textit{Weak form for $\bfv$ }
    
    As described in \cite{Arora_zhang_acharya}, the finite element mesh is moved to update the geometry using the material velocity field $\bfv$. Balance of linear momentum equation \eqref{eq:RMFDM_d} is solved on the current configuration. Given the stresses and material velocity in the domain $\Omega^n$, the velocity at the next time step, $\bfv^{n+1}$, is obtained using the Forward Euler method as follows:  

    \begin{equation*}
    \label{eq:weak_v}
        \int_{\Omega^n} \rho (v_i^{n+1} -  v_i^n) \delta v_i \, dV = 
        \Delta t_n \left( \int_{\partial \Omega^n} t_i \delta v_i \, dA 
        - \int_{\Omega^n} T_{ij} \delta v_{i,j} \, dV \right).
    \end{equation*}
    
    To improve computational efficiency, the mass matrix is lumped. It is verified that the lumping process does not significantly impact the accuracy of the results.

   \item \textit{Weak form for $\bfalpha$ }
   
    As discussed in \cite{Arora_zhang_acharya}, the $\bfalpha$ transport equation \eqref{eq:alpha_conservation_dy} exhibits nonlinear wave type solutions. An \emph{explicit} Galerkin-Least-Squares FEM approach is adopted here to solve \eqref{eq:alpha_conservation_dy} (cf.~\cite{VBAF06} for an algorithm in the small-deformation setting), with diagonal lumping for the `mass' term explicitly displayed in \eqref{eq:mass_term}. Using $\bfL^p = \hat{\bfL}^P + \beta \curl \bfalpha$, \eqref{eq:alpha_conservation_dy} becomes 
\begin{equation*}
\label{eq:resi_alpha}
\text{tr}(\bfL) \bfalpha + \dot{\bfalpha} - \bfalpha \bfL^T = -\curl \left(\bfalpha\times \bfV + \hat{\bfL}^p + \beta \curl \bfalpha \right).
\end{equation*}

\begin{equation}
\label{eq:mass_term}
    \int_{\Omega^n} \delta \alpha_{ij} \left( \alpha^{n+1}_{ij} -  \alpha^n_{ij} \right) dV + R = 0 \\
\end{equation}

where,
\begin{equation}
\allowdisplaybreaks
    \beta = \left( \frac{l^2}{n_{sl}} \sum_{k}^{n_{sl}} |\hat{\gamma}^k|\right)
\end{equation}
    \begin{subequations}
    \label{eq:weak_alpha}
    \allowdisplaybreaks
\begin{align}
    R &= \int_{\Omega^n} \delta \alpha_{ij} \left( \Delta t^n L_{pp} \alpha_{ij}^n - \Delta t^n \alpha_{ip}^n L_{jp} \right) dV  \notag\\
    &+ \Delta t^n\int_{\Omega^n} \varepsilon_{jqp} \varepsilon_{jab} \alpha_{ia} V_b \delta \alpha_{ip,q} dV
    + \Delta t^n\int_{\Omega^n} \hat{L}^{p}_{ij} \varepsilon_{jqp} \delta \alpha_{ip,q}^n dV \notag\\
    &+ \Delta t^n \int_{\Omega^n} \beta \varepsilon_{jab} \alpha_{ib,a}^n \varepsilon_{jqp} \delta \alpha_{ip,q} dV 
    + \Delta t^n\int_{\partial \Omega^n_i} B_{ij} \delta \alpha_{ij} dA \notag\\
    &+ \Delta t^n \int_{\partial \Omega^n_o} \alpha^n_{ij} V_p n_p \delta \alpha_{ij} dA 
    - \Delta t^n \int_{\partial \Omega^n} \alpha^n_{iq}V_j n_q\delta \alpha_{ij} dA
    - \Delta t^n \int_{\partial \Omega^n} \epsilon_{jpq}\hat{L}_{ip}^{p}n_q \delta\alpha_{ij} dA \notag\\
    &- \Delta t^n \int_{\partial \Omega^n} \beta \varepsilon_{jpq} \varepsilon_{pba} \alpha^n_{ia,b} n_q \delta \alpha_{ij} dA \notag\\
    &+ c \left[ \int_{\Omega^n_e} A_{ri} \delta \alpha_{ri} dV 
    + \Delta t^n \int_{\Omega^n_e} L_{pp} A_{ri} \delta \alpha_{ri} dV 
    - \Delta t^n \int_{\Omega^n_e} A_{ri} \delta \alpha_{rp} L_{ip} dV \right. \notag\\
    & \qquad +\Delta t^n  \int_{\Omega_e^n} A_{ri} \left( \delta \alpha_{ri,q} V_q - \delta \alpha_{rq,q} V_i + \delta \alpha_{ri} V_{q,q} - \delta \alpha_{rq} V_{i,q} \right) dV \notag\\
    &\qquad +\Delta t^n \int_{\Omega^n_e} A_{ri} \left( \beta_{,p} \delta \alpha_{rp,i} + \beta \delta \alpha_{rp,ip} - \beta_{,p} \delta \alpha_{ri,p} - \beta \delta \alpha_{ri,pp} \right) dV \Bigg] \tag{\ref{eq:weak_alpha}}.
\end{align}
\end{subequations}
where, 
\begin{equation*}
\begin{aligned}
    A_{ri} &= \Delta t_n \Big[ \alpha^n_{ri} L_{pp} - \alpha^n_{rp} L_{ip} 
    + \alpha^n_{ri,q} V_q - \alpha^n_{rq,q} V_i 
    + \alpha^n_{ri} V_{q,q} - \alpha^n_{rq} V_{i,q} \\
    &\quad + \varepsilon_{ipq} L^p_{rq,p} 
    + \beta_{,p} \alpha^n_{rp,i} + \beta \alpha^n_{rp,ip} 
    - \beta_{,p} \alpha^n_{ri,p} - \beta \alpha^n_{ri,pp} \Big].
\end{aligned}
\end{equation*}

Here, $\bfL, \bfV$ and $ \hat{\bfL}^p $ are considered known data. The boundaries $ \partial\Omega^n_i $ and $ \partial\Omega^n_o $ denote the inflow and outflow portions of $ \partial\Omega^n $, respectively. The term $ \bfB $ represents the input dislocation flux, given by $\bfalpha \left(\bfV.\bfn \right)$ on $ \partial\Omega^n $. $ \Omega^n_e $ indicates the interiors of the elements. The gradient of $\beta$ is ignored in the Least-Square stabilization due to degrading computational approximation observed practically. We choose $c =1$ for MFDM calculations, unless stated otherwise. The reasoning is elaborated in \cite{Arora_zhang_acharya}.

\item \textit{Updating inverse elastic deformation}

The new contribution in this work is the algorithm for the evolution of the inverse elastic distortion field, \eqref{eq:w_evol}, discretely evolved as follows: $\dot{\bfW}$ is not directly discretized in time. Noting that
\begin{equation}
\dot{\bs{W}} + \bs{W} \bs{L} = \bs{\alpha} \times \bfV + \bfL^p \Longleftrightarrow \dot{\overline{\bfW \bfF}}\bfF^{-1} =  \bs{\alpha} \times \bfV + \bfL^p
\label{eq:W_fomru}
\end{equation}
with $\bfF$ representing the deformation gradient (w.r.t an arbitrarily chosen, but fixed, reference configuration), $\dot{\overline{\bfW \bfF}}$ is discretized in time so that when the right-hand side of \eqref{eq:W_fomru} vanishes, $\bfW=\bs F^{-1}$ is recovered in discrete time. The exact update scheme is shown in \eqref{eq:W_update}, where $a^b$ denotes the field $a$ at time $b$:

\begin{equation}
\frac{\bfW ^{t+\Delta t} \bfF^{t+\Delta t} - \bfW^t \bfF^t}{\Delta t} =\left(\bfalpha \times \bfV + \bfL^p \right)^t \left(\bfF^t\right).
\label{eq:W_update}
\end{equation}

\end{itemize}

\subsection{Algorithm}

Based on the schemes discussed above, we evolve the coupled system \eqref{eq:RMFDM}. A robust and efficient time stepping criterion used for this model is explained in detail in \cite{Arora_zhang_acharya}. Briefly, these criteria are based on plastic relaxation, elastic wave speed in the material and a $0.2 \%$ yield strain threshold. A cut-back algorithm is used to ensure stable and robust evolution of state variables. This carefully controls the magnitude of plastic strain in each increment. The algorithm used for the simulations in this work is presented in Table.~\ref{tab:algo_dy}.

\begin{table}[H]
    \centering
    \renewcommand{\arraystretch}{1.2}
 \caption{Algorithm for dynamic MFDM}
    \begin{tabular}{p{15cm}}
        \midrule
        \textbf{Given:} Material properties, initial conditions, boundary conditions, and applied loading conditions. \\
        \midrule
        \textbf{ Evolution of the system}: Assume that $\bfv^{n-1}$the state at time $t^n$ is known: $\bfx^n, \bfalpha^n, \bfV^n, \bfL^p, \Delta t^n, \bfW^n$. To get the state at time step $t^{n+1}$ the following is done: \\
        \hspace{5mm} $\bullet$ Material velocity $\bfv^n$ is obtained by solving the balance of linear momentum Eq.~\eqref{eq:RMFDM_d} $\Omega^n$.\\
        \hspace{5mm} $\bullet$ $\bfalpha^{n+1}$ on $\Omega^{n+1}$ is obtained by solving weak form of $\bfalpha$ evolution equation \eqref{eq:weak_alpha} on $\Omega^n$. \\
        \hspace{5mm} $\bullet$ The configuration of the body is discretely updated, i.e., $\bfx^{n+1} = \bfx^n + \bfv^n \Delta t^n$. \\
        \hspace{5mm} $\bullet$ $\bfW^{n+1}$ on $\Omega^{n+1}$ is obtained by using Eq.~\ref{eq:W_update}. \\
        \midrule
        \textit{State acceptance criteria:} Let $PSR = (\max(|\bfF^e \bfalpha \times \bfV|^{n+1}) + \max(\hat{\gamma}^{n+1}))$. If $PSR \times \Delta t^n \leq 0.002$, the state is accepted. $\Delta t^{n+1}$, based on the new state, is calculated as discussed in \cite{Arora_zhang_acharya} Table 6 and this algorithm is repeated to get state at increment $t^{n+2}$. If the condition is not satisfied: \\
        \hspace{5mm} $\bullet$ Go back to the state at time $t^n$. \\
        \hspace{5mm} $\bullet$ Use $\Delta t^{n,new} = \min \left(\frac{0.002}{PSR}, 0.5 \Delta t^n \right)$ and repeat the algorithm to obtain a new state at $t^{n+1}$. \\
        \bottomrule
    \end{tabular}
    \label{tab:algo_dy}
\end{table}
\subsection{Verification}
\label{sub_sec:validation}
For verification, we have applied a constant velocity in the uniaxial compression test as shown in Fig.~\ref{fig:valid_sc}. As expected, this constant loading produces an elastic wave first, followed a the plastic wave - we use the $J_2$-MFDM  plasticity model for this exercise. The domain selected for this simulation is $10 \times 15 ~\mu m^2$. We calculate the speed of the elastic and plastic waves separately, as discussed below:

\begin{figure}[htbp]
    \centering
    \includegraphics[scale=0.6]{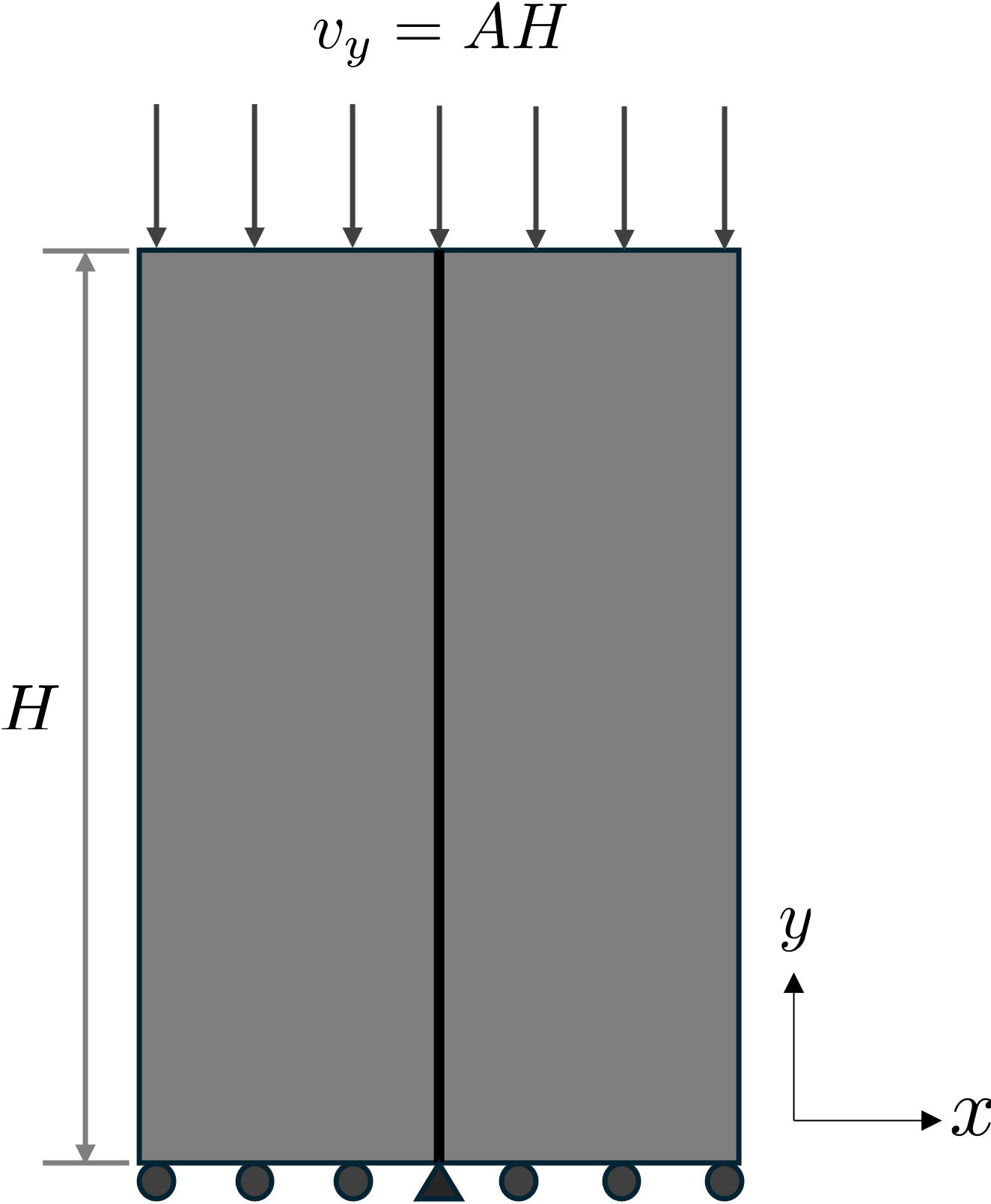}
    \caption{Schematic showing the boundary condition used for verification. A uniform velocity is applied at the top surface, where $A$ is the applied strain rate. The displacement profile and the stress deviator are discussed along the black line. The origin is at the center of the domain.}
    \label{fig:valid_sc}
\end{figure}

\begin{itemize}
    \item \textit{Elastic wave speed}: This speed is calculated based on the displacement profile. Specifically, we plot the displacement along the vertical line $x = 0$, as shown by the black line in Fig.~\ref{fig:valid_sc}. The obtained result is presented in Fig.~\ref{fig:disp_elas_plot}. It is observed that the ratio of the analytical linear elastic dilatational wave speed and the calculated speed is nearly 1. This verifies that the simulation is producing the expected elastic wave speed.
    \begin{figure}[H]
	\centering
    \begin{minipage}{.485\textwidth}
		\centering
		{\includegraphics[width=1\linewidth]{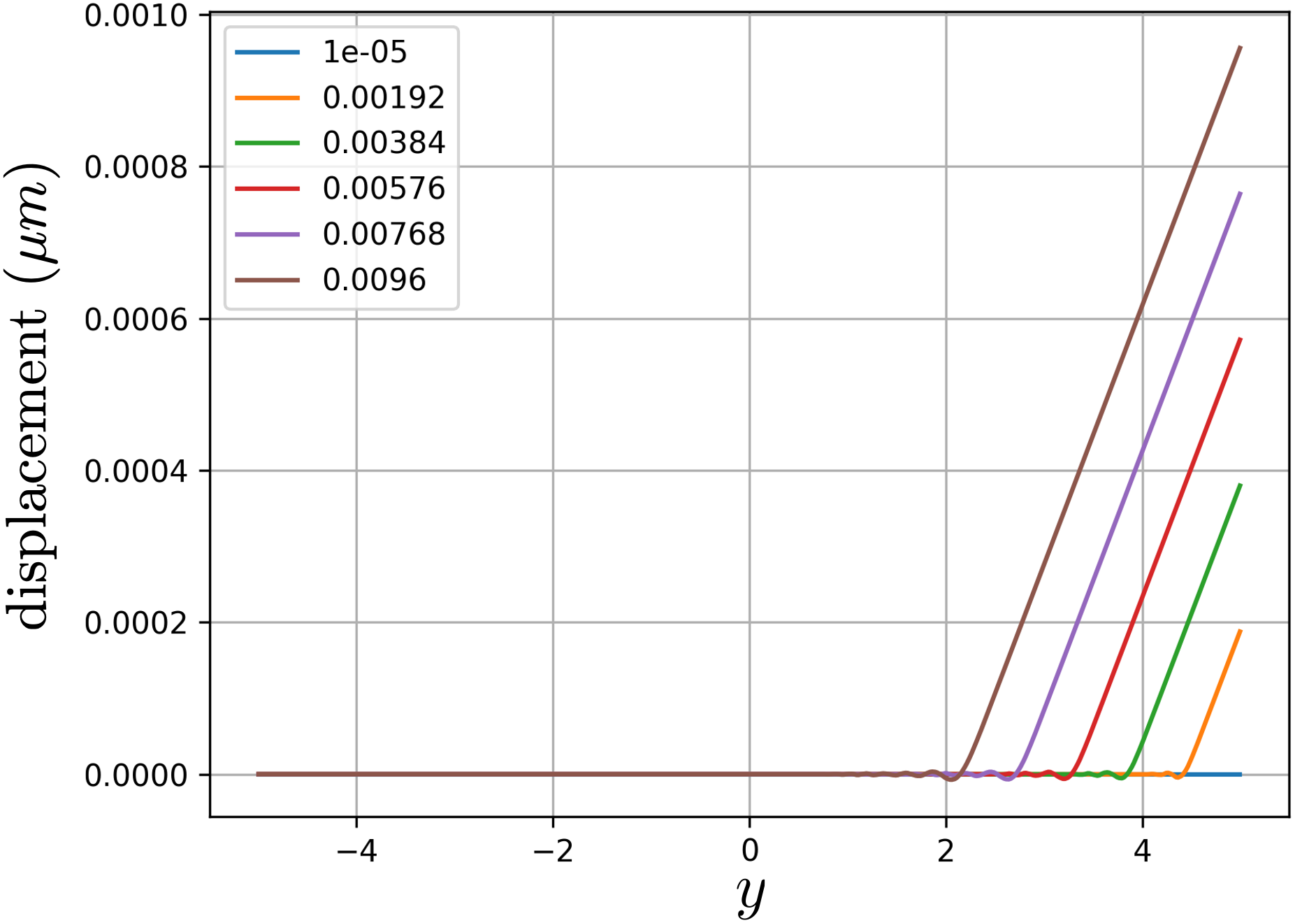}}
		\caption{Displacement along the $x = 0$ line, plotted at different times to calculate the elastic wave speed.}
		\label{fig:disp_elas_plot}
	\end{minipage}\hfill
    \begin{minipage}{.485\textwidth}
		\centering
{\includegraphics[width=1\linewidth]{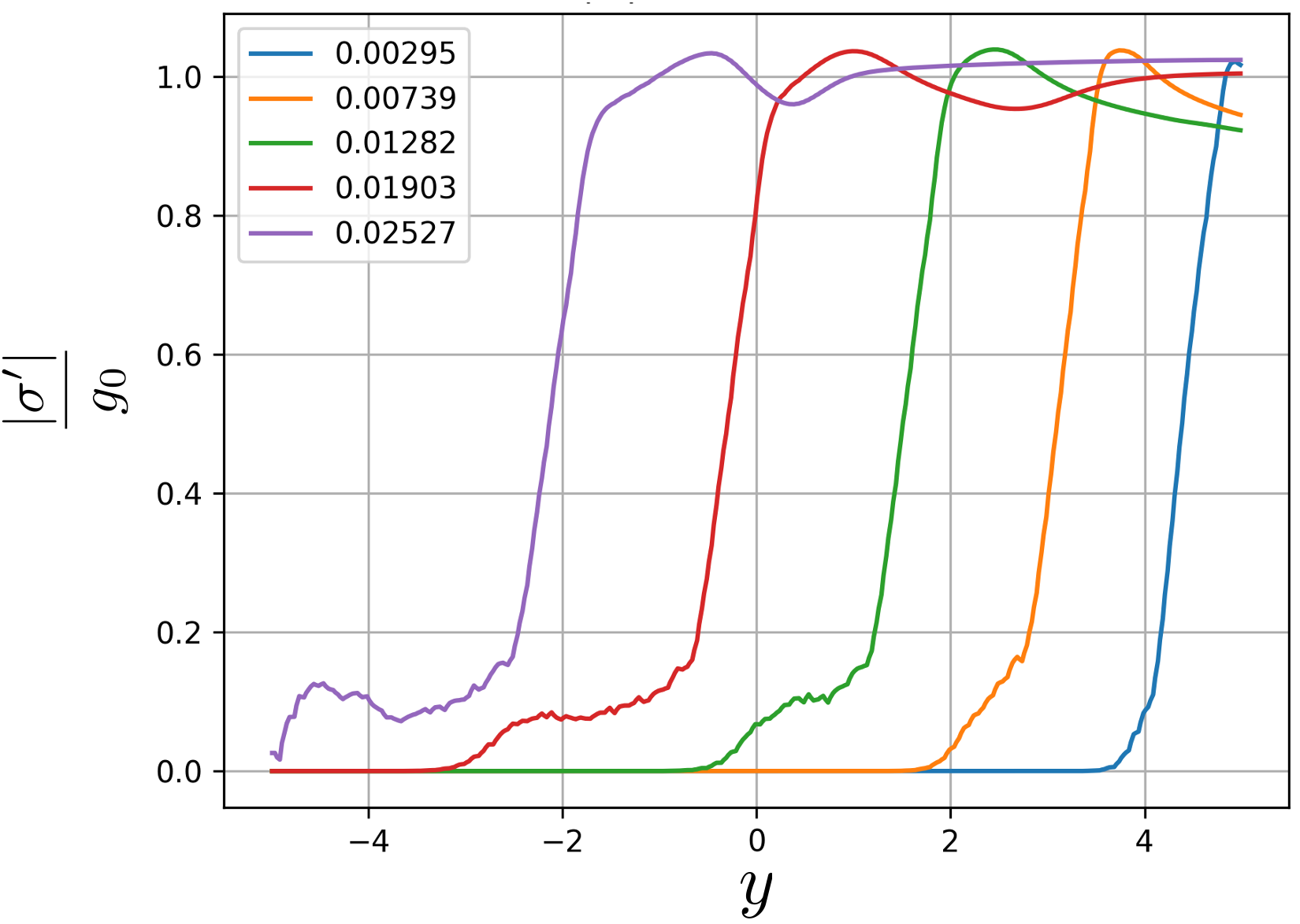}}
		\caption{Norm of deviatoric stress ($\sigma ^{'}$) normalized by yield stress ($g_0$) along the $x = 0$ line.}
				\label{fig:disp_plas_plot}
	\end{minipage}\\
\end{figure}
\item \textit{Plastic wave speed:} We define the plastic wave speed as the rate at which the plastically yielded material front is moving through the domain as explained in the following. The wave-front of $\sigma' = |\bfsigma'|$, the magnitude of the deviatoric stress tensor, is tracked in the simulation and its speed calculated. This measure serves as an effective indicator of plasticity. The deviatoric stress normalized by the yield stress is plotted along a vertical line (the black line in Fig.~\ref{fig:valid_sc}). Fig.~\ref{fig:disp_plas_plot} shows smoothed curves to enable clearer analysis. Smoothing is performed by taking a moving average, over a fixed size, of the original curve. 

The elastic wave speed calculated in this case is approximately 1.8 times the linear elastic dilatation wave speed. This is reasonable because a nonlinear elastic material has a linearized elastic modulus which depends on elastic strain and typical nonlinear elastic materials become stiffer with increasing compression \cite{ZAWB15}. Moreover, the plastic wave speed, approximately half the elastic wave speed, aligns with experimental findings that plastic wave speeds in steels typically range between 0.2 and 0.6 times the elastic wave speed \cite{meyers_chawla}.
    
\end{itemize}

\section{Motivating experiments and general considerations for the simulation studies} \label{sec:Exp_analysis}
SHPB, also known as the Kolsky bar \cite{Kolsky1963}, is a well-established experimental technique for characterizing material behavior at high strain rates, typically in the range of $10^2$--$10^4~\mathrm{s}^{-1}$. A schematic of the apparatus is shown in Fig.~\ref{fig:SHPBT_stup}. The setup consists of three main components: a striker bar (not shown in the figure), an incident bar, and a transmission bar, with the specimen sandwiched between the incident and transmission bars. Upon impact of the striker bar, a compressive stress wave propagates through the incident bar. At the specimen interface, part of this wave is reflected back into the incident bar while the remainder is transmitted through the specimen into the transmission bar. Strain gauges mounted on the incident and transmission bars record the incident ($\epsilon_i$), reflected ($\epsilon_r$), and transmitted ($\epsilon_t$) strain pulses, from which specimen stress, strain, and strain rate can be inferred. Details of the experimental protocol and underlying formalism can be found in \cite{SHPBT}.

\begin{figure}[htbp]
    \centering
    \includegraphics[scale=0.5]{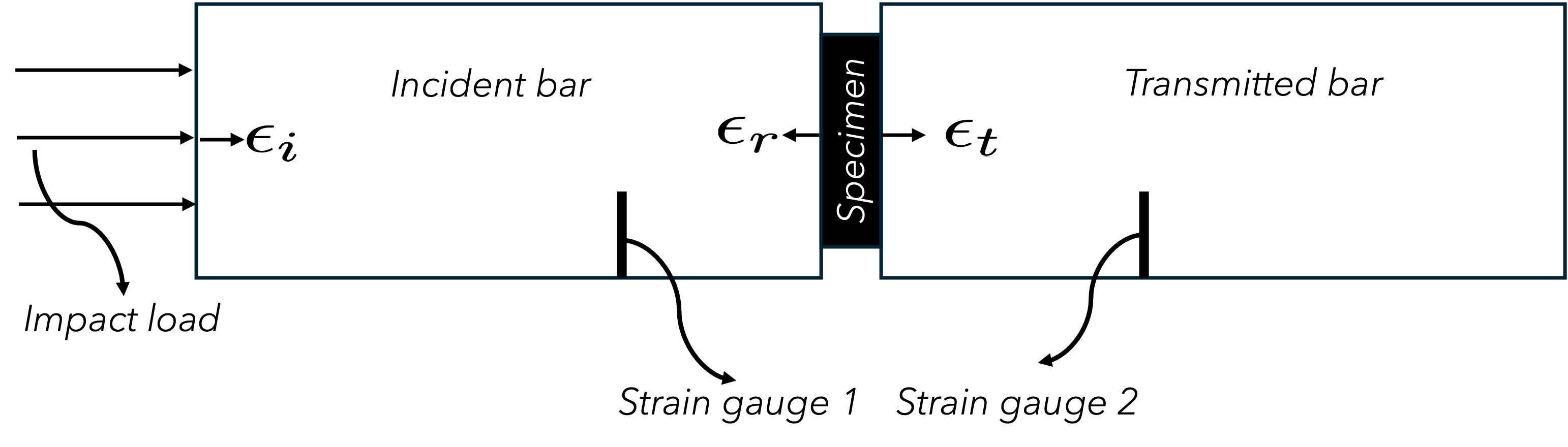}
    \caption{Schematic showing the components of the Split Hopkinson Pressure Bar apparatus: striker bar, incident bar, specimen, and transmission bar.}
    \label{fig:SHPBT_stup}
\end{figure}

Cylindrical specimens ($6~\mathrm{mm}$ diameter $\times$ $6~\mathrm{mm}$ height) machined from an $\frac{1}{2}$~in. steel plate along three orthogonal directions were characterized by loading at several different initial temperatures and strain rates \cite{Rasmussen2026}. Tests were conducted at nominal strain rates exceeding $1000~\mathrm{s}^{-1}$, and the measured stress--strain response was used to calibrate our model parameters, as discussed in Sec.~\ref{sec:ESD}. The cylindrical geometry, in combination with uniaxial compression boundary conditions, produces a nominally homogeneous deformation field throughout the specimen. We perform simulations on a statistically representative volume element (RVE) \cite{Behnoudfar2026} subjected to boundary conditions derived from the experimental loading, rather than simulating the full macroscale specimen. 

To study adiabatic shear band formation, SHPB experiments were performed using a top-hat specimen geometry. The top-hat geometry features a reduced section (shown in Fig.~\ref{fig:top_hat_specimen}) which concentrates deformation into a predetermined shear zone, enabling controlled and reproducible shear band initiation \cite{meyer1986hat}. This design allows for targeted post-mortem characterization of the shear band region.

Rather than modeling the entire specimen, we focus on a statistically representative RVEs in the size range of $10\,\mu m$ - $160\,\mu m$, which encompasses the range in which adiabatic shear bands are experimentally observed, as shown in Fig.~\ref{fig:top_hat_specimen}c and Fig.~\ref{fig:th_combined}.

The polycrystalline microstructure used in the simulations here is described in \cite{Behnoudfar2026}. Different boundary conditions, initial conditions, and geometries are employed to probe distinct aspects of localization and thermal softening in dynamic shear banding phenomena.

\begin{figure}
    \centering
    \includegraphics[scale=0.8]{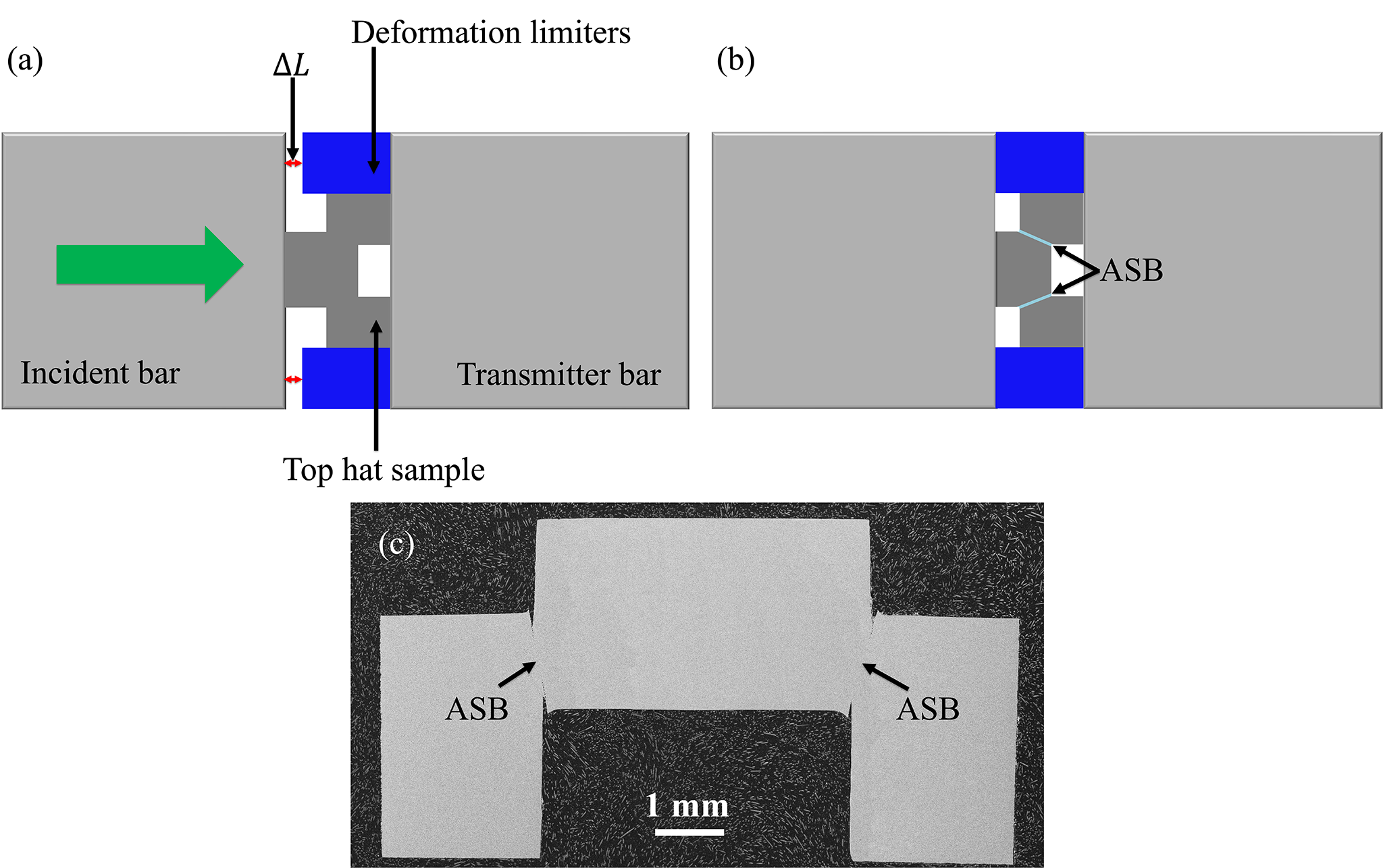}
    \caption{Top-hat specimen configuration in the SHPB setup: (a) undeformed assembly with deformation limiters, (b) deformed configuration showing 
localized shear regions, and (c) Scanning electron micrograph of a recovered specimen cross-section.}
    \label{fig:top_hat_specimen}
\end{figure}

\subsection{Experimental observations}

Experimental observations provide critical insight into the observed phenomenology of shear band formation and evolution, and guide the selection of appropriate test problems studied in our simulations. 

Localization behavior was observed under dynamic loading (Fig.~\ref{fig:th_combined}) conditions. The polycrystalline microstructure used for analysis was informed by high-resolution electron backscatter diffraction (EBSD) measurements of the undeformed material.  A ZEIS GEMINI microscope equipped with Oxford Symmetry S3 EBSD detector was used.  An EBSD step size of $15\,nm.$ was used to collect data. Dictionary indexing (DI) was used to index the EBSD patterns.

In this dynamic test configuration, the incident bar strikes the specimen at high impact velocity (Fig.~\ref{fig:top_hat_specimen}a), initiating deformation localization in the region highlighted in Fig.~\ref{fig:top_hat_specimen}b. The deformation localizes into a narrow adiabatic shear band at the hat–brim interface of the specimen. The band maintains a finite width on the order of $10 - 40 \,\mu m.$ and remains sharply defined throughout deformation.

This finite-width character of the shear band is a key experimental observation that forms one of the primary questions of our modeling studies.

A particularly noteworthy observation from the EBSD analysis (Fig.~\ref{fig:th_combined}) is the development of intragranular substructure within the shear band region, with the grains exhibiting a slender, elongated morphology approximately parallel to the shear band. The inverse pole figure (IPF) maps reveal subgrain formation at large deformation. These microstructural features suggest that dislocation density evolution plays a central role in the process of shear banding.

\begin{figure}[htbp]
    \centering
        \includegraphics[scale=0.48]{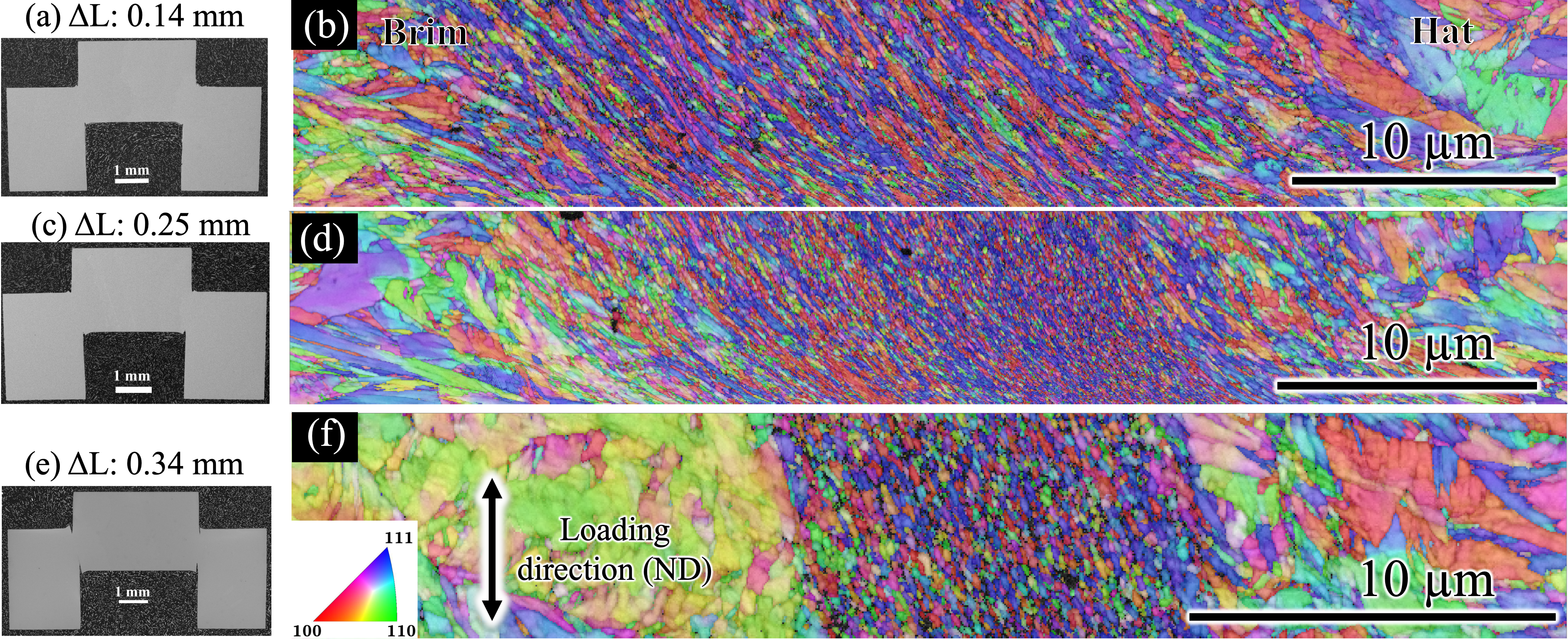}
        \label{fig:th_del_L_14}
    \caption{Scanning electron micrographs (left) and IPF maps (IPF parallel to the loading direction $y$) via EBSD (right) of the top-hat specimen 
    under continuous dynamic loading at displacement increments of 
    (a,b) $\Delta L = 0.14~mm$, 
    (c,d) $\Delta L = 0.25~mm$, and 
    (e,f) $\Delta L = 0.34~mm$. 
    Progressive deformation localization (left) and grain refinement with subgrain 
    formation (right) are evident.}
    \label{fig:th_combined}
\end{figure}

\subsection{A scaling argument to speed up mesoscale simulations} \label{sec:ESD}

Results from a SHPB test \cite{Rasmussen2026} on a cylindrical specimen are used to determine the boundary conditions that are applied to an RVE of the steel and the material parameters of the model. A schematic of the applied boundary conditions is shown in Fig.~\ref{fig:2d_comp}, with details given in Table.~\ref{tab:boundary_conditions_2}.

The procedure for determining the velocity boundary conditions on the specimen from experimental data obtained from the test is first outlined. A detailed description of the approach can be found in \cite{SHPBT}.
\subsection*{Velocity boundary conditions for SHPB simulations} \label{sss:SHPB_vel}

\begin{figure}[htbp]
    \centering
    \begin{subfigure}{0.48\textwidth}  
        \centering
        \includegraphics[scale=0.5]{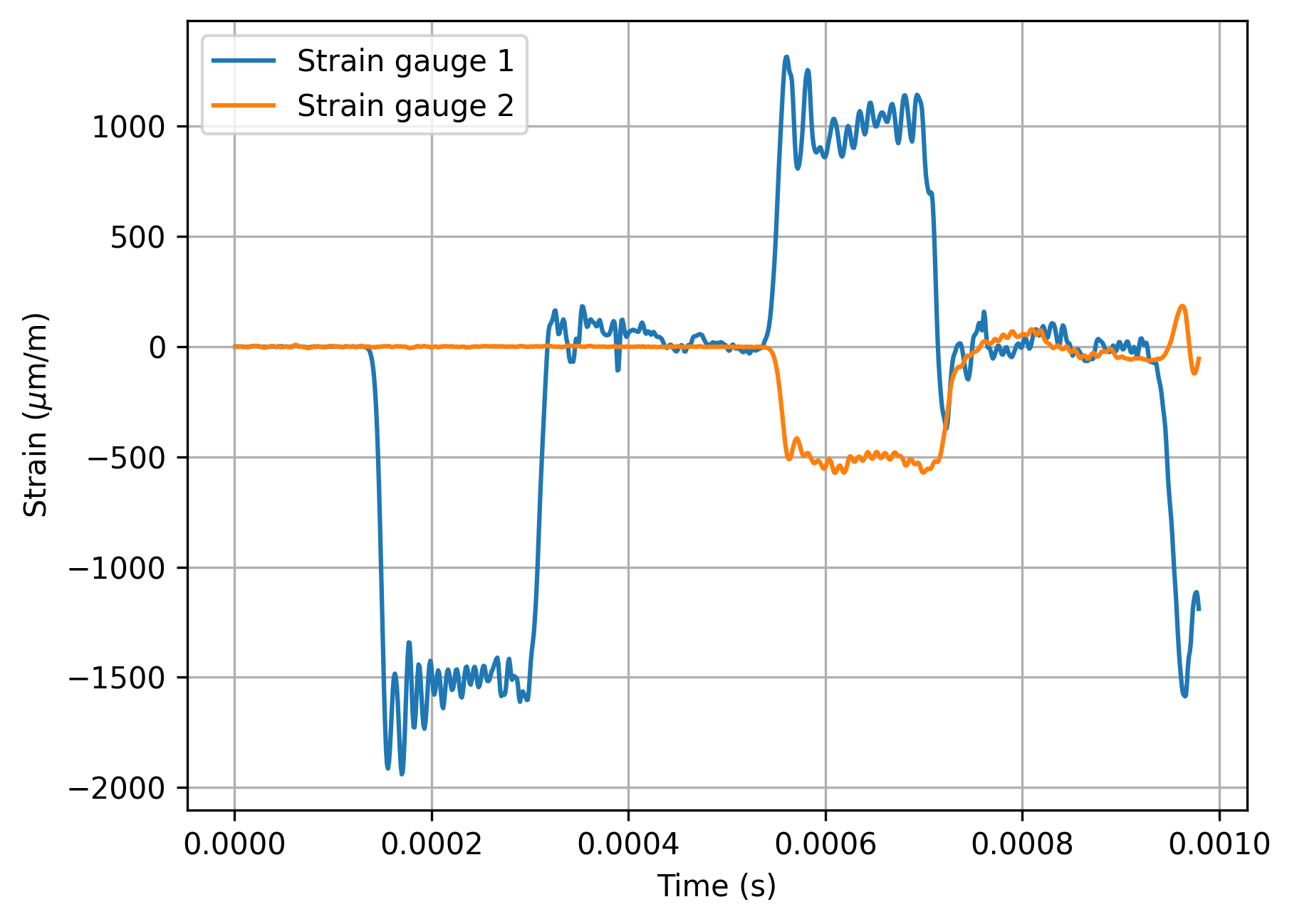}
        \caption{}
        \label{fig:exp_obs_shpbt}
    \end{subfigure}
    \hfill
    \begin{subfigure}{0.48\textwidth}  
        \centering
        \includegraphics[scale=0.5]{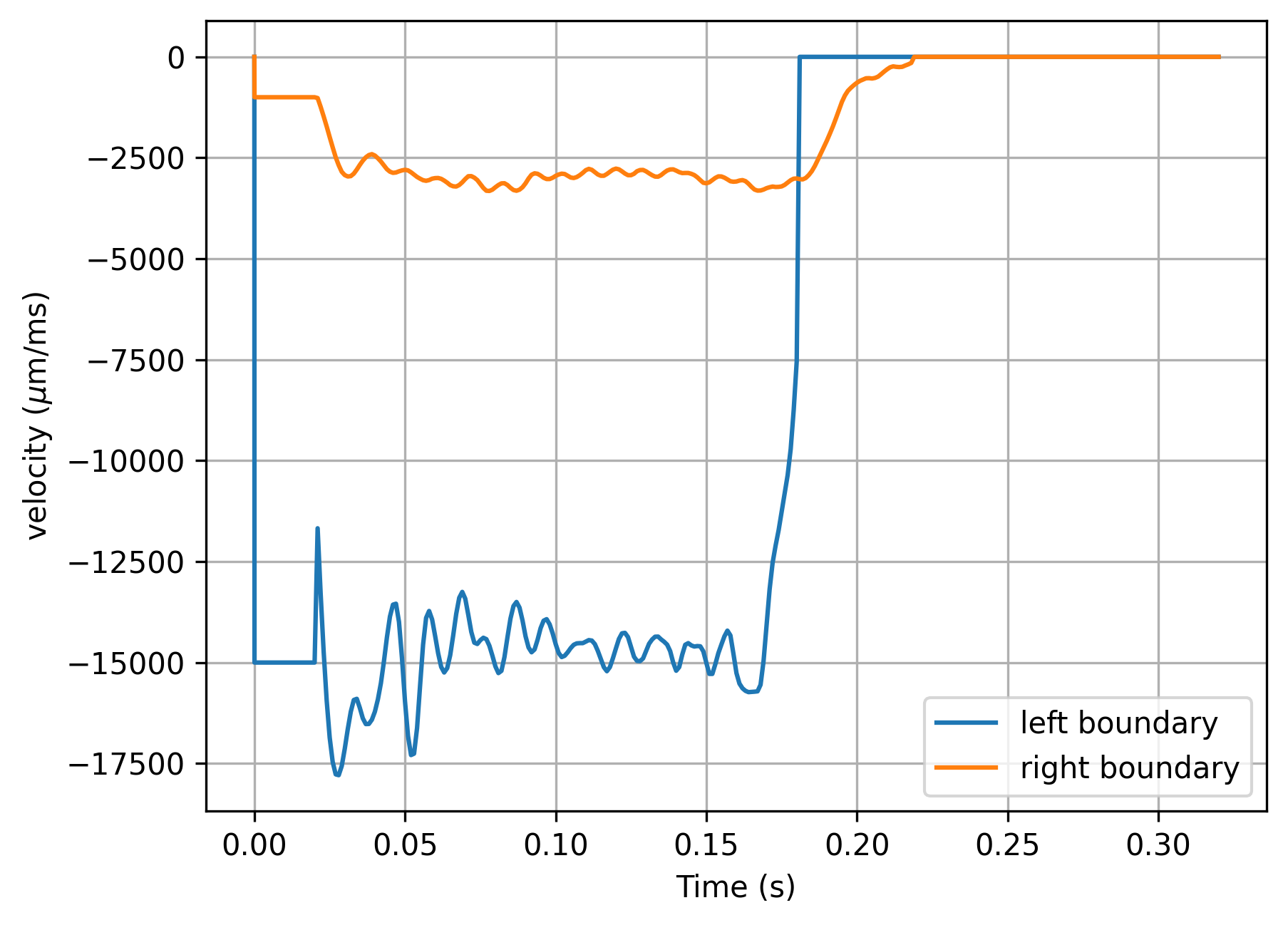}
        \caption{}
        \label{fig:app_bc}
    \end{subfigure}
    \caption{(a) Experimental strain gauge readings from the SHPB test, and (b) velocity boundary conditions derived for the simulations.}
\end{figure}

The experimental data available from the SHPB test are the strain gauge readings shown in Fig.~\ref{fig:exp_obs_shpbt}. To obtain the velocity boundary conditions for our calibration simulations, one-dimensional wave analysis is employed. The resulting velocity histories applied at the left and right specimen boundaries are shown in Fig.~\ref{fig:app_bc}. In this figure, time zero corresponds to the instant when the incident wave reaches the left interface. The boundary condition on the right interface is applied with a time delay equal to the wave transit time through the specimen, calculated using the elastic wave speed of the material. The velocities at the left ($v_l$) and right ($v_r$) boundaries are computed from the strain data as:
\begin{equation}
    \label{eq:v_1d_anly}
    \begin{aligned}
    v_{l} &= c_b \left( \epsilon_i - \epsilon_r \right), \\
    v_{r} &= c_b \left( \epsilon_t \right),
    \end{aligned}
\end{equation}
where $\epsilon_i$, $\epsilon_r$, and $\epsilon_t$ are the incident, reflected, and transmitted strain pulses, respectively, and $c_b$ is the elastic wave speed in the pressure bars.

This approach ensures that the simulation boundary conditions faithfully represent the loading history experienced by the specimen during the experiment, capturing the initial stress wave arrival as well as the subsequent wave reflections within the specimen. In the simulation configuration (Fig.~\ref{fig:2d_comp}), the specimen is oriented vertically: the left experimental boundary is mapped to the top boundary and the right experimental boundary to the bottom boundary. The velocity boundary conditions $v_l$ and $v_r$ are accordingly applied.

\subsection*{Dimensional analysis to reduce wall-clock time for mesoscale simulations}\label{ss_sec:da}

As mentioned, we consider RVEs of linear dimension in the $\sim 10 -100 ~\mu m$ range. The experimental specimen has a characteristic dimension of $H_{\text{exp}} = 6~\text{mm}$, and the velocity boundary conditions shown in Fig.~\ref{fig:app_bc} correspond to this full-scale specimen. For the RVE of size $H_{\text{RVE}}$, we make the assumption of calibrating the behavior of the RVE at the same applied nominal strain rate as the macroscopic SHPB cylindrical sample for which homogeneous deformation conditions are expected to set in after a few wave reflections. Hence,
\begin{equation}
A = \frac{v_{\text{exp}}}{H_{\text{exp}}} = \frac{v_{\text{RVE}}}{H_{\text{RVE}}}, \quad \implies \quad v_{\text{RVE}} = v_{\text{exp}} \cdot \frac{H_{\text{RVE}}}{H_{\text{exp}}}
\end{equation}
where $A$ is the applied strain rate, and $v_{\text{RVE}}$ and $v_{\text{exp}}$ are the velocity differences between the top and bottom faces of the RVE and the experimental domain, respectively ($v_{exp} = v_l - v_r \eqref{eq:v_1d_anly}$). Thus the physical problem that is simulated for the $H_{\text{RVE}}$ sized sample for the purposes of calibration uses the above scaled magnitude for the velocity-time profile shown in Fig.~\ref{fig:app_bc}.

The main obstacle when simulating smaller domains and/or the finer meshes (on the scale $\mu m$) is related to wall clock time taken by simulation for an adequate mesh refinement. As the time stepping is directly proportional to the element size (Courant-Friedrichs-Lewy (CFL) condition) and inversely proportional to the elastic wave speed, when the element size is reduced by a factor with elastic wave speed remaining constant, the simulation time increment decreases and the number of time-steps required for reaching a prescribed strain level at a fixed strain rate, increases by the same factor. 

We use Dimensional Analysis \cite{Barenblatt1996} to overcome the above bottleneck. This approach is fundamentally based on the idea that physics does not depend on the physical units used to describe any chosen natural phenomena. Relevant observables of the modeling and simulation setup can be written as a function of dimensionless ratios as shown in \eqref{eqn:dim_anly} (example shown for the true stress $\sigma_t$):
\begin{equation}    \label{eqn:dim_anly}
    \frac{\sigma_t}{g_0} = \phi\left(\frac{E}{g_0}, \frac{\hat{\gamma}_0}{fA}, \frac{V_e}{H(fA)}, \frac{b}{H}, f, \kappa, k_0  \right),
\end{equation}
where $\sigma_t$ is the true stress, $g_0$ is the initial yield strength, $E$ is the elastic modulus, $\hat{\gamma}_0$ is the reference plastic strain rate, $A$ is the applied strain rate, $H$ is the length of the domain, $V_e = \sqrt{E/\rho}$ is the elastic wave speed, $\rho$ is the mass density, $b$ is the magnitude of the Burgers vector, $f, \kappa, k_0$ are non-dimensional parameters representing the scaling factor by which the applied strain rate is increased, the Taylor-Quinney coefficient \eqref{eq:T_evolve}, and the GND hardening coefficient (Table.~\ref{tab:constitutive_relation_g}), respectively.

The main idea is to reach the same strain in a smaller number of steps and if we can increase the amount of strain in each step, keeping the same time stepping, we can resolve the issue. This idea translates to increasing the applied strain rate by a factor $f$ (i.e., $A \to fA$), while ideally maintaining the ratios in \eqref{eqn:dim_anly} in which $A$ appears fixed. Thus, we increase the reference plastic strain rate, $\hat{\gamma}_0$, by the same factor $f$. The other ratio, $\frac{V_e/H}{fA}$ representing the ratio of an elastic strain-rate defined by an elastic wave speed and specimen geometry to the applied strain rate, is assumed to have no significant effect on the results pertaining to plasticity effects, within the strain rate range of interest. This assumption is verified a-posteriori. As shown in Fig.~\ref{fig:da_ss_curve}, the assumption holds within the given strain range. Here, both $A$ and $\hat{\gamma}_0$ have been increased by the same factor. The factor used for our simulations is $f = 150$ to balance the computational gain with making sure that the differences in stress-strain response between the $f = 1$ and $f = 150$ profiles in Fig.~\ref{fig:da_ss_curve} are minimal.

Using this approach, the simulation time for a given problem was reduced to nearly $1/10^{th}$ when $A$ was increased by a factor of 25. Initially, the simulation for the given domain was estimated to take around 120 hours (based on extrapolation) using 128 cores. After applying the dimensional analysis approach, the same strain was achieved in just 12 hours by scaling $A$ and $\hat{\gamma}_0$ appropriately. Using more cores results in a speedup. For the same problem, if 128 cores take 12 hours, the task can be completed in approximately 8 hours with 256 cores, without any further attention paid to code efficiencies.

\begin{figure}[htbp]
    \centering
    \begin{tikzpicture}
        \node[anchor=south west, inner sep=0] (img) 
            {\includegraphics[scale=0.50]{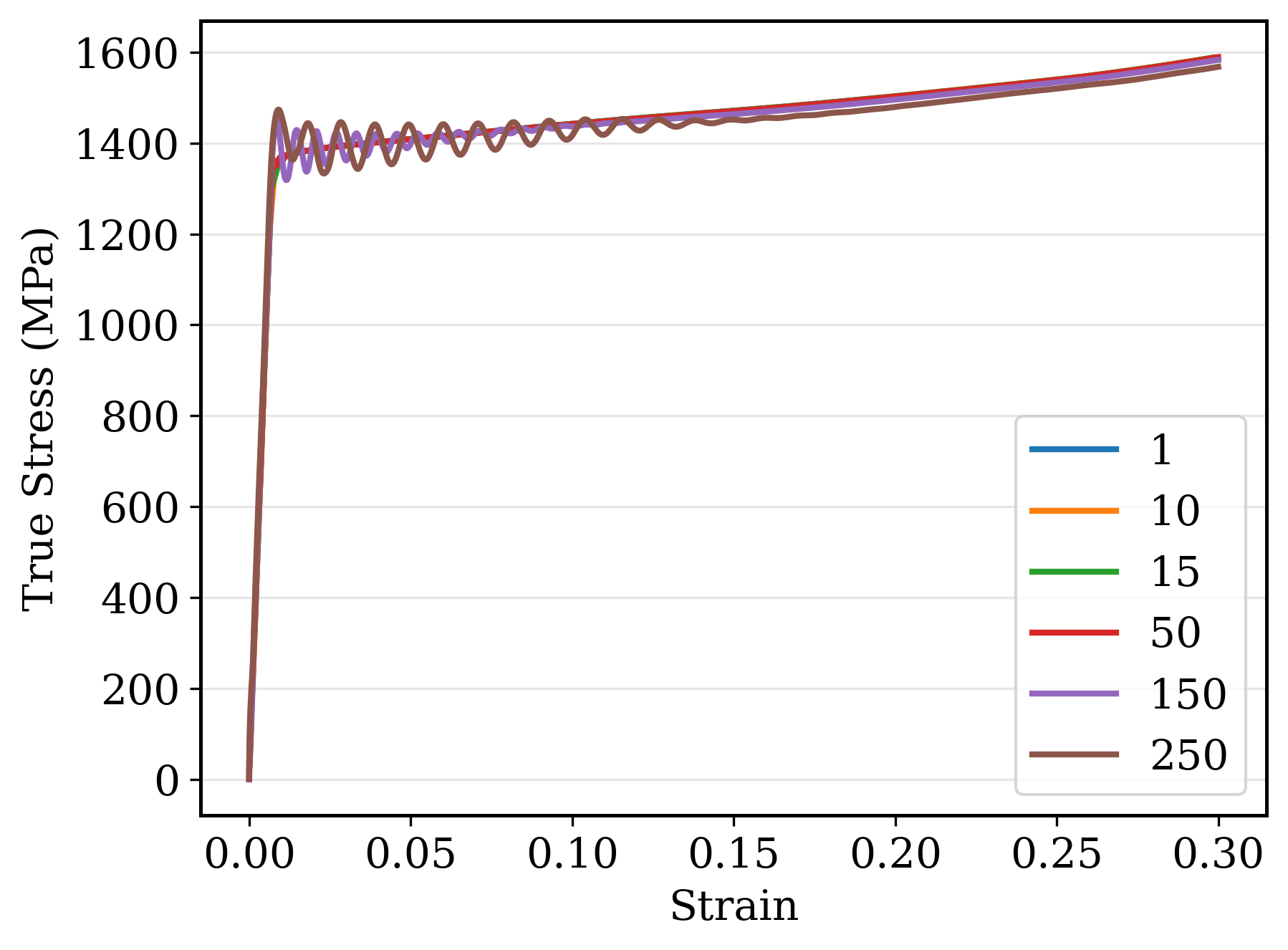}};
    \end{tikzpicture}
    \caption{Comparison of stress-strain curve for different values of applied strain rate but keeping the non-dimensional ratio, $\frac{\hat{\gamma}_0}{fA}$, fixed. Different values of $f$ are shown in legend.}
    \label{fig:da_ss_curve}
\end{figure}

\begin{figure}[htbp]
    \centering
    \begin{tikzpicture}
        \node[anchor=south west, inner sep=0] (img) 
            {\includegraphics[scale=0.48]{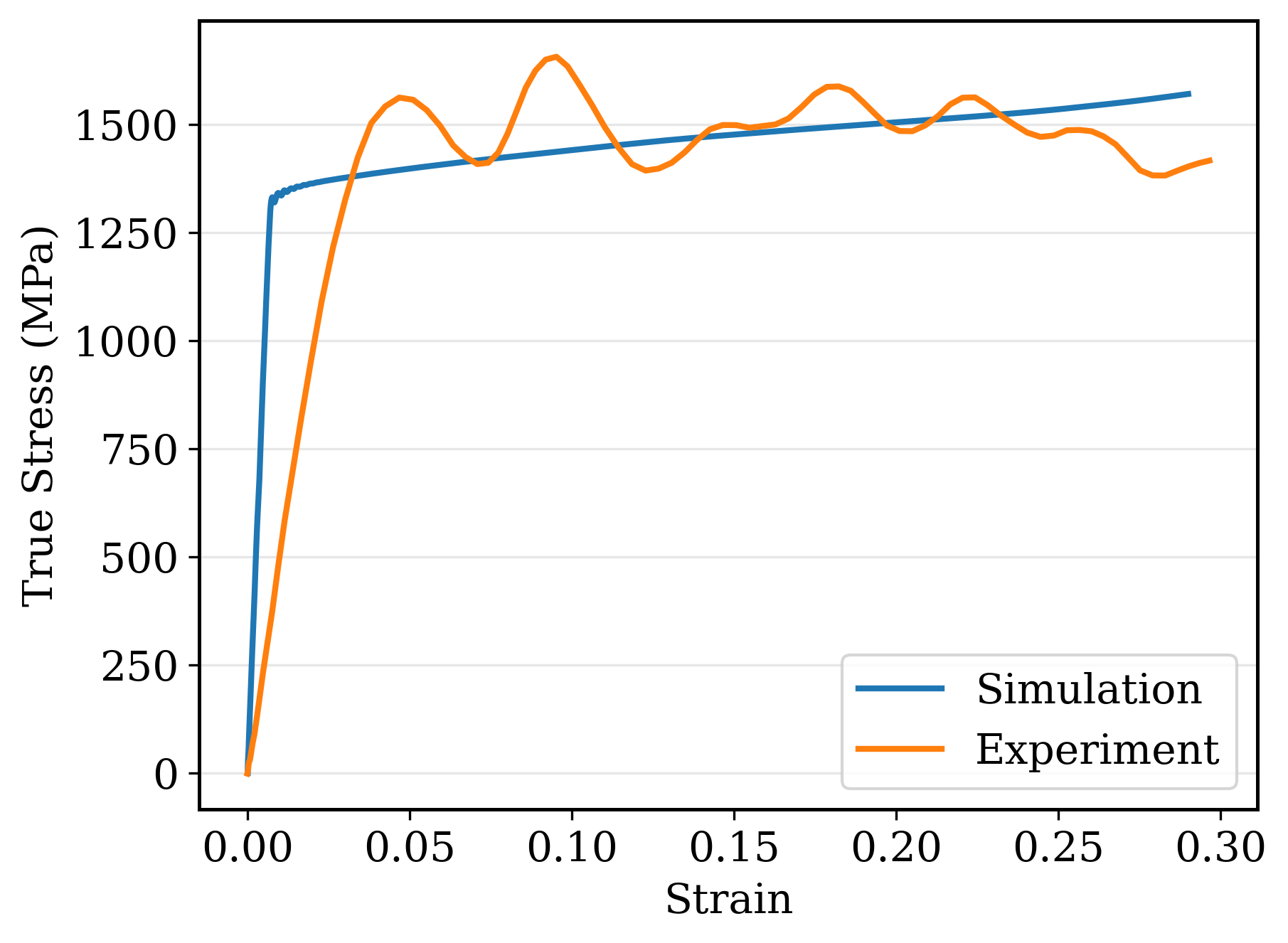}};
    \end{tikzpicture}
    \caption{Stress-strain response from the SHPB experiment compared with the simulation.}
    \label{fig:SS_calib_2}
\end{figure}

\begin{figure}[htbp]
    \centering
    \begin{subfigure}{0.48\textwidth}  
        \centering
        \includegraphics[scale=0.5]{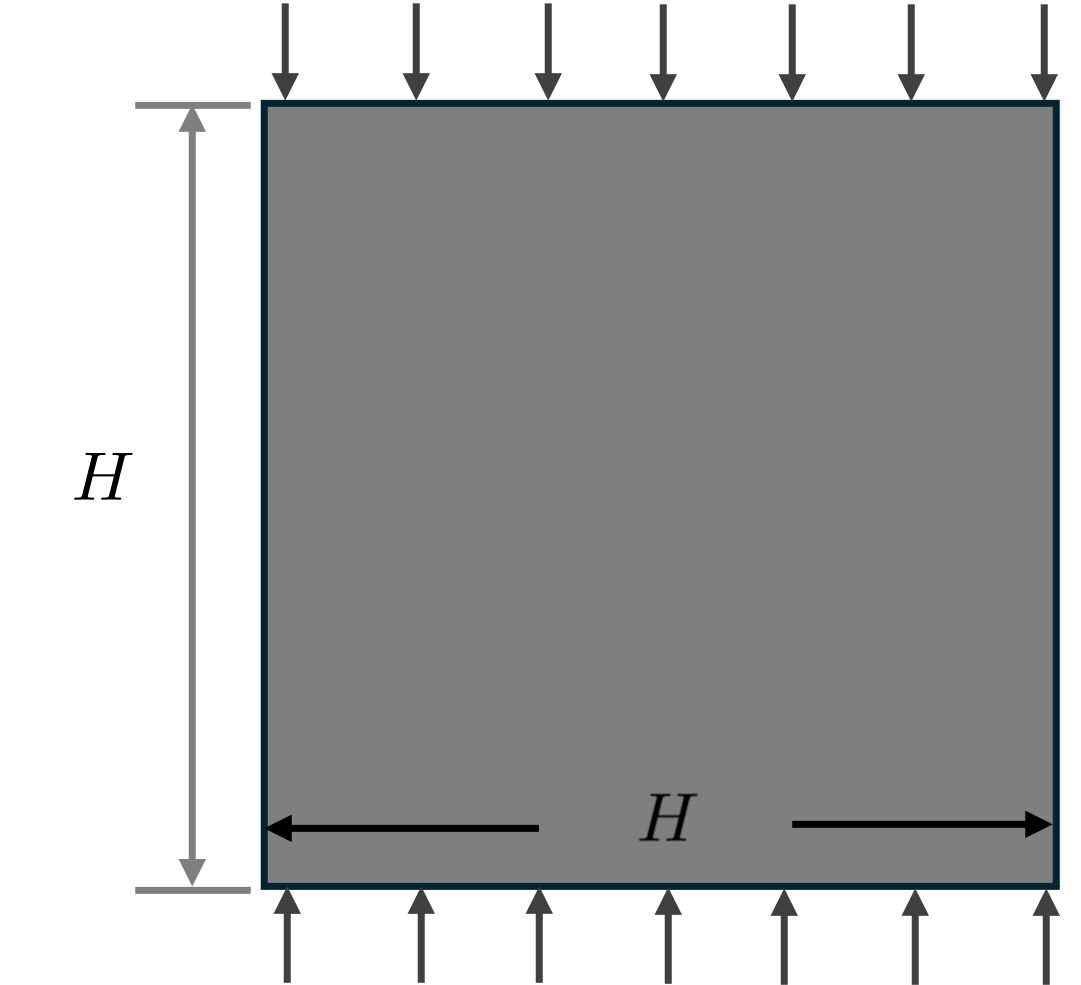}
        \caption{}
        \label{fig:2d_comp}
    \end{subfigure}
    \hfill
    \begin{subfigure}{0.48\textwidth}  
        \centering
        \includegraphics[scale=0.52]{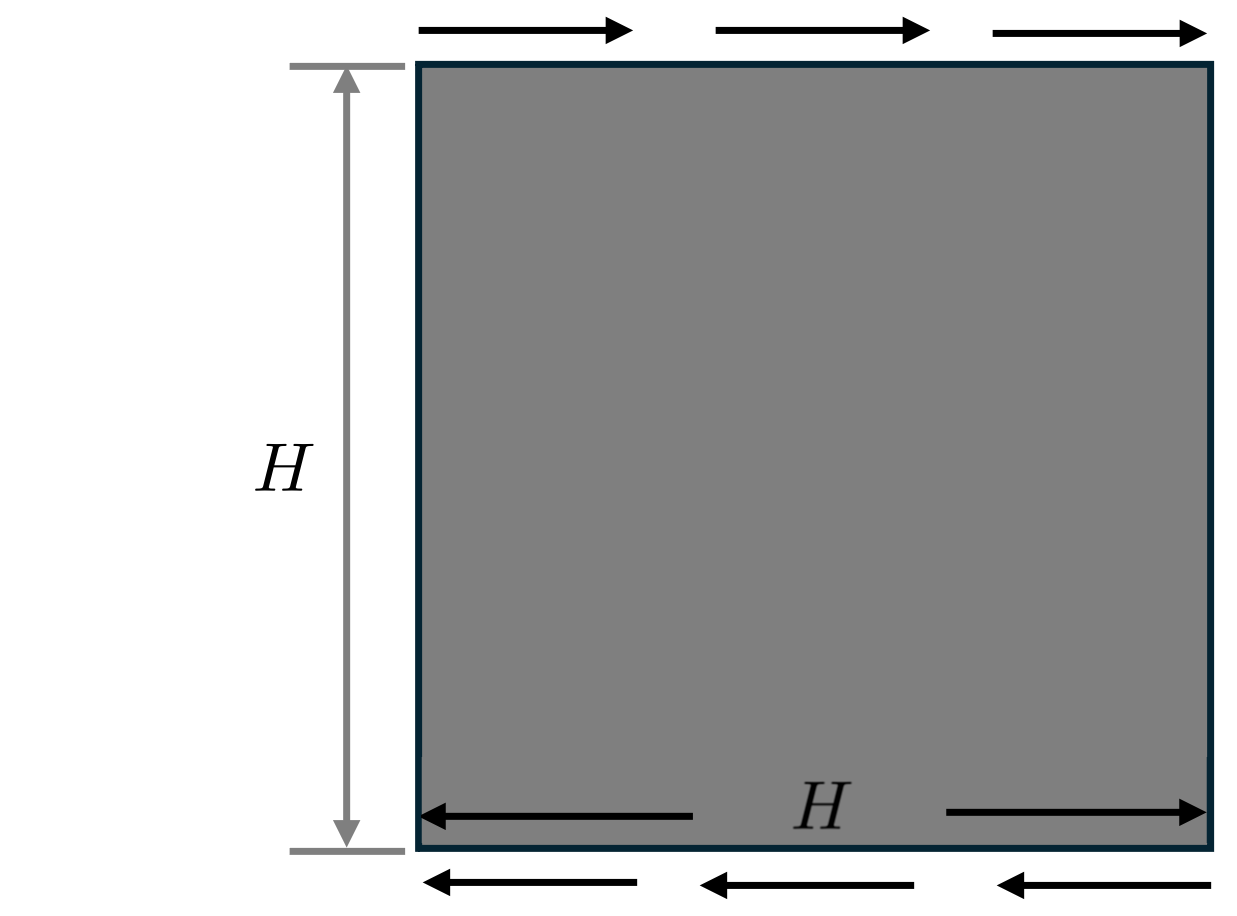}
        \caption{}
        \label{fig:2d_shear}
    \end{subfigure}
    \caption{$ 10 \leq H \leq 160 \,(\mu m)$. (a) Boundary conditions for compression and (b) boundary conditions used for simulating a shear band.  Details of these boundary conditions are listed in Tables~\ref{tab:boundary_conditions_2} and \ref{tab:boundary_conditions_2d_shear}.}
\end{figure}

\begin{table}[htbp]
\centering
\caption{Boundary conditions for Compression.}
\label{tab:boundary_conditions_2}
\begin{adjustbox}{max width=\textwidth}
\begin{tabular}{lll}
\hline
Boundary & Type & Condition \\
\hline
Top & Traction and velocity & $t_x = 0$, $v_y = v_l$ \\
Bottom & Traction and velocity & $t_x = 0$, $v_y = v_r$ \\
Left and Right & Traction-free & $\bft = \mathbf{0}$ \\
\hline
\multicolumn{3}{p{0.9\textwidth}}{\footnotesize $v_y$: $y$-component of velocity; $v_l$, $v_r$: prescribed velocities at the top and bottom boundaries from SHPB data; $t_x$: traction component in the $x$-direction; $\bft$: traction vector. Note: the simulations do not include the unloading phase of the SHPB data.} \\
\hline
\end{tabular}
\end{adjustbox}
\end{table}

\begin{table}[htbp]
\centering
\caption{Boundary conditions applied for shear band studies in 2-d.}
\label{tab:boundary_conditions_2d_shear}
\begin{adjustbox}{max width=\textwidth}
\begin{tabular}{lll}
\hline
Boundary & Type & Condition \\
\hline
Top & Prescribed velocity & $v_x = v_1 = AH/2$, uniform $v_y$ via MPC \\
Bottom & Prescribed velocity & $v_x = v_1 = -AH/2$, uniform $v_y$ via MPC \\
Left and Right & Periodic BC & Velocity periodicity in $x$ and $y$ \\
\hline
\multicolumn{3}{p{0.9\textwidth}}{\footnotesize $v_x$, $v_y$: components of the velocity vector; $A$: applied strain rate; $H$: height of the RVE; MPC: multi-point constraint enforcing uniform vertical displacement. The loading is applied as a step in time, with the velocity equal to zero at $t = 0$ and held constant for all subsequent times.} \\
\hline
\end{tabular}
\end{adjustbox}
\end{table}

\noindent \underline{Calibration for 2-d simulations}: The experimental stress-strain response shown in Fig.~\ref{fig:SS_calib_2} is used to define the initial yield stress, $g_0$, and the Stage II hardening rate $h_0$ of the material model. Elastic properties are obtained from independent material characterization tests.

\section{Results and Discussion}\label{Sec:RnD}

We present and discuss results of various 2-d and 3-d simulations that were performed to understand the behavior of the model. Behavior under boundary and initial conditions that would induce both nominally homogeneous and heterogeneous deformation in quasi-static motions of a homogeneous material are employed in our dynamic simulations. Comparisons with predictions of classical plasticity theory are provided where appropriate. Details of the evolution of the microstructural state of the material are also shown.

We take as an estimate of the initial statistical dislocation density, $\bar{\rho}_s$ (see \cite[Eqn.~5]{Arora_acharya_2} for the definition of the statistical density $\rho_s$ at any time), as 

\begin{equation}
    \label{eq:rho_s_init}
    \bar{\rho}_s = \left(\frac{g_0}{\eta \mu b}\right)^2.
\end{equation}

\subsection{Model setup and material parameters: 2-d simulations}\label{sec:mat_mod_sim}

The RMFDM model requires initial conditions for four field variables: velocity ($\bfv$), GND density ($\bfalpha$), inverse elastic distortion ($\bfW$), and the motion map ($\bfx$). The velocity field is initialized to zero throughout the domain, except at the boundaries where loading is applied. The GND density is initialized to zero. The initial inverse elastic distortion $\bfW$ is set to the transpose of the rotation field that characterizes the polycrystalline microstructure, thereby encoding the crystallographic orientation of each grain. In principle, such an (strain-free) inverse elastic distortion field corresponds to a localized GND distribution at the grain boundaries  through its  $curl$ field. We intentionally do not include this contribution in the initialization of $\bfalpha$ - thus, all subsequent GND accumulation at grain boundaries is an (physically realistic) outcome of our framework.

The polycrystal microstructure used is shown in Fig.~\ref{fig:grain_strct}. In this 2-d setting, a synthetic microstructure is constructed, with crystallographic orientations arbitrarily assigned to each grain\footnote{The synthetic orientation distribution is generated by arbitrarily assigning the in-plane rotation angle for each 2-d grain to be the first angle of its corresponding experimentally determined orientation field.}, and uniform within it. In the 2-d setting, the orientation of each grain is characterized by a single rotation angle. The grain shapes and sizes are obtained from a cross-sectional plane through the center of the full 3-d polycrystal described below.

\begin{figure}
        \centering
        \begin{overpic}[scale=0.165]{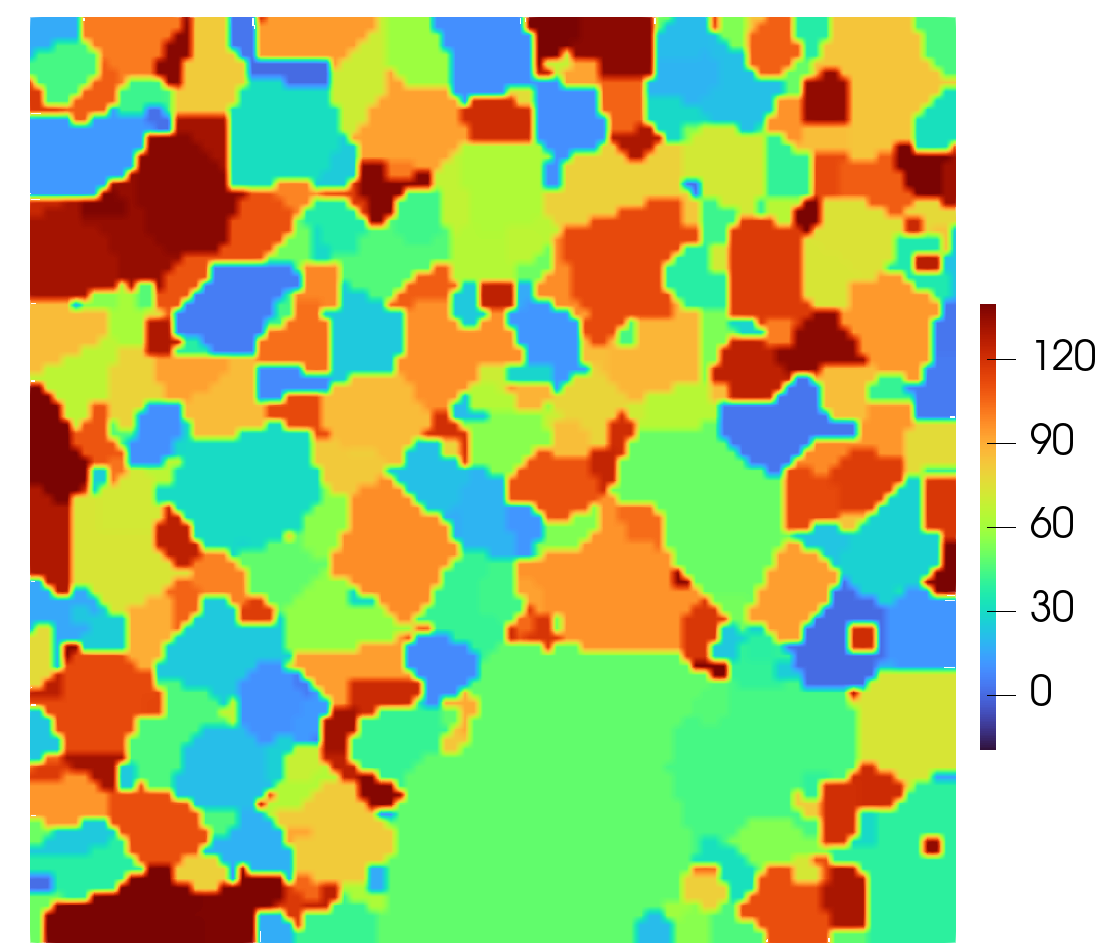} 
        \put(89, 62){{\fontsize{8}{5}\selectfont $(deg.)$}}
        \end{overpic}
        \caption{Rotation angle field in the undeformed configuration. Different colorbar scales are used to highlight the grains.}
    \label{fig:grain_strct}
\end{figure}

Boundary conditions are imposed on both $\bfv$ and $\bfalpha$. For the velocity, periodic boundary conditions, as shown in Fig.~\ref{fig:2d_shear}, are applied on the lateral (side) faces in both directions (Table ~\ref{tab:boundary_conditions_2d_shear}). On the top and bottom faces, a prescribed shear velocity is applied in the horizontal direction, while in the vertical direction, multi-point constraints (MPCs) are used to enforce uniform displacement of the boundary nodes. This constraint ensures that the top and bottom surfaces remain planar throughout the deformation. For the GND density, unconstrained boundary condition (as described in Sec.~\ref{sec:boun_con}) is imposed.

The simulation settings and material parameters are listed in Table.~\ref{table:metal}.

\begin{table}[H]
\centering
\begin{adjustbox}{max width=\textwidth}
\begin{tabular}{|l|l|c|l|}
\hline
\textbf{Model} & \textbf{Parameter} & \textbf{Value} & \textbf{Description} \\[6pt]
\hline
\multirow{12}{*}{\shortstack{Classical Crystal \\ Plasticity}}
& $\hat{\gamma}_0$ (s$^{-1}$) & 0.001 & Reference plastic strain rate (Table.~\ref{tab:constitutive_relation_Lp}) \\
& $m$              & 0.03  & Rate sensitivity exponent (Table.~\ref{tab:constitutive_relation_Lp})\\
& $g_0$ (GPa)      & 0.450 & Initial yield strength (Table.~\ref{tab:constitutive_relation_g})\\
& $g_s$ (GPa)      & 0.450 & Saturation yield strength (Table.~\ref{tab:constitutive_relation_g})\\
& $h_0$ (GPa)       & 0     & Stage 2 hardening rate (Table.~\ref{tab:constitutive_relation_g}) \\
& $E$ (GPa)        & 200   & Young's modulus \\
& $\nu$            & 0.30  & Poisson's ratio \\
& $\rho$ (kg/m$^3$) & 7830 & Mass density \eqref{eq:RMFDM_d} \\
& $b$ (\AA)        & 4.05  & Burgers vector magnitude (Table.~\ref{tab:constitutive_relation_Lp}) \\
& $\eta$        & $\frac{1}{3}$  & Taylor relationship constant for macroscopic strength vs. dislocation density (Table.~\ref{tab:constitutive_relation_Lp}) \\
\hline
\multirow{2}{*}{MFDM (additional)}
& $k_0$            & 0.5   & Constant in coefficient defining GND hardening (Table.~\ref{tab:constitutive_relation_g})\\
& $l$ ($\mu m$)            & 0.1732 & Constant in coefficient defining core energy density (Table.~\ref{tab:constitutive_relation_Lp}) \\

\hline
\multirow{4}{*}{\shortstack{Thermal Parameters \\ (Classical plasticity)}}
& $\theta_0 ~(K)$            & 293   & Reference temperature \eqref{eqn:g_theta} \\
& $p$              & 0.38 & Exponent in thermal softening law \eqref{eqn:g_theta}\\
& $\kappa$              & 0.55 & Taylor-Quinney coefficient \eqref{eq:Beta_form} \\
& $C_p$ (J/Kg .$K$)              & 490 & Specific heat capacity \eqref{eq:Beta_form} \\
& $D_0$              & $\frac{1}{6}$ & Parameter related to elastic moduli softening \eqref{eq:ES_evolve} \\
\hline
\end{tabular}
\end{adjustbox}
\caption{Material parameters used for 2-d simulations, unless mentioned otherwise. $k_0$ has been varied for the size-effect investigation.}
\label{table:metal}
\end{table}
\begin{table}[htbp]
\centering
\caption{Computational details for the 2-d MFDM simulations.}
\label{tab:comp_details_2d}
\begin{tabular}{lcc}
\hline
 & 2-d \\
\hline
Domain size & $80 \times 80~\mu m^2$ \\
Elements & $25 \times 10^4$ \\
Strain reached & 1.80\\
Slip systems & 3 \\
Nominal strain rate & $4 \times 10^3~\text{s}^{-1}$ \\
Cores & 512 \\
Wall-clock time & 8~hrs\\
\hline
\end{tabular}
\end{table}
Planar slip systems are required for 2-d simulations. However, most, if not all, crystal structures (including BCC Iron) lack planes containing multiple slip systems. To approximately accommodate this feature within a 2-d ansatz, we used a least squares projection of the two predominant deformation slip systems onto a single analysis plane (as shown in Fig.~\ref{fig:analyse_slip} (a)).

\begin{figure}[htbp]
    \centering
    \begin{subfigure}{0.45\textwidth}
        \centering
        \includegraphics[scale=0.5]{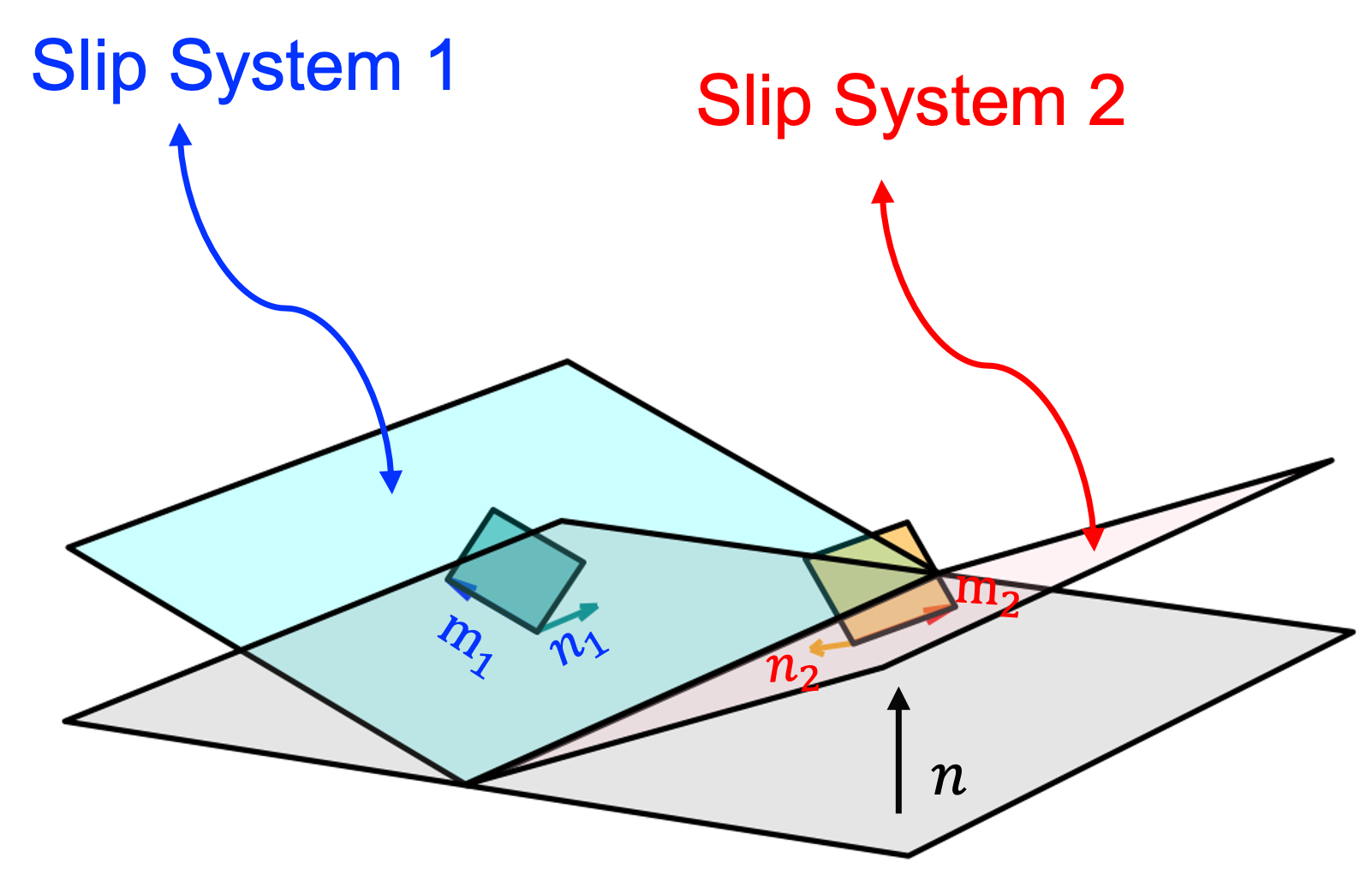}
        \caption{}
    \end{subfigure}
    \begin{subfigure}{0.45\textwidth}
        \centering
        \includegraphics[scale=0.5]{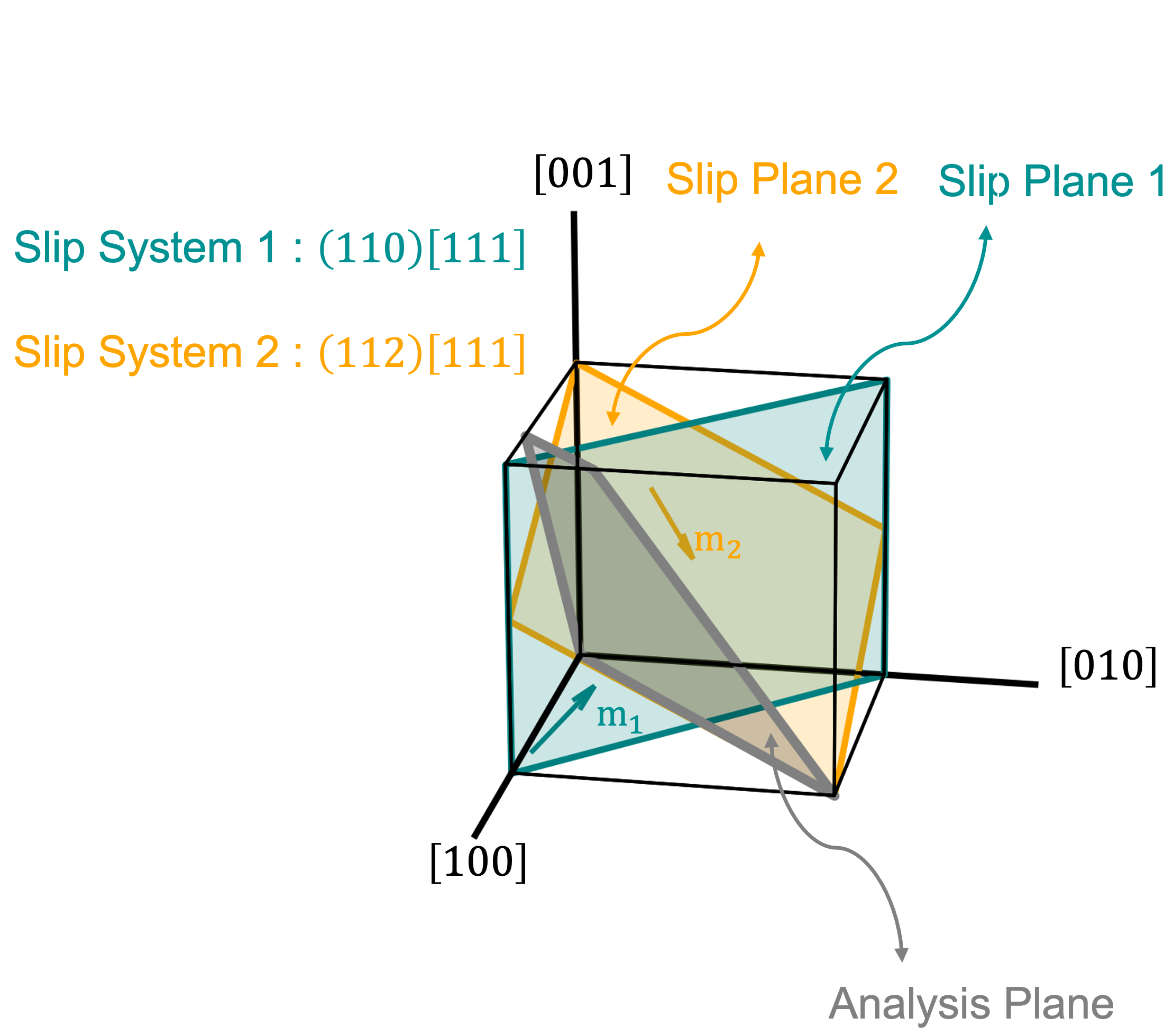}
        \caption{}
    \end{subfigure}
    \caption{\textbf{(a)} Schematic showing the slip systems considered and the corresponding analysis plane. \textbf{(b)} Crystallographic view of the slip systems and the analysis plane.}
    \label{fig:analyse_slip}%
\end{figure}

The analysis plane is defined by the normal, $\bfn$, determined by solving the following optimization,
\begin{eqnarray*}
\begin{aligned}
F(\bfz)&= (\bfn_1 \cdot \bfz)^2+(\bfm_1 \cdot \bfz)^2+(\bfn_2 \cdot \bfz)^2+(\bfm_2\cdot \bfz)^2 \\
    \bfn&=\operatorname*{argmin}_{\bfz, |\bfz| =1} F(\bfz)\\
\end{aligned}
\end{eqnarray*}
where $\bfm_i$, $\bfn_i$ are, respectively, the slip direction and the normal to the slip plane for the $i^{th}$ slip system.

The two projected slip systems capture the dominant deformation modes of the crystal within the analysis plane. A third slip system is then introduced along the remaining in-plane direction to ensure a reasonable level of plasticity.

\subsection{Modeling adiabatic shear bands}\label{subsec:ASB_Comp}

We analyze shear band formation using two models -- Classical Crystal Plasticity (CCP) and MFDM. A key feature of MFDM is its explicit treatment of GND density evolution via the Nye tensor, which naturally incorporates an intrinsic material length scale and GND-induced hardening -- factors of some significance in the shear localization problem. The two models are compared with respect to convergence of results upon mesh refinement and the predicted shear band width against experimental observations, keeping in mind that the deformation of an RVE is studied under boundary conditions suited more for inducing nominally homogeneous deformation (in a homogeneous material under quasi-static conditions).

\begin{figure}[htbp]
    \centering
    \begin{subfigure}{0.48\textwidth}  
        \centering
        \includegraphics[scale=0.5]{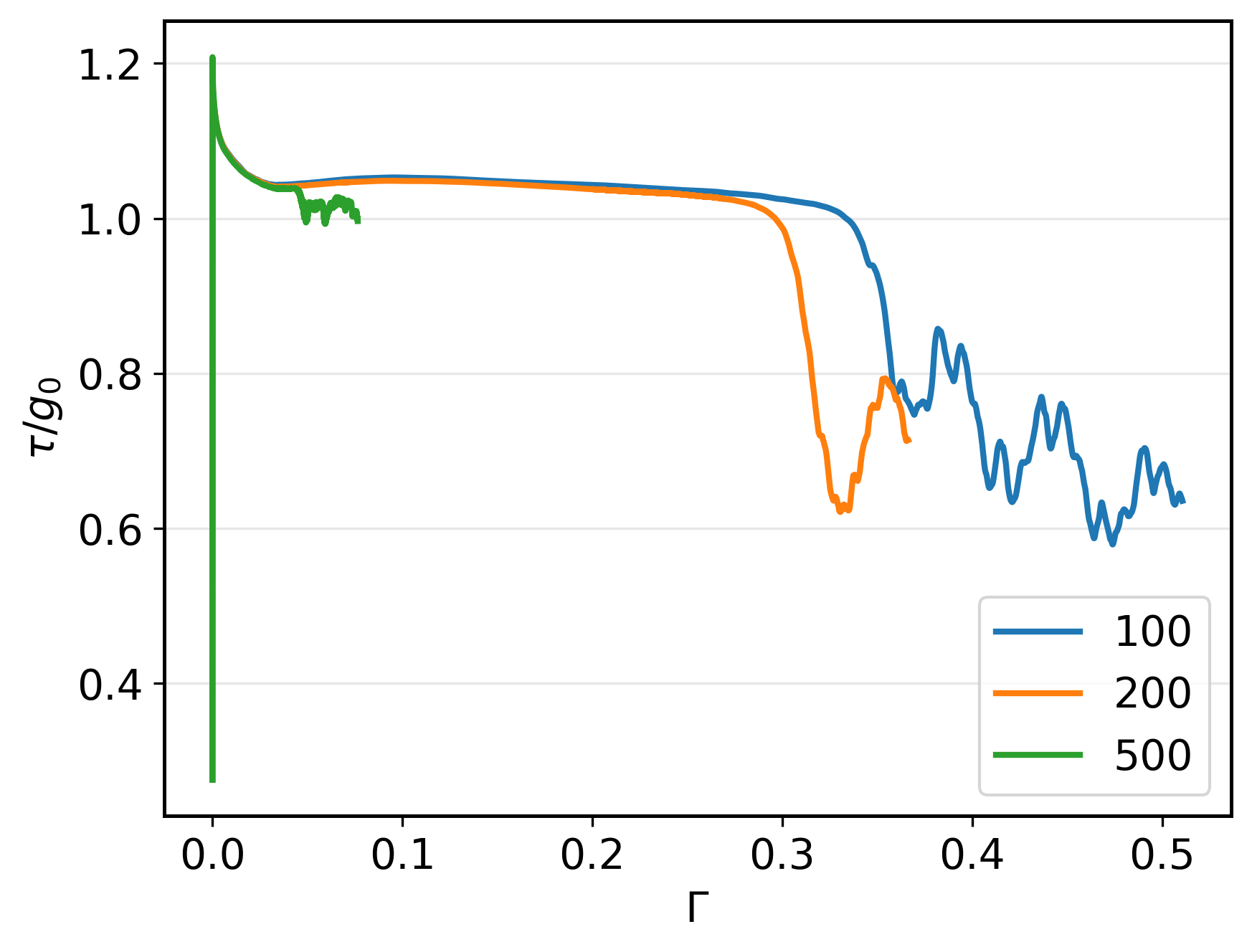}
        \caption{}
        \label{fig:ss_CCP}
    \end{subfigure}
    \hfill
    \begin{subfigure}{0.48\textwidth}  
        \centering
        \includegraphics[scale=0.5]{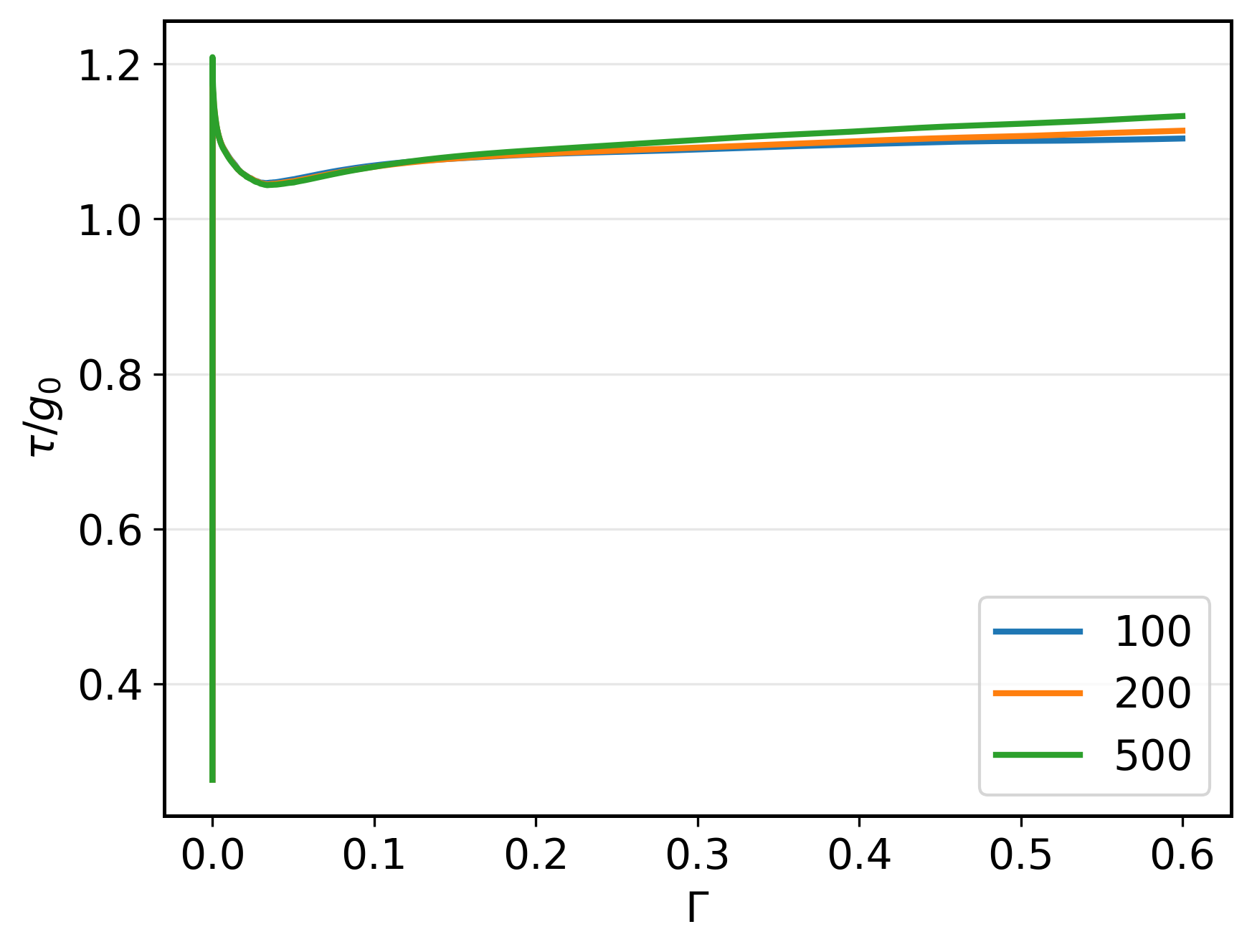}
        \caption{}
        \label{fig:SS_MFDM_3}
    \end{subfigure}
    \caption{Stress-strain curve for (a) Classical crystal plasticity (CCP) and (b) MFDM for mesh refinements of $100^2, 200^2, 500^2$ elements.}
    \label{fig:conv_stress_strain}
\end{figure}

\begin{figure}
    \centering
    \begin{subfigure}{0.48\textwidth}  
        \centering
        \begin{overpic}[scale=0.2]{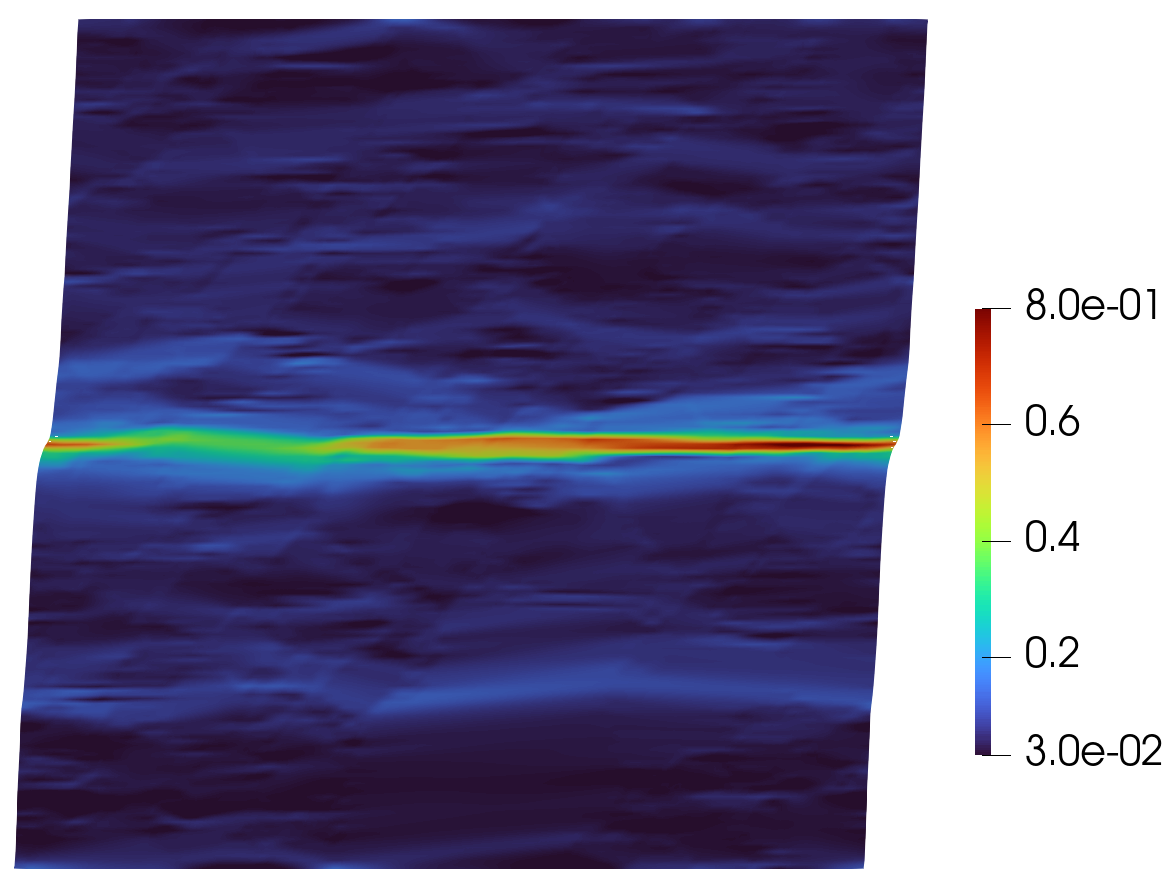}
            \put(82, 56){{\fontsize{8}{5}\selectfont $F_{12}$}}
        \end{overpic}
        \caption{}
        \label{fig:CCP_LTQ_4k}
    \end{subfigure}
    \hfill
    \begin{subfigure}{0.48\textwidth}  
        \centering
        \begin{overpic}[scale=0.2]{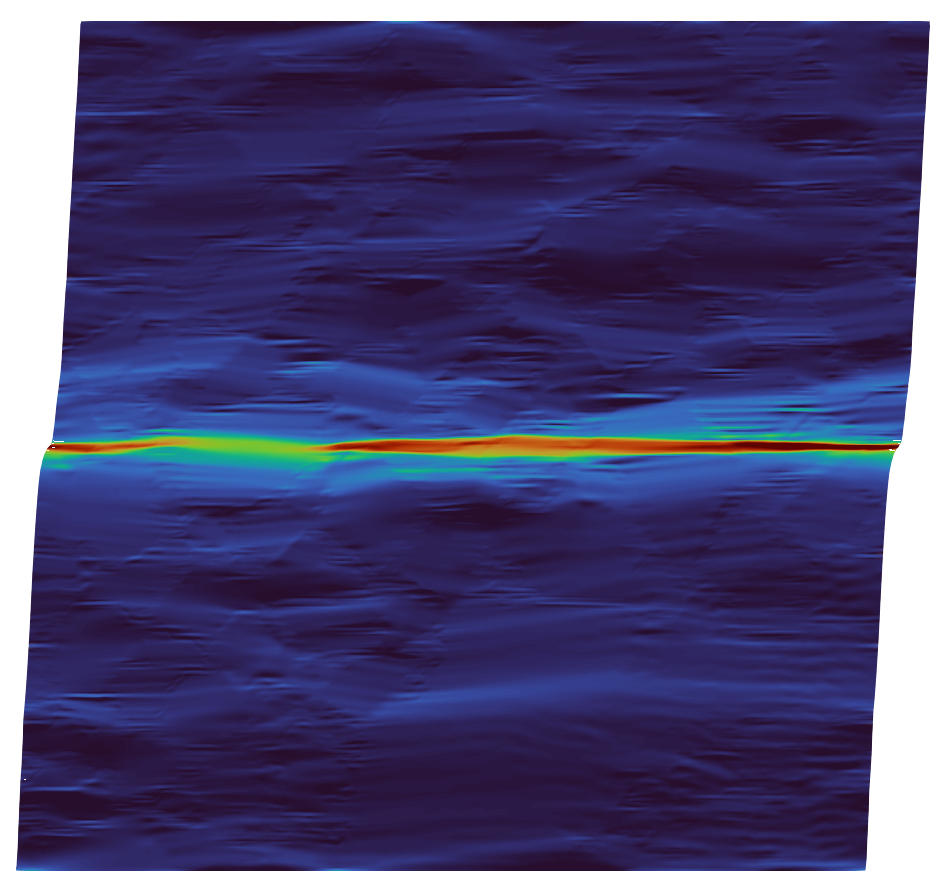}  
        \end{overpic}
        \caption{}
        \label{fig:CCP_}
    \end{subfigure}
    \caption{
    The (shear) $F_{12}$ component of the deformation gradient using the Classical crystal plasticity formulation for (a) a coarse mesh ($200^2$ elements) at an applied strain of 0.3 and (b) a fine mesh ($500^2$ elements) at an applied strain of 0.08. The simulation on the fine mesh could not progress further.}
    \label{fig:F_12_CCP}
\end{figure}

\begin{figure}
    \centering
    \begin{subfigure}{0.48\textwidth}  
        \centering
        \begin{overpic}[scale=0.175]{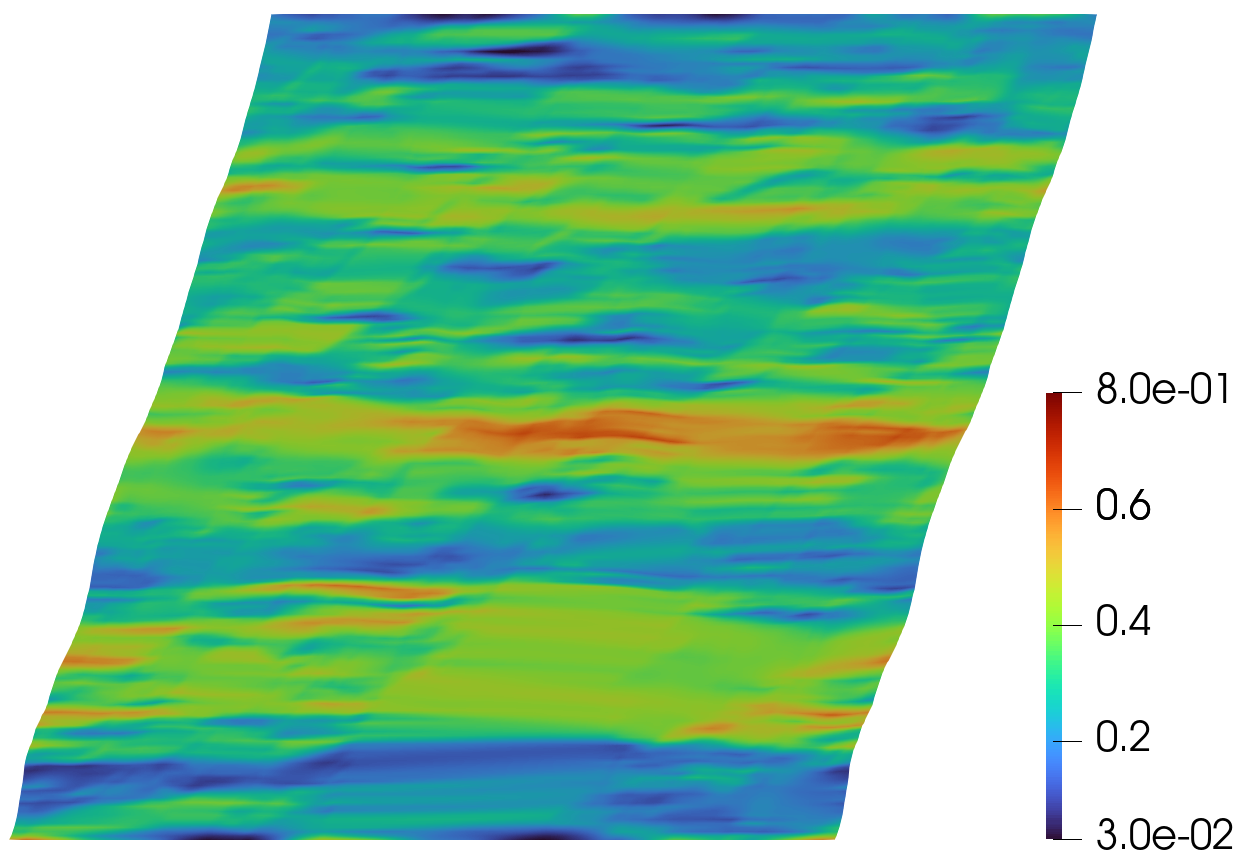}
            \put(82, 42){{\fontsize{8}{5}\selectfont $F_{12}$}}
        \end{overpic}
        \caption{}
        \label{fig:MFDM_LTQ_13k}
    \end{subfigure}
    \hfill
    \begin{subfigure}{0.48\textwidth}  
        \centering
         \begin{overpic}[scale=0.175]{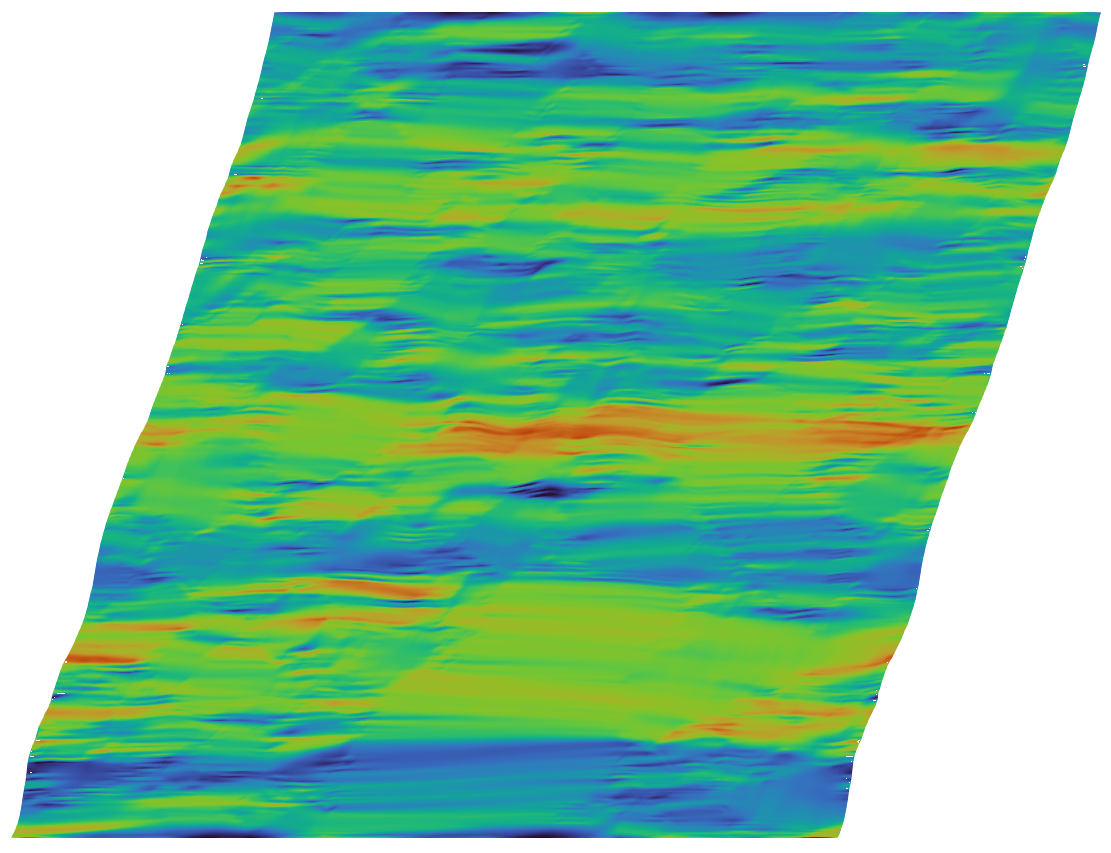}
        \end{overpic}
        \caption{}
        \label{fig:MFM_fine}
    \end{subfigure}
    \caption{The (shear) $F_{12}$ component of the deformation gradient at an applied strain of 0.3 using the MFDM formulation for (a) a coarse mesh ($200^2$ elements)  and (b) a fine mesh ($500^2$ elements).}
    \label{fig:F_12_MFDM}
\end{figure}

The stress-strain curve is obtained by computing the engineering shear strain from the relative lateral displacement between the top and bottom boundaries and the shear stress from the $x$-component of the reaction force on the top boundary.
\begin{equation*}
    \begin{aligned}
    \Gamma &= \frac{\Delta u}{H}, \\
    \tau &= \frac{r_{x}}{S},
    \end{aligned}
\end{equation*}
where $\Gamma$ is the engineering shear strain, $\Delta u$ is the relative lateral displacement between the top and bottom boundaries, $H$ is the height of the domain, $\tau$ is the shear stress, $r_{x}$ is the $x$-component of the reaction force on the top boundary, and $S$ is the area of the top boundary.

We first examine the convergence of the stress-strain response in Fig.~\ref{fig:conv_stress_strain} under progressive mesh refinement, with mesh densities of 100, 200, and 500 elements per direction. The oscillations seen in Fig.~\ref{fig:ss_CCP} are the result of elastic wave interactions with the boundary when signficant portions of the domain have undergone elastic unloading, observed here primarily due to the (small) size of the domain. The size involved also prevents a direct comparison of the stress-strain response with that of an experiment.

 As shown in Fig.~\ref{fig:ss_CCP}, the stress-strain curve for the CCP case diverges with refinement, and the simulation eventually fails due to the localization of shear in a narrow band (Fig.~\ref{fig:F_12_CCP}). In contrast, the stress-strain behavior in Fig.~\ref{fig:SS_MFDM_3} obtained using the MFDM approach remains nearly identical for all mesh sizes as shown. 

The non-convergence in the stress-strain response of CCP (Fig.~\ref{fig:ss_CCP}) is also observed in the deformation fields in the domain (Fig.~\ref{fig:F_12_CCP}). MFDM produces a significantly more homogeneous deformation field in comparison (Fig.~\ref{fig:F_12_MFDM}) with no localization (as defined in Sec.~\ref{sec:intro}), and the field remains approximately consistent across both mesh resolutions. The field produced by CCP localizes into a narrow band across the domain, accompanied by increased strain concentration within the band with refinement (Fig.~\ref{fig:F_12_CCP}): the maximum value of $F_{12}$ inside the band increases from approximately $0.88$ on the $200^2$ mesh to approximately $1.20$ for the finer mesh, a roughly $32\%$ increase in strain. This difference arises because classical crystal plasticity lacks an intrinsic material length scale\footnote{The ratio of the grain size to the domain size introduces a non-dimensional parameter related to lenght scales, but this is not sufficient to influence the non-convergence of the strain field.}. As shown in \cite{NEEDLEMAN1988}, the strain within the band will keep increasing with refinement, even though with rate-dependence and inertia, the band width can attain a converged value. MFDM, however, has an intrinsic length scale through the Burgers' vector involved in defining the GND hardening term (Table.~\ref{tab:constitutive_relation_g}) resulting in mesh-independent fields. 

\subsection{Size effect} \label{sec:size_2d}

\begin{figure}[htbp]
    \centering
    \begin{subfigure}[t]{0.32\textwidth}  
        \centering
        \includegraphics[width=\textwidth]{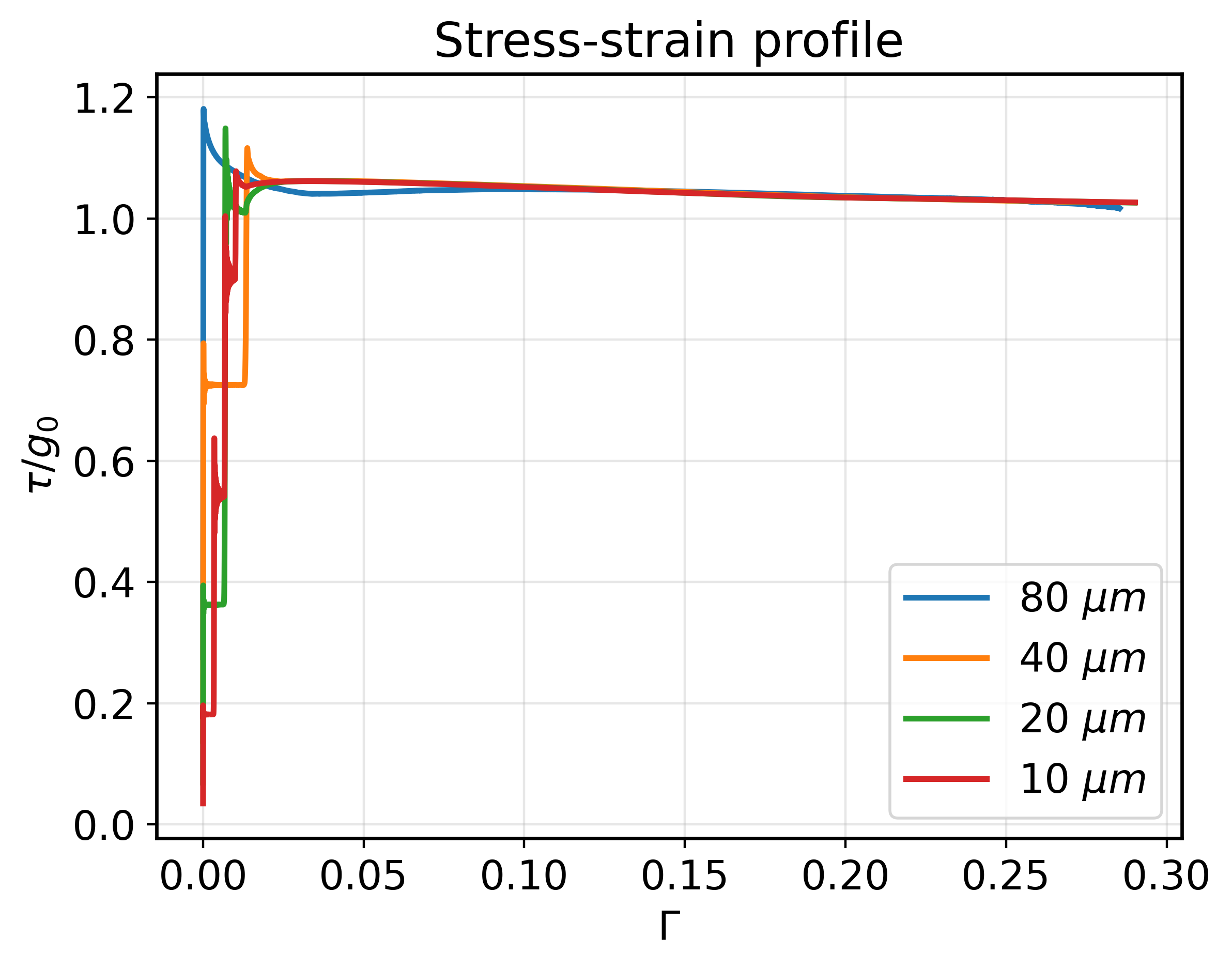}
        \caption{}
        \label{fig:Size_effect_CCP}
    \end{subfigure}
    \hfill
    \begin{subfigure}[t]{0.32\textwidth}  
        \centering
        \includegraphics[width=\textwidth]{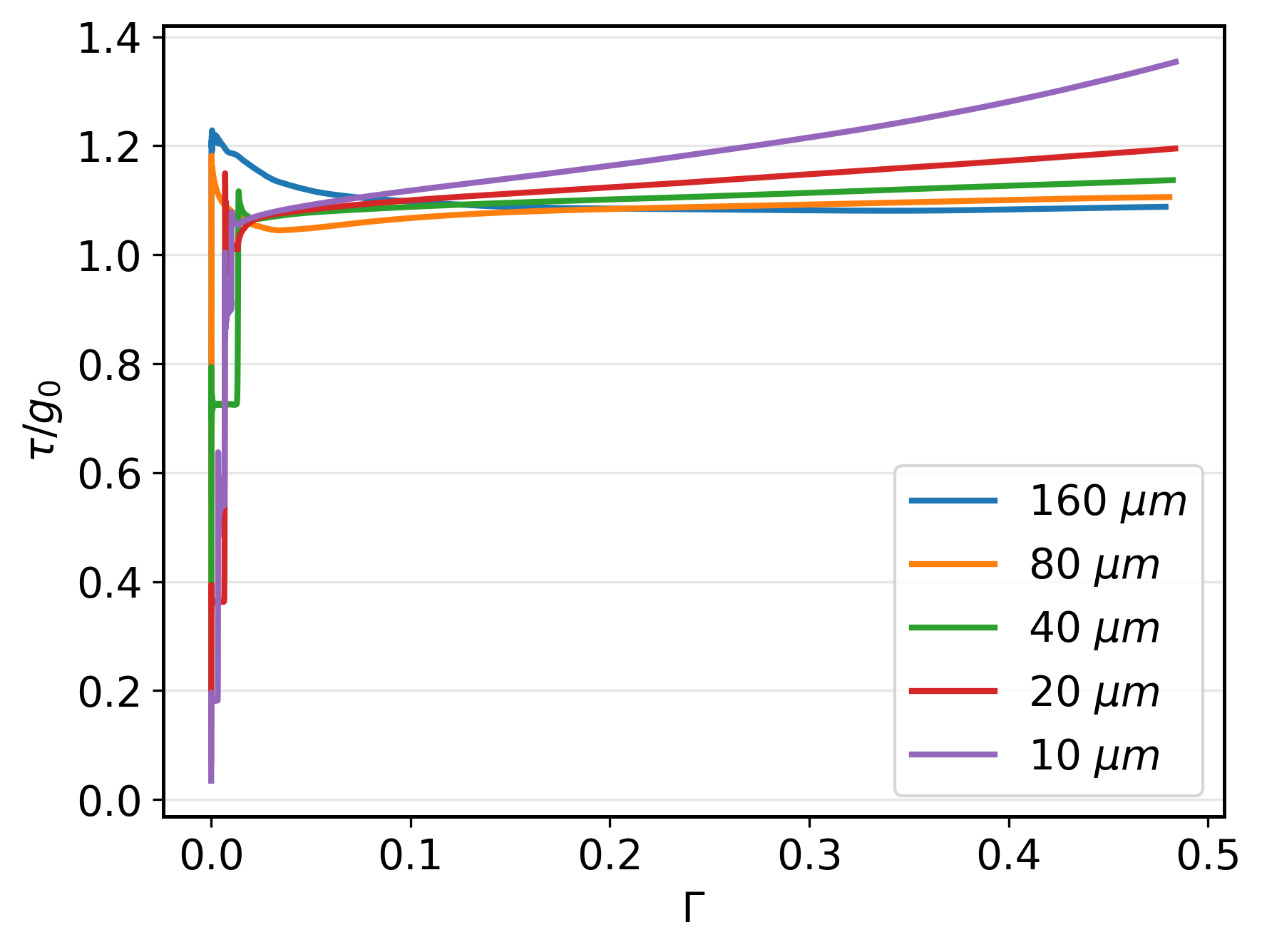}
        \caption{}
        \label{fig:Size_Effect_MFDM}
    \end{subfigure}
    \hfill
    \begin{subfigure}[t]{0.32\textwidth}  
        \centering
        \includegraphics[width=\textwidth]{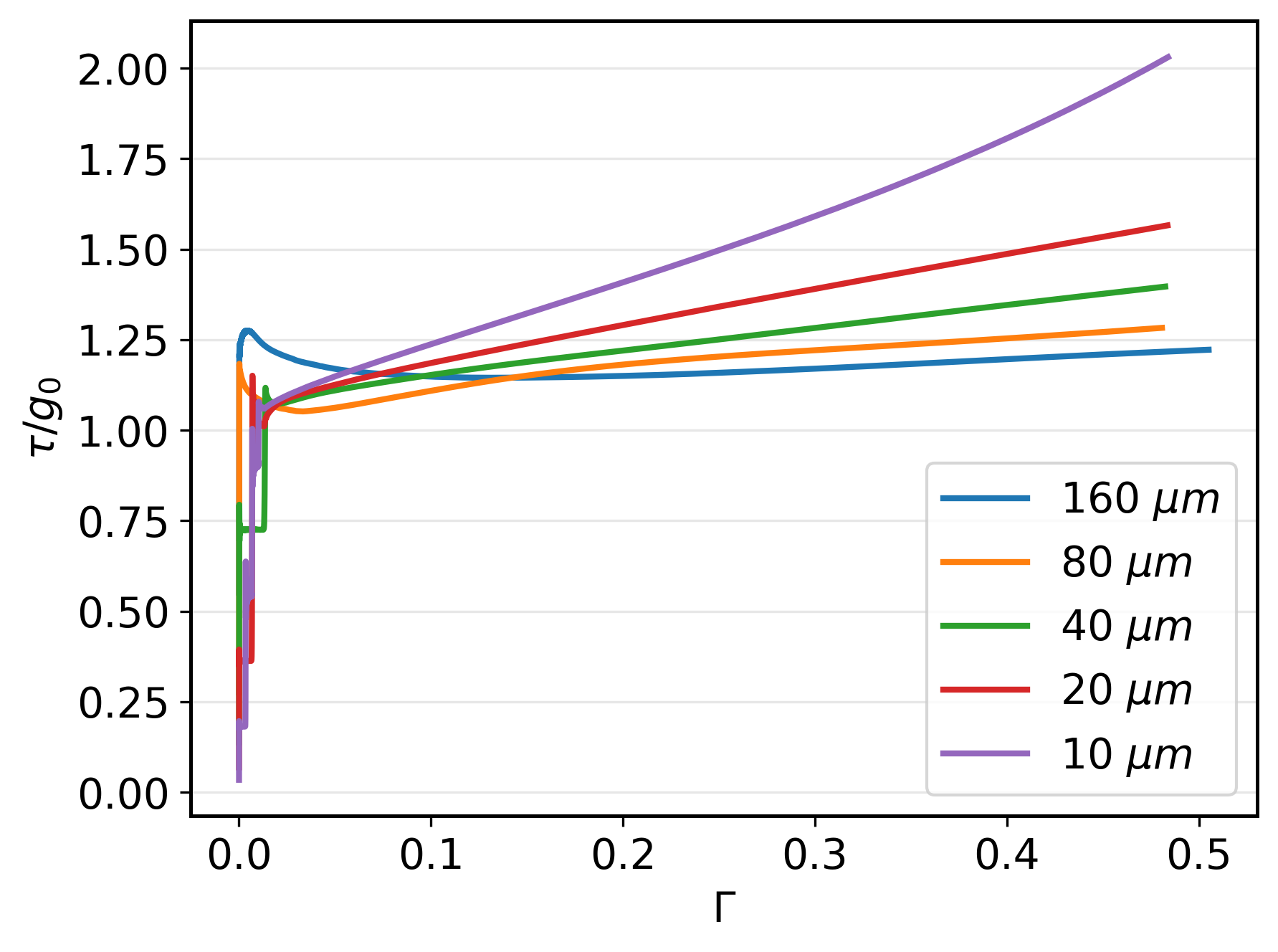}
        \caption{}
        \label{fig:Size_effect_k0_vary}
    \end{subfigure}
    \caption{Stress-strain curves for varying domain sizes using (a) classical crystal plasticity (CCP), (b) MFDM, and (c) MFDM with higher $k_0$ value. The average grain size is $\sim$ one-tenth of the domain size mentioned in the plots.}
    \label{fig:Size_effect}
\end{figure}

Next, we briefly present results that demonstrate the size effect captured at this length scale. Experiments have established that metallic materials exhibit a pronounced size-dependent strengthening response: as specimen dimensions decrease, the material becomes harder, providing greater resistance to plastic flow \cite{Exp_Size_effect,Size_effect_2}. To explore this effect in the present context, we perform a series of simulations in which the domain size is scaled from 10 to 160 $\mu m$ in a self-similar manner, maintaining a constant ratio of grain size to domain size of approximately $1/10$.

Classical crystal plasticity predicts size-independent behavior and thus cannot capture experimentally observed phenomena, as shown in Fig.~\ref{fig:Size_effect}. MFDM, by incorporating the Burgers vector as an intrinsic length scale, and accounting for the contributions of GNDs and their effect in hardening, enables the prediction of size-dependent mechanical response. This effect is more pronounced as we increase the material parameter, $k_0$; an example for the same is shown in the Fig.~\ref{fig:Size_effect_k0_vary}. This observation aligns with the understanding that, accounting more for GND hardening results in more pronounced size effects. Size effects have been simulated in different problems using MFDM. Here, we have shown it in the context of shear banding. As alluded to earlier, in all stress-strain traces shown in Fig.~\ref{fig:Size_effect}, the small plateaus are the result of elastic waves from the interior impinging on the top surface where the reaction force is calculated. 

Additionally, Fig.~\ref{fig:Size_Effect_MFDM} shows that as the domain size increases (with grain size maintained at one-tenth of the domain size), the size effect diminishes and eventually saturates. This behavior is consistent with experimental observations, where size effects become negligible in the limit of large grain sizes \cite{Exp_Size_effect}.

\subsection{Microstructure evolution}

We present results characterizing the possible microstructure evolution in the steel microstructure, with the aim of studying the dynamic behavior and failure mechanisms within the shear band. Adiabatic shear bands are often precursors to fracture in high-strain-rate applications such as ballistic impact, machining, crashworthiness. Understanding microstructure evolution spatially can help identify early signatures of localization before catastrophic failure, enabling better predictive models for structural integrity. The state variables used to describe this evolution are GND density ($\rho_g$), material strength ($g_\theta$), the magnitude of the stress deviator (${\sigma}'$), and temperature ($\theta$). Fields are shown at overall strain levels of $0$, $0.15$, and $0.30$, corresponding to the stress-strain response in Fig.~\ref{fig:SS_MFDM_3}. We will discuss these results.

\subsection*{GND evolution}

The GND density is calculated using the measure 
\[
\rho_g = \sqrt{\frac{|\bfalpha|}{b}},
\]
where $b$ is the magnitude of the Burgers vector. Fig.~\ref{fig:SE_GND} shows the spatial distribution of the GND density on a logarithmic scale. As shown in Fig.~\ref{fig:SE_GND}(a) and mentioned earlier, the simulation begins with zero initial GND density throughout the polycrystal, i.e., no pre-existing GNDs are assumed within the grains or at the grain boundaries. Of course, the initial statistical dislocation density in the material is non-vanishing \eqref{eq:rho_s_init} and is given, for the 2-d simulations,
by
\[
\bar{\rho}_s \approx 1.995 \times 10^{15} \, m^{-2}.
\]
As deformation proceeds, GND density accumulates near grain boundaries. This occurs because incompatibility in the SD  plastic strain rate, $\bfL^p$, acts as a source for the evolution equation for GND density \eqref{alpha_conservation}. MFDM theory imposes a jump condition on plastic strain rate ($\bfalpha \times \bfV + \bfL^p$) at interfaces where it may be discontinuous \cite{Ach_07,PDA11}. This jump condition depends on all 5 grain boundary parameters and naturally produces a greater degree of blocking of plastic flow at interfaces with higher misorientation \cite{PDA11}. With continued shearing, this leads to greater GND accumulation within grains near grain boundaries, and patterned structures emerge within grain interiors.

\begin{figure}[htbp]
    \centering
    \begin{subfigure}{0.32\textwidth}  
        \centering
        \begin{overpic}[scale=0.13]{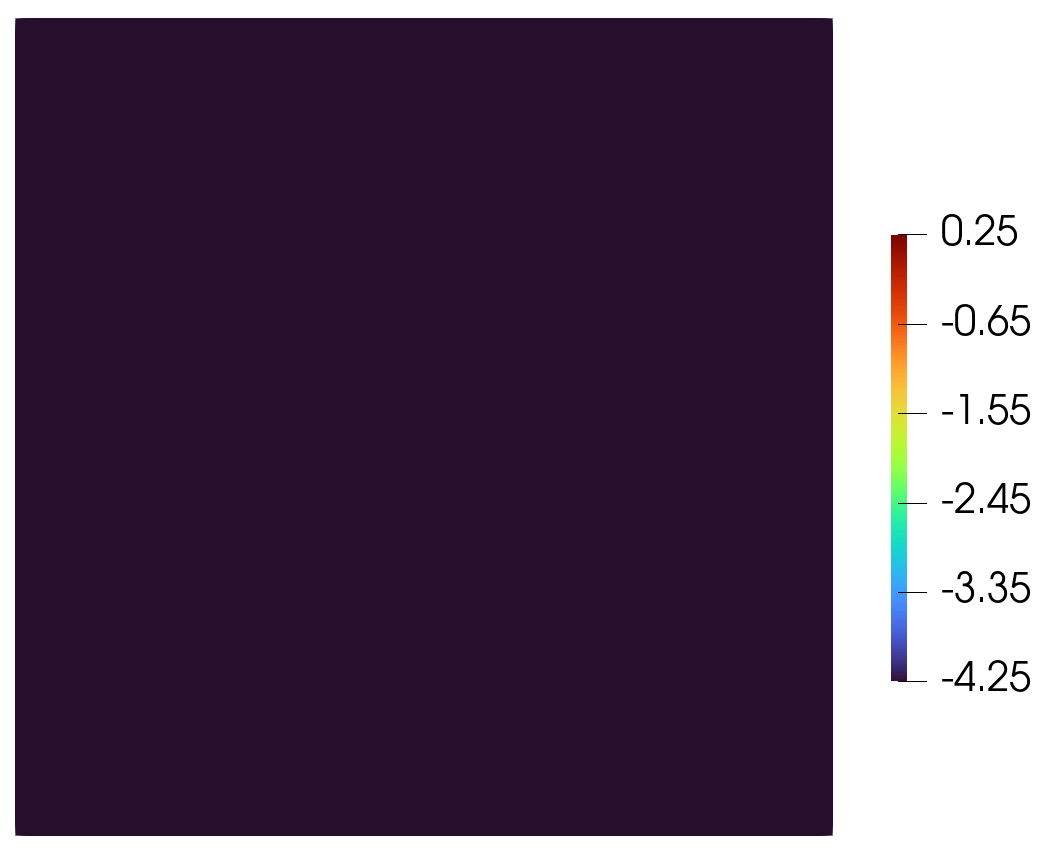}
            \put(83, 64){\fontsize{8}{7}\selectfont $\log(\rho_g/\bar{\rho}_s)$}
        \end{overpic}
        \caption{}
        \label{fig:gnd_0}
    \end{subfigure}
    \hfill
    \begin{subfigure}{0.32\textwidth}  
        \centering
        \begin{overpic}[scale=0.13]{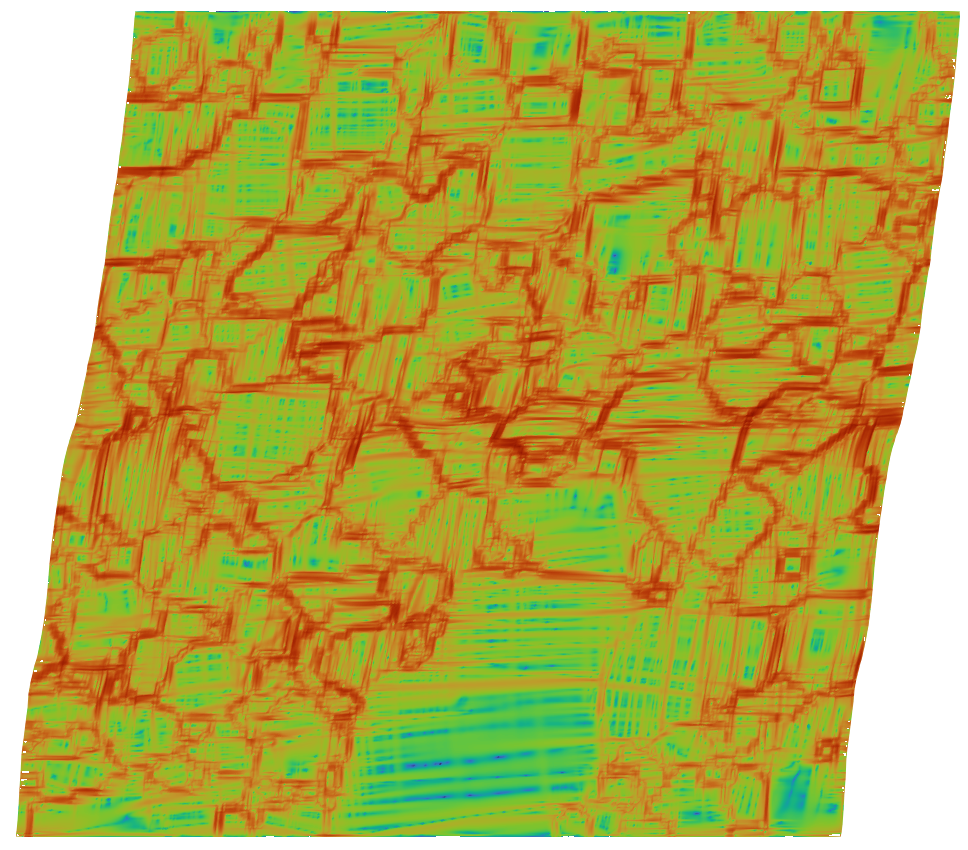}
        \end{overpic}
        \caption{}
        \label{fig:gnd_1}
    \end{subfigure}
    \hfill
    \begin{subfigure}{0.32\textwidth}  
        \centering
        \begin{overpic}[scale=0.13]{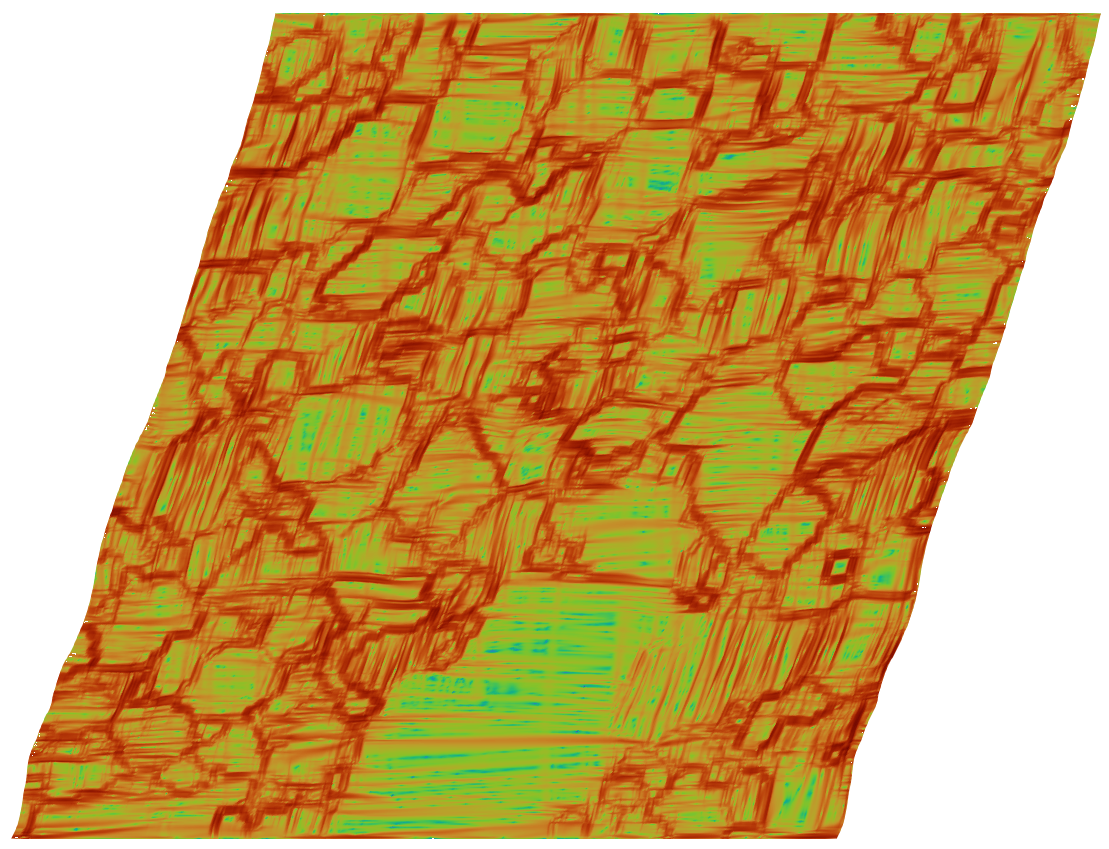}
        \end{overpic}
        \caption{}
        \label{fig:gnd_2}
    \end{subfigure}
    \caption{Evolution of GND density plotted on a logarithmic scale ($\log(\rho_g/\bar{\rho}_s)$) at applied strains of a) 0.0 b) 0.15 and c) 0.30.}
    \label{fig:SE_GND}
\end{figure}

\subsection*{Evolution of material strength}
Fig.~\ref{fig:SE_g} shows the evolution of material strength in the simulation. Initially, the strength field is uniform across the domain and remains so until the onset of plastic deformation as expected. Recall that, based on the macroscopic experimental results, Figs.~\ref{fig:SS_calib_2} and \ref{fig:SS_calib_3}, we do not consider any hardening due to the SDs, i.e., $h_0 = 0$, Table~\ref{tab:constitutive_relation_Lp}. MFDM captures hardening arising from GND accumulation, which serves as the sole source of spatial variation in strength. As the simulation proceeds, grain boundaries become progressively harder relative to grain interiors, consistent with the higher GND densities generated at these interfaces Fig.~\ref{fig:SE_GND}.

The physical implication of this grain boundary hardening is that it creates a mechanical heterogeneity where boundaries act as barriers to dislocation motion, impeding plastic flow across grains. Nevertheless, this spatially localized hardening in concert with thermal softening produces no net hardening in the stress-strain response of the RVE (see Fig.~\ref{fig:conv_stress_strain}b, and \ref{fig:Size_effect}b) for domain sizes $40 \, \mu m.$ and above.

\begin{figure}[htbp]
    \centering
    \begin{subfigure}{0.32\textwidth}  
        \centering
        \begin{overpic}[scale=0.13]{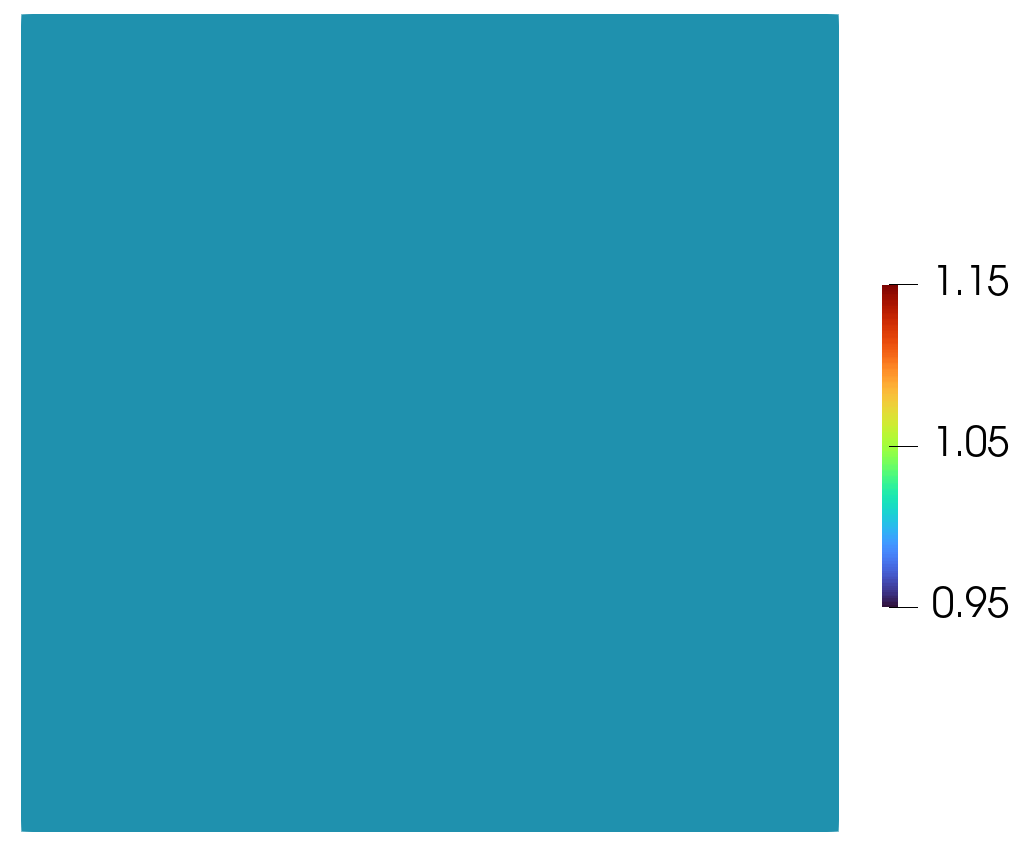}
            \put(86, 64){\fontsize{8}{7}\selectfont $g_{\theta}/g_0$}
        \end{overpic}
        \caption{}
        \label{fig:g_0}
    \end{subfigure}
    \hfill
    \begin{subfigure}{0.32\textwidth}  
        \centering
        \begin{overpic}[scale=0.13]{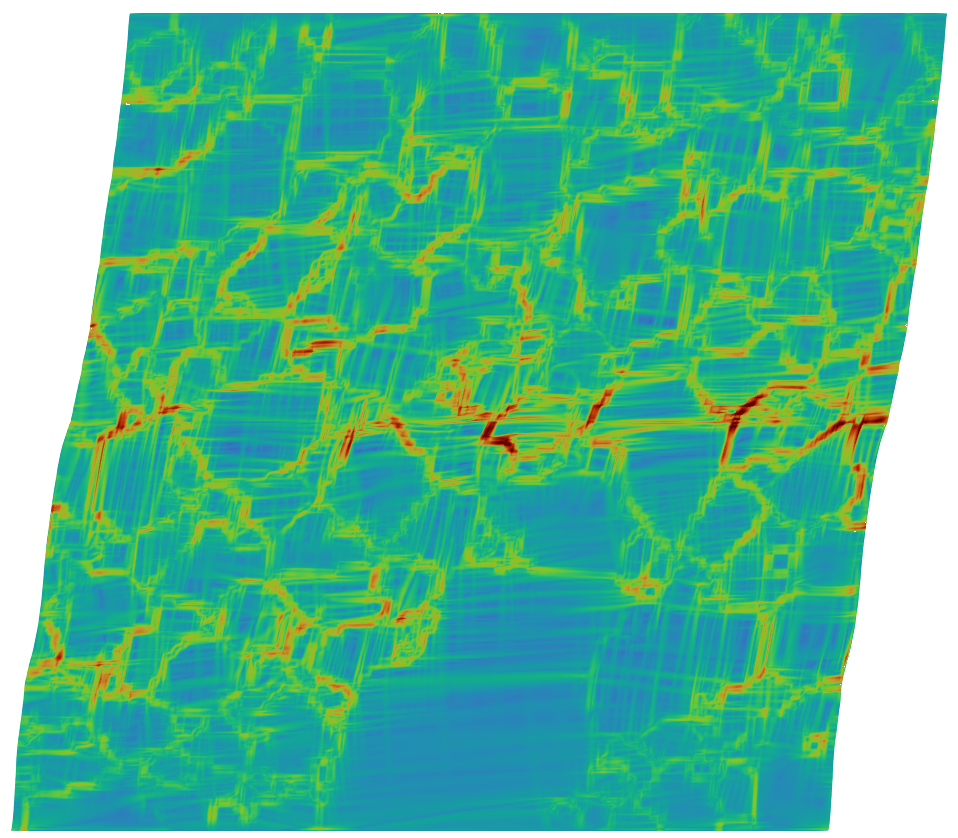}
        \end{overpic}
        \caption{}
        \label{fig:b_SE_g}
    \end{subfigure}
    \hfill
    \begin{subfigure}{0.32\textwidth}  
        \centering
        \begin{overpic}[scale=0.13]{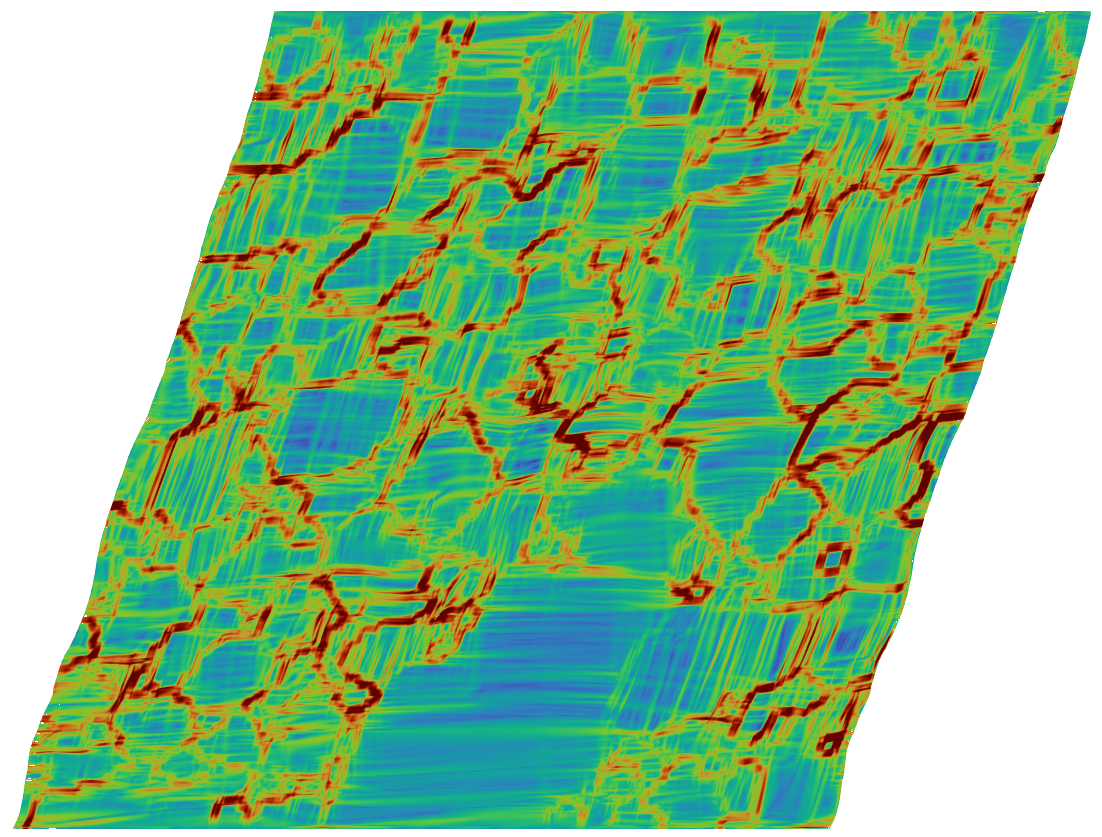}
        \end{overpic}
        \caption{}
        \label{fig:g_2}
    \end{subfigure}
    \caption{Evolution of material strength ($g_{\theta}/g_0$) plotted at applied strains of a) 0.0 b) 0.15 and c) 0.30, where $g_0$ is the initial yield strength.}
    \label{fig:SE_g}
\end{figure}

\subsection*{Evolution of stress deviator}

Subsequently, we present the microstructural evolution in terms of stress deviator, normalized by the material strength, as shown in Fig.~\ref{fig:SE_dev_stress}. Since plasticity is driven by resolved shear stress, which is directly related to the stress deviator, studying its evolution reveals which grains are more favorably oriented for slip and may act as potential failure sites. This also helps us understand differences in plastic strain rate within and across grains. Preferential plastic activity in certain grains reduces the local deviatoric stress due to stress relaxation from plastic deformation (when unconstrained by other kinematic constraints arising from grain boundaries or external boundary conditions).

\begin{figure}[htbp]
    \centering
    \begin{subfigure}{0.32\textwidth}  
        \centering
        \begin{overpic}[scale=0.13]{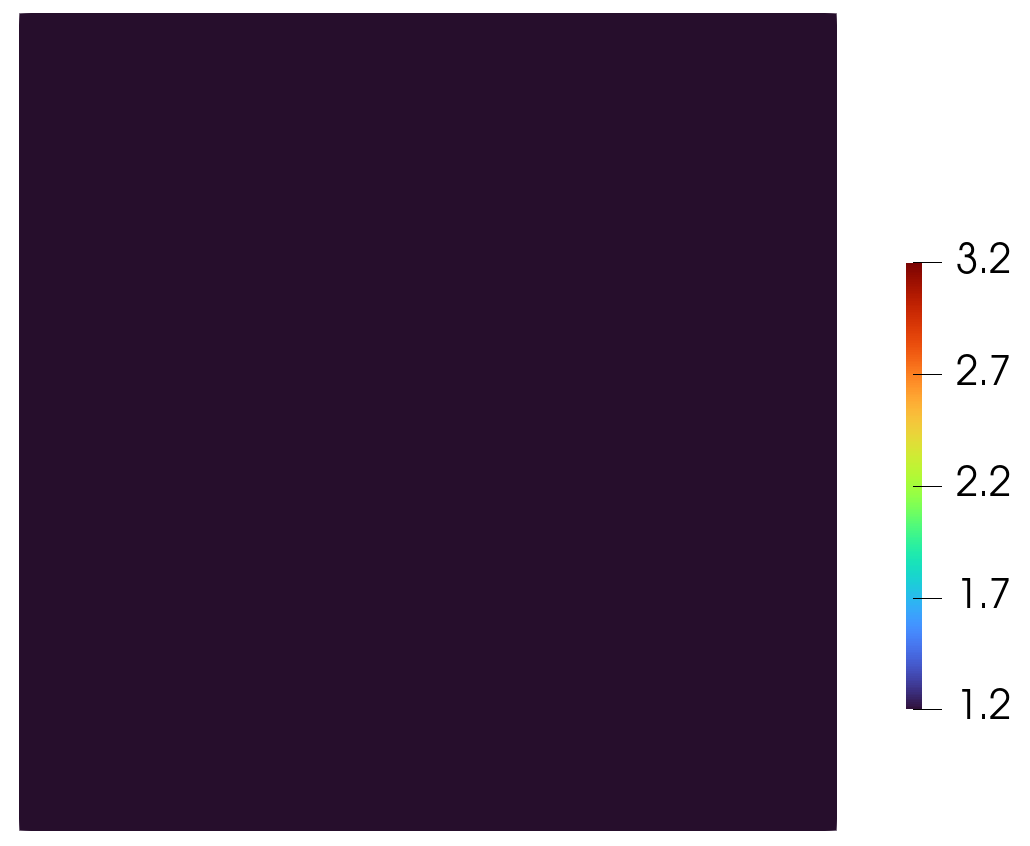}
            \put(85, 65){\fontsize{8}{7}\selectfont $\sigma^{'}\!\!/g_{0}$}
        \end{overpic}
        \caption{}
        \label{fig:S12_0}
    \end{subfigure}
    \hfill
    \begin{subfigure}{0.28\textwidth}  
        \centering
        \includegraphics[width=\textwidth]{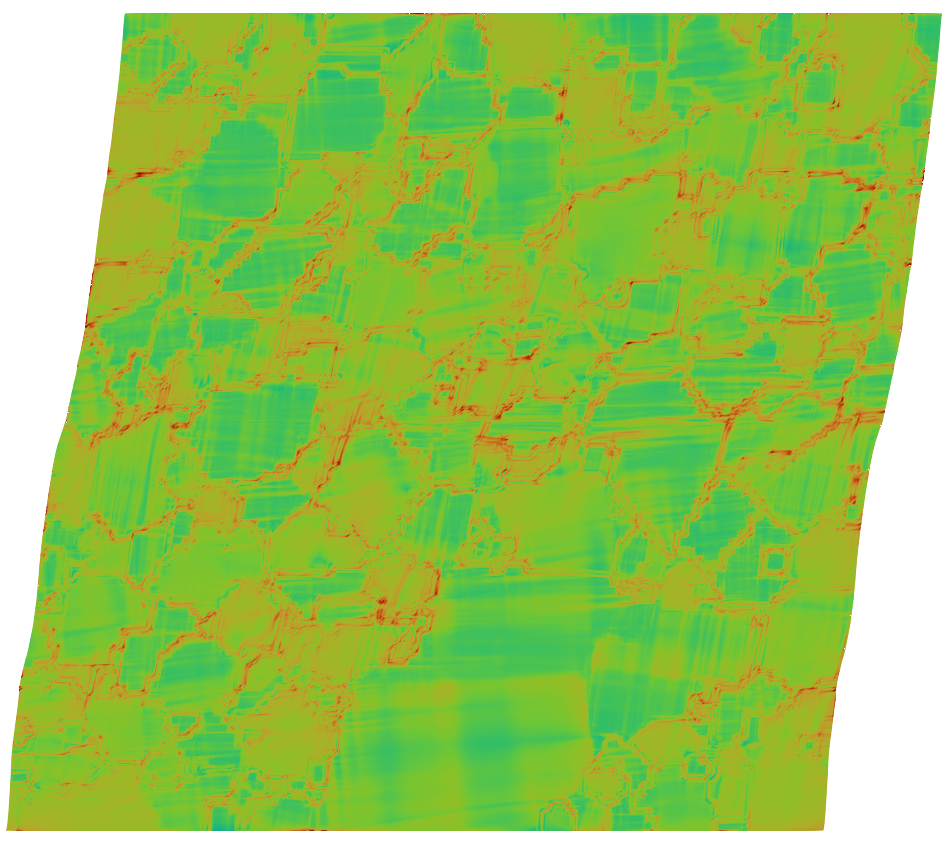}
        \caption{}
        \label{fig:S12_1}
    \end{subfigure}
    \hfill
    \begin{subfigure}{0.32\textwidth}  
        \centering
        \includegraphics[width=\textwidth]{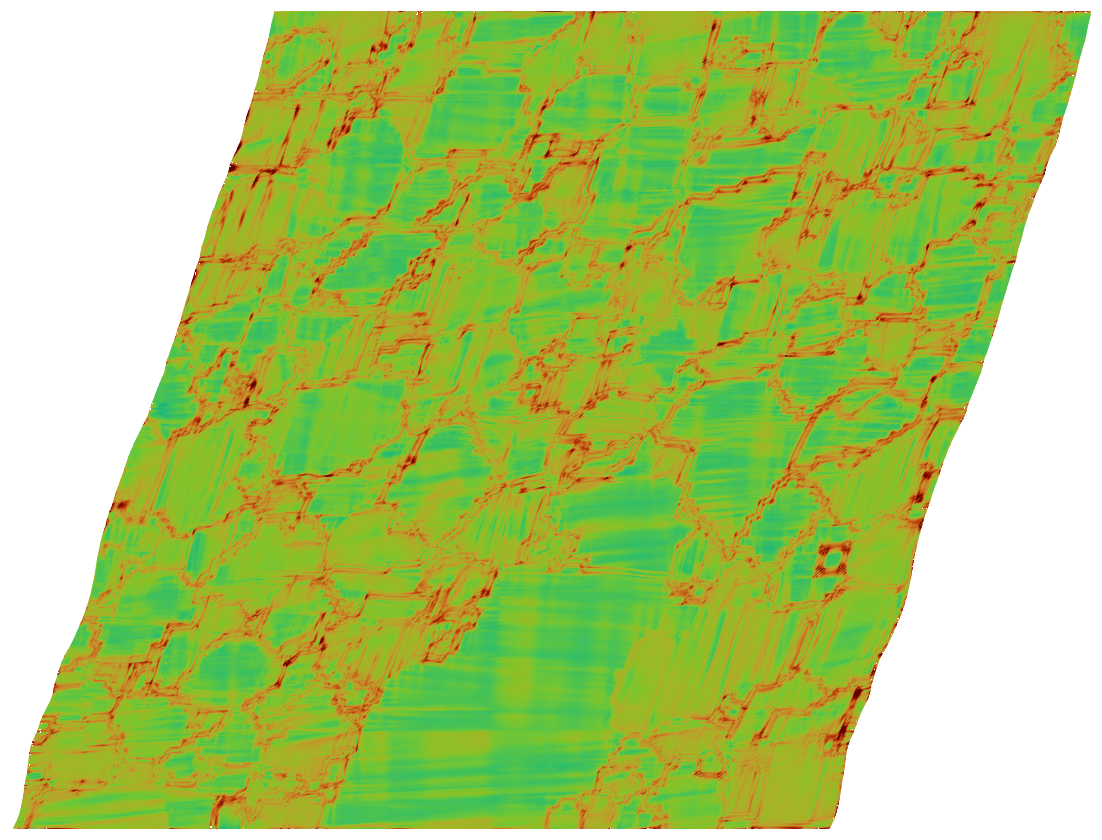}
        \caption{}
        \label{fig:S12_2}
    \end{subfigure}
    \caption{Evolution of stress deviator plotted at applied strains of a) 0.0 b) 0.15 and c) 0.30.}
    \label{fig:SE_dev_stress}
\end{figure}

\subsection*{Evolution of temperature}

Fig.~\ref{fig:SE_temp} shows the evolution of the temperature field at applied shear strains of $0.0$, $0.15$, and $0.30$. At the onset of deformation (Fig.~\ref{fig:T_2d_0}), the temperature is uniformly distributed at the initial value of 290~$K$ across the undeformed domain. At 0.15 applied shear strain (Fig.~\ref{fig:T_2d_15}), the domain has undergone some shearing and the temperature field begins to exhibit heterogeneity, with localized regions of elevated temperature emerging in band-like patterns oriented roughly in the direction of shearing. By 0.30 applied strain (Fig.~\ref{fig:T_2d_30}), these bands have intensified significantly, with peak temperatures reaching approximately 410~$K$. The temperature field at this stage displays a clear banded morphology, with zones of elevated temperature separated by regions that remain closer to the initial temperature. This spatially heterogeneous increase of temperature is a hallmark of adiabatic shear band formation: under the high strain rates considered here, conditions are effectively adiabatic, and the heat generated by plastic dissipation remains confined to the regions of intense shearing. The resulting local thermal softening further concentrates deformation into these bands, producing the characteristic feedback loop between plastic work, temperature rise, and strain localization that contributes adiabatic shear band development.

\begin{figure}[htbp]
    \centering
    \begin{subfigure}{0.28\textwidth}  
        \centering
        \begin{overpic}[scale=0.14]{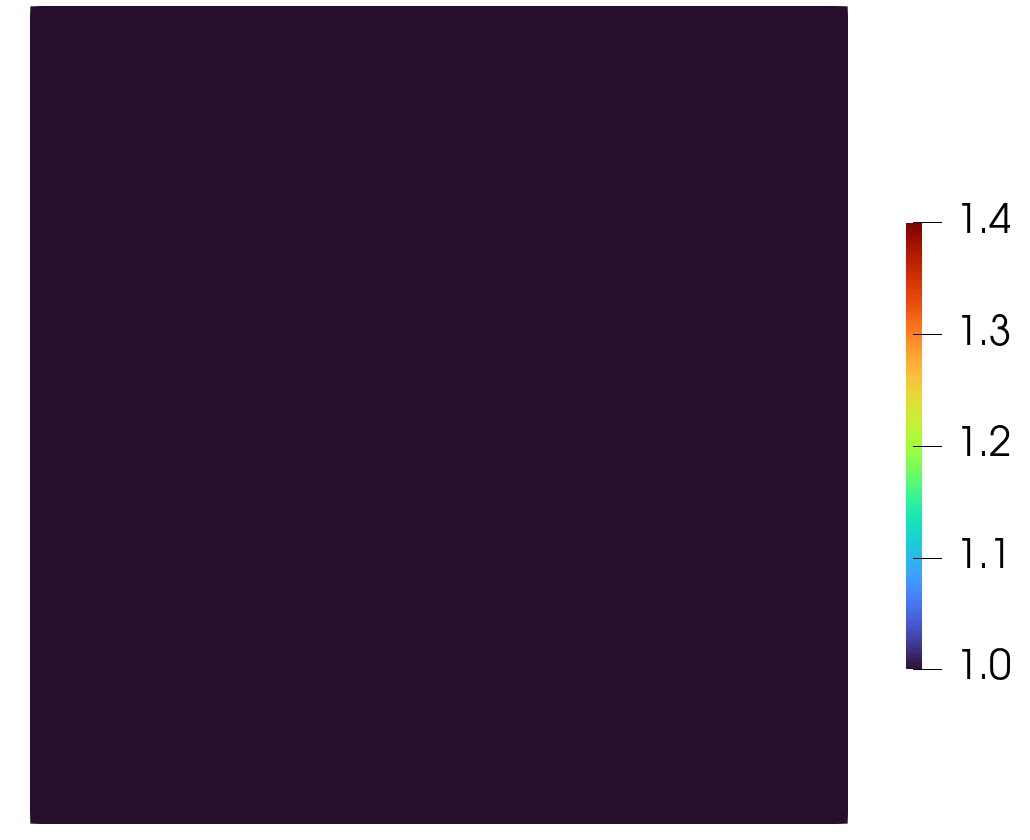}
            \put(88, 65){\fontsize{8}{7}\selectfont $\theta/\theta_{0}$}
        \end{overpic}
        \caption{}
        \label{fig:T_2d_0}
    \end{subfigure}
    \hfill
    \begin{subfigure}{0.28\textwidth}  
        \centering
        \includegraphics[width=\textwidth]{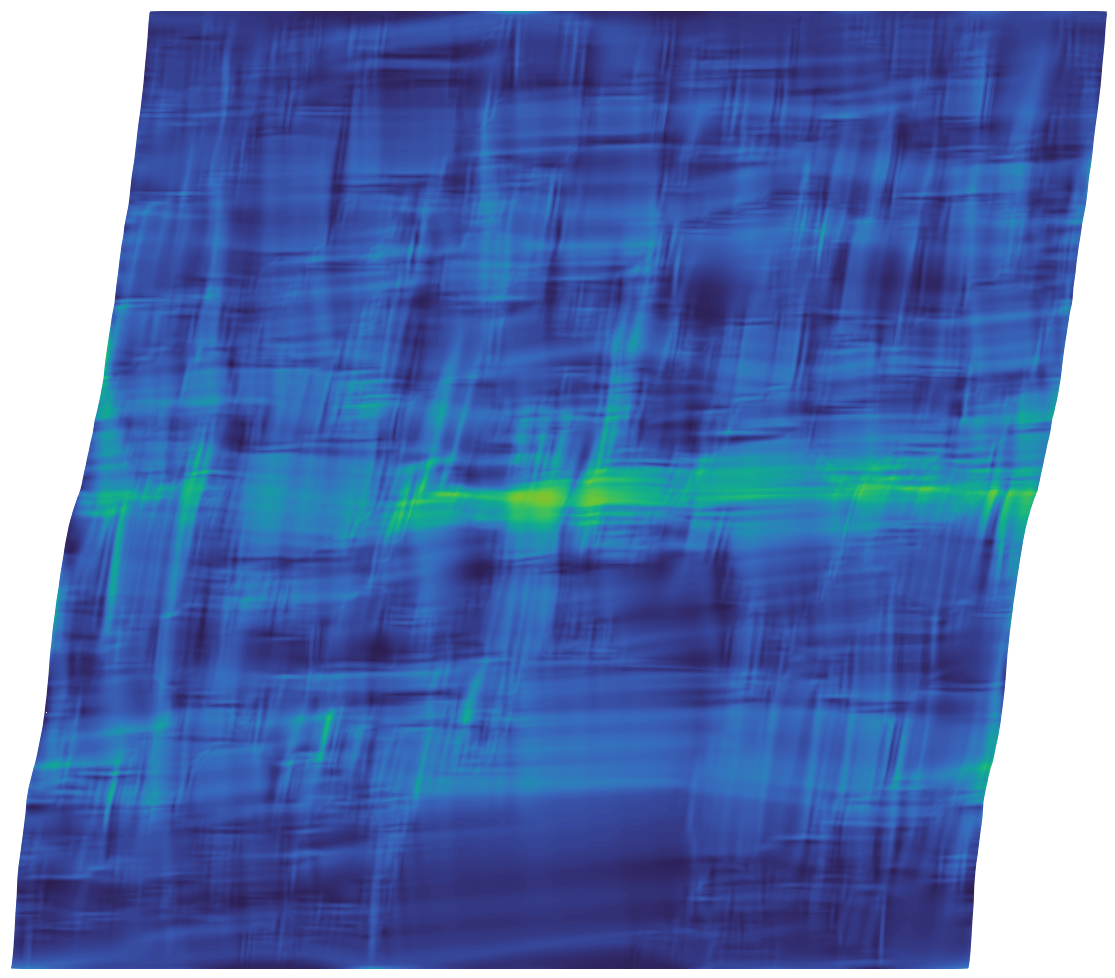}
        \caption{}
        \label{fig:T_2d_15}
    \end{subfigure}
    \hfill
    \begin{subfigure}{0.32\textwidth}  
        \centering
        \includegraphics[width=\textwidth]{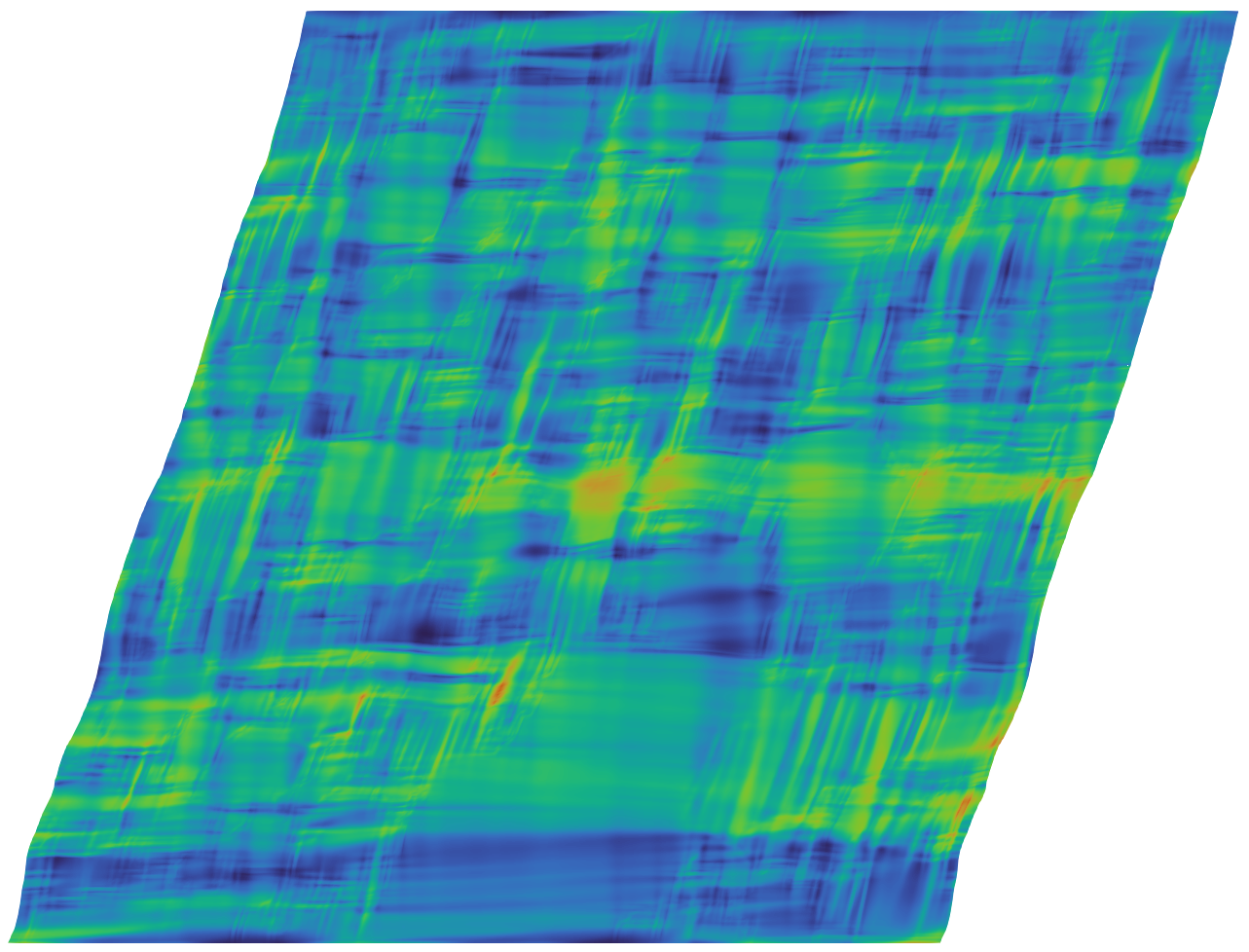}
        \caption{}
        \label{fig:T_2d_30}
    \end{subfigure}
    \caption{Evolution of normalized temperature ($\theta/\theta_0$) plotted at applied strains of a) 0.0 b) 0.15 and c) 0.30, where $\theta_0 = 293 \, \mathrm{K}$ is the reference temperature.}
    \label{fig:SE_temp}
\end{figure}

\begin{figure}
    \centering
        \begin{overpic}[scale=0.165]{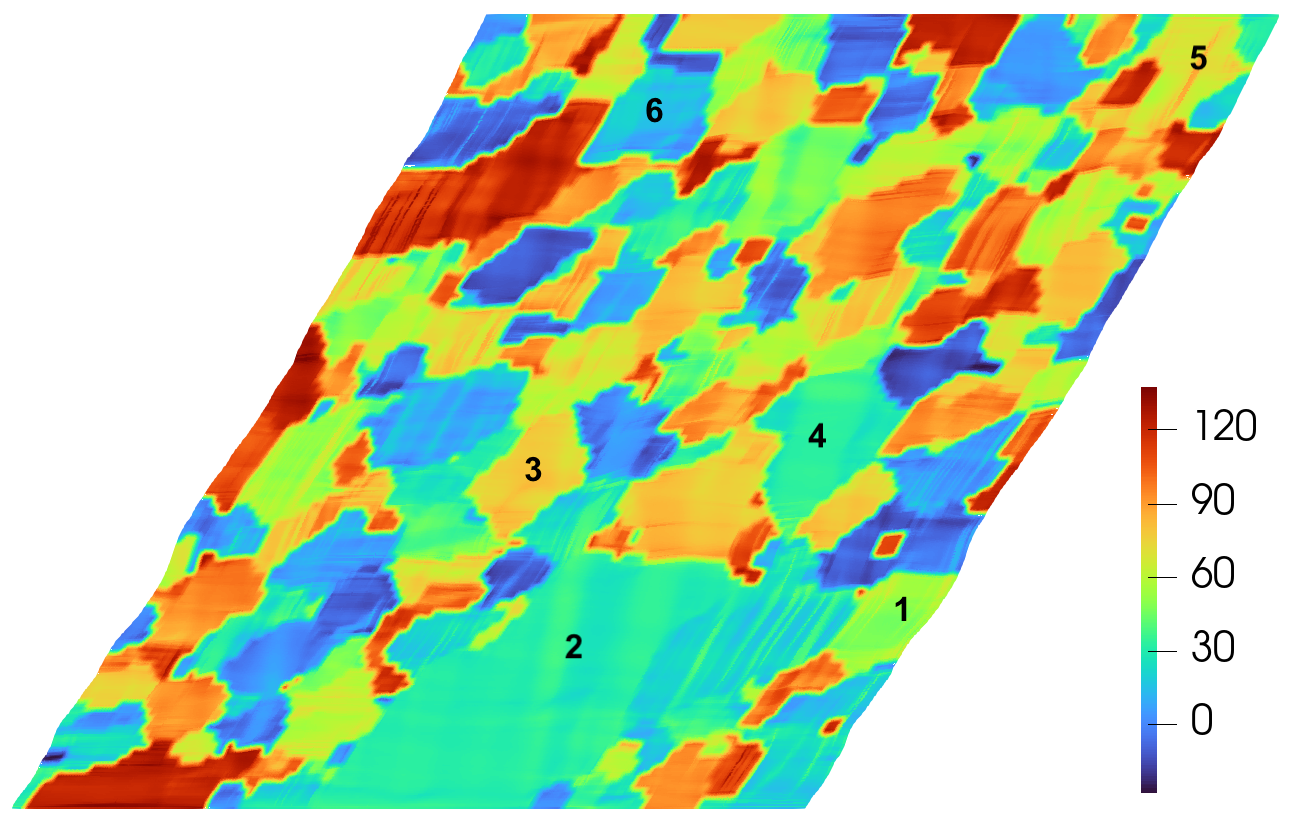}
            \put(87, 37){\fontsize{8}{7}\selectfont $(deg.)$}
        \end{overpic}
    \caption{Rotation angle field in the deformed configuration at an applied strain of 0.3. Grains analyzed for misorientation within the grain are marked.}
    \label{fig:angle_def}
\end{figure}

\begin{figure}[htbp]
    \centering
    \begin{subfigure}{0.48\textwidth}  
        \centering
        \includegraphics[scale=0.5]{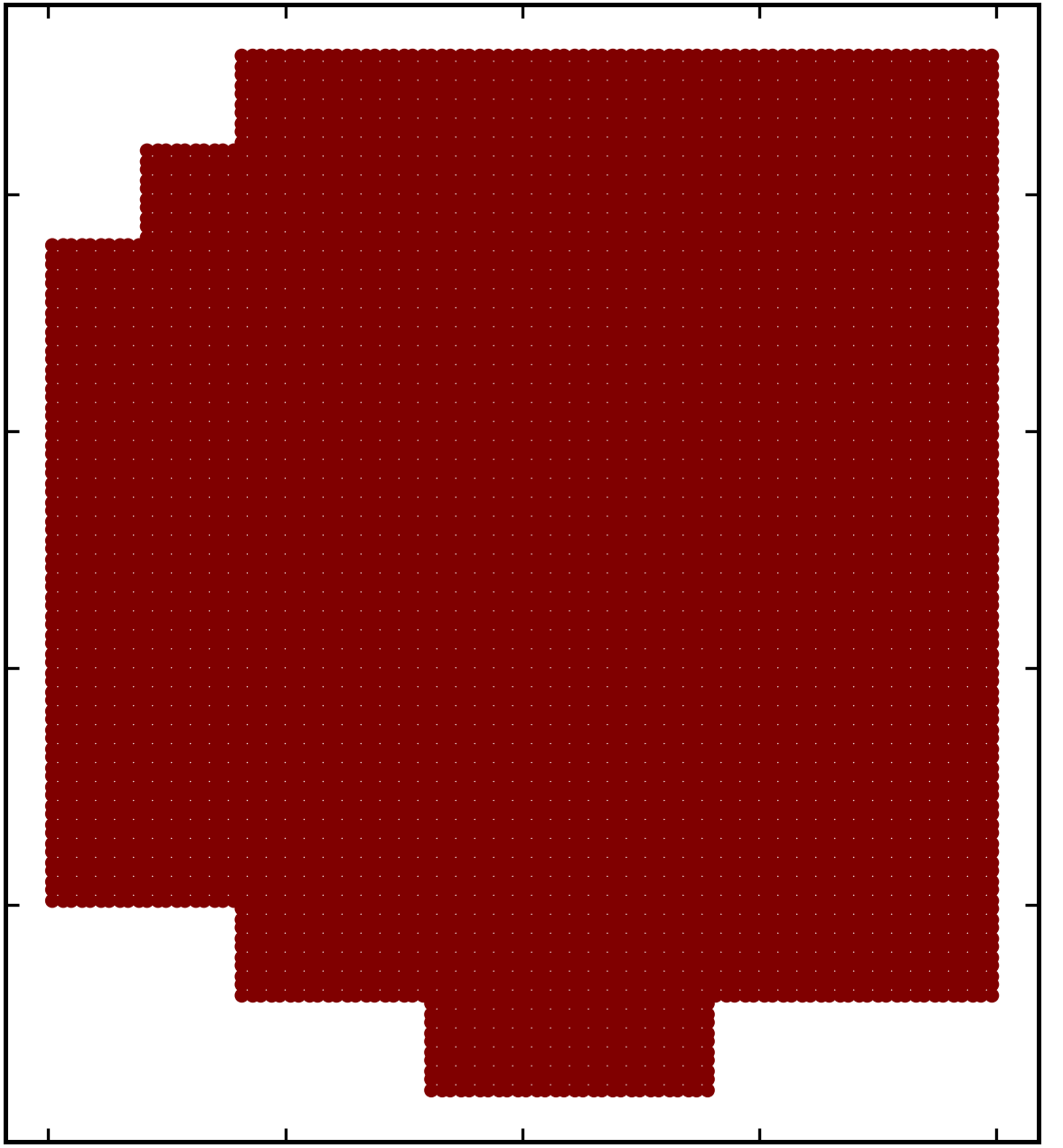}
        \caption{}
        \label{fig:grain}
    \end{subfigure}
    \hfill
    \begin{subfigure}{0.48\textwidth}  
        \centering
        \begin{overpic}[scale=0.60]{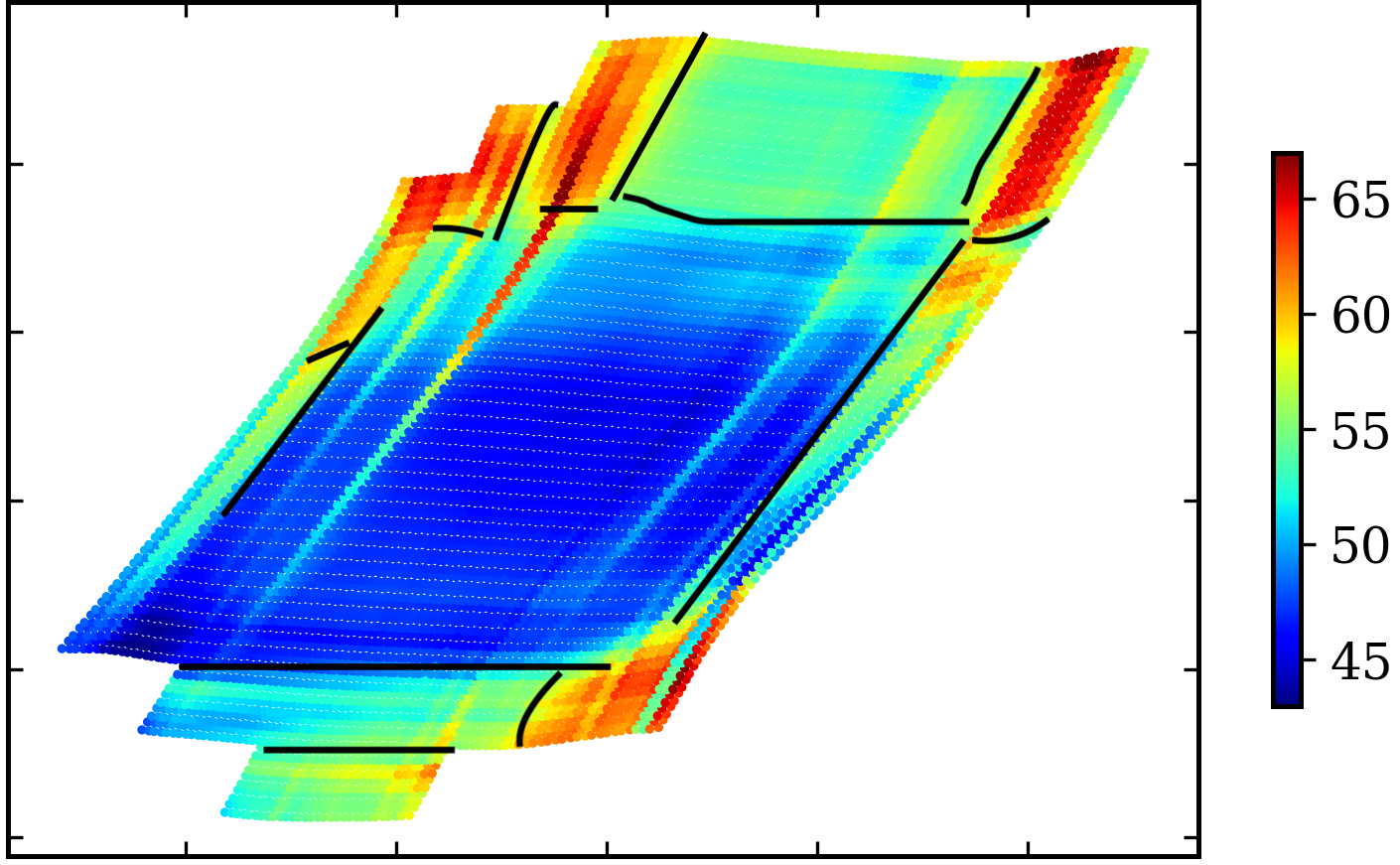}
            \put(91, 53){\fontsize{8}{7}\selectfont $ (deg.) $}
        \end{overpic}
        \caption{}
        \label{fig:grain_def}
    \end{subfigure}
    \caption{Spatial distribution of the lattice orientation angle in degrees for grain 1 (marked in the Fig.~\ref{fig:angle_def}). (a) The initial undeformed configuration showing a uniform orientation. (b) The deformed configuration at an applied strain of 0.3 exhibiting significant lattice rotation and geometric distortion. The overlaid black lines indicate locations of subgrain boundaries, indicating formation of substructure within the grain.}
    \label{fig:sub_grain_calc}
\end{figure}
\begin{table}[htbp]
    \centering
    \caption{Subgrain boundary statistics for selected grains.}
    \label{tab:subgrain_stats}
    \begin{tabular}{|c|c|c|c|c|c|}
        \hline
        \multirow{2}{*}{Grain} & \multirow{2}{*}{\# Subgrain boundaries} & \multicolumn{4}{c|}{Misorientation distribution} \\
        \cline{3-6}
        & & 0--2° & 2--5° & 5--11° & $>$11° \\
        \hline
        1 & 13 & 2 & 3 & 6 & 2  \\
        \hline
        2 & 8 & 1 & 3 & 3 & 1 \\
        \hline
        3 & 9  & 2 & 5 & 2 & 0  \\
        \hline
        4 & 6 & 1 & 3 & 2 & 0 \\
        \hline
        5 & 8  & 2 & 3& 2 & 1  \\
        \hline
        6 & 6  & 1 & 3 & 2 & 0  \\
        \hline
    \end{tabular}
\end{table}

\subsection*{Subgrain dislocation microstructure formation}
Finally, we examine the formation of subgrain structures. Fig.~\ref{fig:SE_GND} reveals localized bands of elevated GND density, suggesting the emergence of intragranular dislocation substructures. To investigate further, we examine a single grain under deformation and find that the spatial distribution of these GND bands closely aligns with the lattice orientation heterogeneity shown in Fig.~\ref{fig:grain_def}. The lattice orientation, initially uniform within the grain, develops significant spatial variation over the course of the simulation. These emergent low-angle features resemble the Incidental Dislocation Boundaries (IDBs) observed experimentally by Hughes and Hansen \cite{HUGHES20002985} in cold-rolled Nickel in a quasistatic setting. To characterize the misorientation associated with these boundaries, we analyze a selected set of grains (labeled 1 - 6 in Fig~.\ref{fig:angle_def}), with the quantitative results summarized in Table.~\ref{tab:subgrain_stats}. Additionally, Fig.~\ref{fig:grain_def} highlights specific interfaces within a representative grain, defined by their misorientation angle and boundary normal, demonstrating substructure formation throughout the grain interior. The formation of such intragranular substructures is a pervasive feature observed throughout the entire polycrystalline domain.

The substructure evolution observed in our simulations is also consistent with the early stages of rotational dynamic recrystallization, a mechanism first identified and extensively characterized by Meyers and coworkers \cite{Andrade1994} in their seminal studies of high-strain-rate deformation. They observed via TEM that shock-loaded copper deformed at strain rates of $\sim 10^4$ s$^{-1}$ developed elongated dislocation cells ($\sim 0.1$ $\mu m$ width) that progressively broke down into equiaxed sub-grains and ultimately transformed into recrystallized grains with high-angle boundaries. Subsequently they demonstrated that this microstructural pathway, proceeding from dislocation tangles to elongated cells to sub-grains to recrystallized micrograins, operates across a range of materials including tantalum~\cite{Nesterenko1997} and titanium~\cite{Meyers1995}. The misorientations and GND localization patterns observed in our simulations appear to align with this well-established precursor stage. Experimental observations also reveal that, at early stages of deformation, the shear band microstructure is dominated by elongated blocks containing a well-defined low-angle boundary substructure with characteristic length scales on the order of $\sim 200–300 \ nm.$, which progressively evolves into an equiaxed nanograin structure with increasing deformation, as shown in Fig.~\ref{fig:th_combined}. Experimental observations also reveal the evolution of elongated grain structures, as shown in Fig.~\ref{fig:th_combined}. The grain structures predicted by our simulations appear to be qualitatively consistent with these experimental observations.

\subsection{3-d simulations}

\begin{table}[htbp]
\centering
\caption{Computational details for the 3-d MFDM simulations.}
\label{tab:comp_details_3d}
\begin{tabular}{lcc}
\hline 
&3-d \\
\hline
Domain size & $40 \times 40 \times 40~\mu m^3$ \\
Elements & $ 1 \times 10^6$ \\
Strain reached & 0.40\\
Slip systems & 48 \\
Nominal strain rate & $4 \times 10^3~\text{s}^{-1}$ \\
Cores & 1024 \\
Wall-clock time & 15~hrs \\
\hline
\end{tabular}
\end{table}

We demonstrate the capability of the computational framework to simulate 3-d polycrystalline domains under dynamic loading. The computational domain is a cubic representative volume element. The simulation parameters are presented in Table.~\ref{tab:comp_details_3d}. The polycrystalline microstructure is generated using a procedure developed and described in \cite{SCHMELZER2025104318,Behnoudfar2026}, producing a statistically representative grain structure with orientations assigned to capture the crystallographic texture of the material. Fig.~\ref{fig:grain_3d} shows the polycrystalline microstructure used in the 3-d simulations, with grains colored by crystallographic orientation.

In 3-d, 48 slip systems are used to characterize the BCC crystal. The slip systems used are:
\begin{itemize}[noitemsep,nolistsep]
    \item 24 \{123\} planes with one $<111>$ direction.
    \item 12 \{112\} planes with one $<111>$ direction.
    \item 6 \{110\} planes with two $<111>$ directions.
\end{itemize}
Boundary conditions for the velocity, shown in Fig.~\ref{fig:3d_shear} (and mentioned in \ref{tab:bc_shear_3d}) are specified as follows: periodic boundary conditions are applied on the lateral faces in the $x$ and $y$ directions to approximate an infinite medium. The front and back faces ($z$-normal) are traction-free in the $x$-$y$ plane, and motion in the $z$-direction is constrained on these faces. On the top and bottom faces, prescribed shear velocities are applied in the horizontal direction, while multi-point constraints (MPCs) enforce uniform vertical displacement to maintain planarity. The initial GND density is set to zero throughout the domain, and the applied nominal strain rate is $4000~\text{s}^{-1}$.

To make these large-scale 3-d simulations computationally tractable, we employ the dimensional analysis approach described in Sec.~\ref{ss_sec:da}. By simultaneously scaling the applied strain rate and reference plastic strain rate by the same factor while keeping their ratio fixed, we achieve the same mechanical response in fewer time steps without altering the underlying physics. This reduced the wall-clock time by approximately an order of magnitude, enabling simulations that would otherwise require several days to be completed in a few of hours. The exact number of cores used along with the time for the simulation is mentioned in Table.~\ref{tab:comp_details_2d} and Table.~\ref{tab:comp_details_3d}.

\begin{figure}[htbp]
    \centering
    \begin{tikzpicture}
        \node[anchor=south west, inner sep=0] (img) 
            {\includegraphics[scale=0.48]{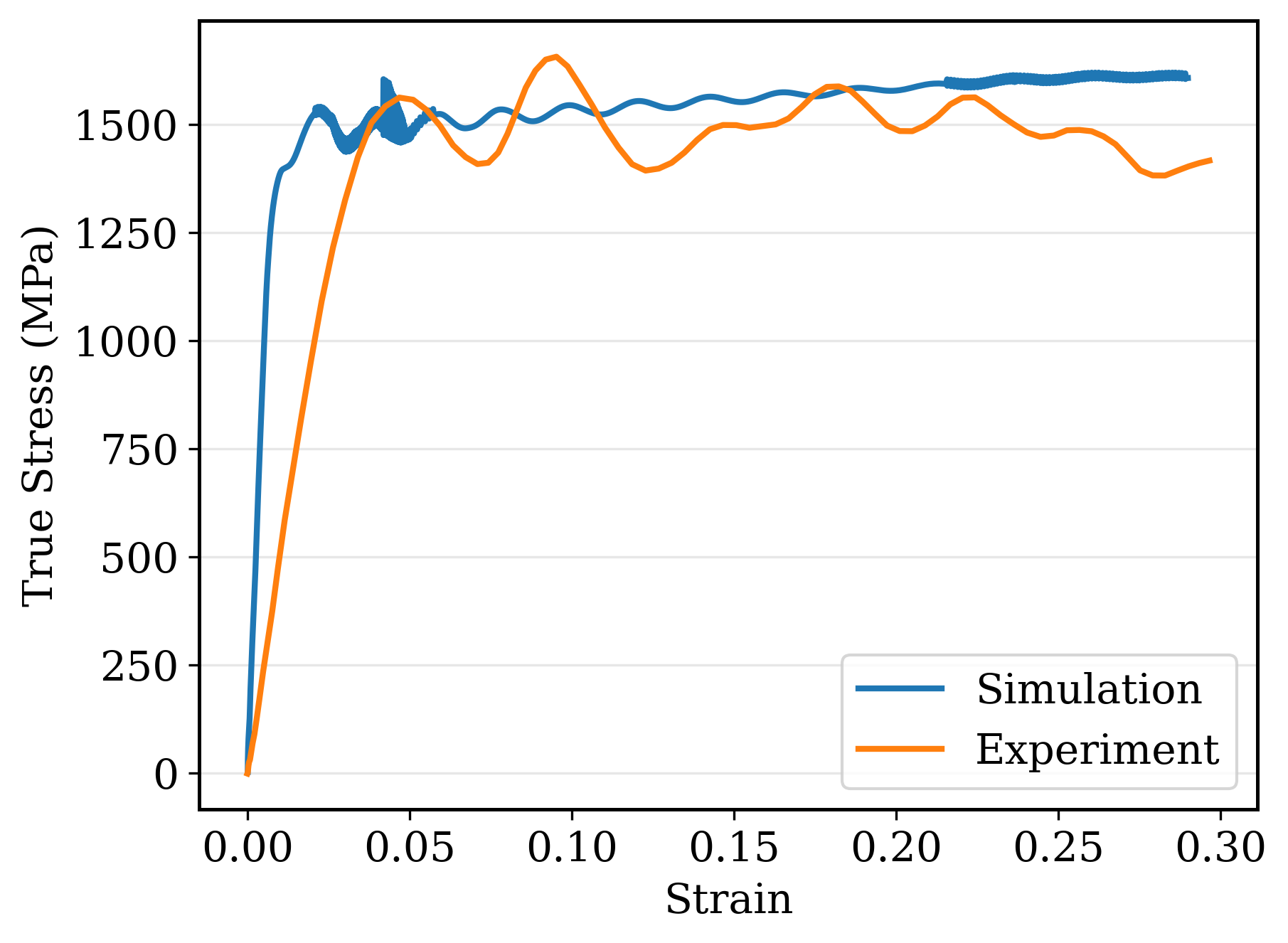}};
    \end{tikzpicture}
    \caption{Stress-strain response from the SHPB experiment and 3-d simulation. The relatively darker portions in the simulation curve arise from the adaptive time-step cutback algorithm. This technique reduces the time-step size when the plastic strain increment exceeds a prescribed threshold, thus resulting in more dense data output.}
    \label{fig:SS_calib_3}
\end{figure}
\noindent \underline{Calibration for 3-d simulations}: As in the 2-d case, we calibrate the model against the experimental stress-strain curve from a dynamic compression test on a cylindrical specimen performed using the SHPB, with boundary conditions as described in Sec.~\ref{sss:SHPB_vel}. Zero traction is imposed on the boundaries where there is no velocity boundary condition, to approximate homogeneous uniaxial compression conditions (Fig.~\ref{fig:3d_comp} and Table.~\ref{tab:bc_cal_3d}). Since the 3-d model includes all 48 BCC slip systems rather than the reduced set used in 2-d, the material parameters required recalibration; the resulting stress-strain curve is shown in Fig.~\ref{fig:SS_calib_3}.

The classical hardening modulus $h_0$ is again set to 0 and the initial strength is taken as  $g_0 = 0.40 GPa$ GPa. The estimate of the initial statistical density for the 3-d simulations is then given by
\[
\bar{\rho}_s = (g_0/\eta \mu b)^2 \approx 1.576 \times 10^{15} \, m^{-2}.
\]

\begin{figure}[htbp]
    \centering
    \begin{subfigure}{0.48\textwidth}  
        \centering
        \includegraphics[scale=0.50]{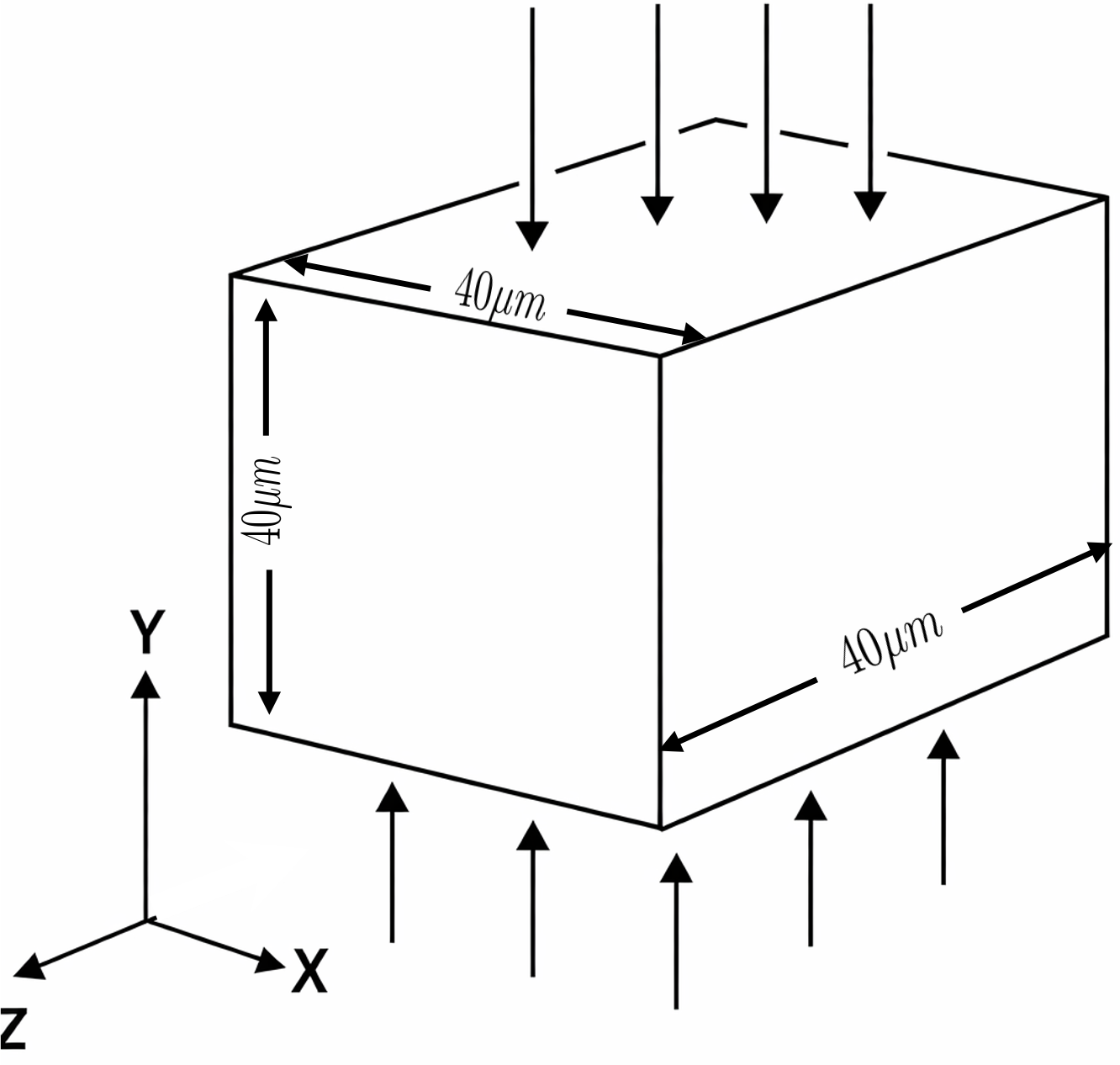}
        \caption{}
        \label{fig:3d_comp}
    \end{subfigure}
    \hfill
    \begin{subfigure}{0.48\textwidth}  
        \centering
        \includegraphics[scale=0.175]{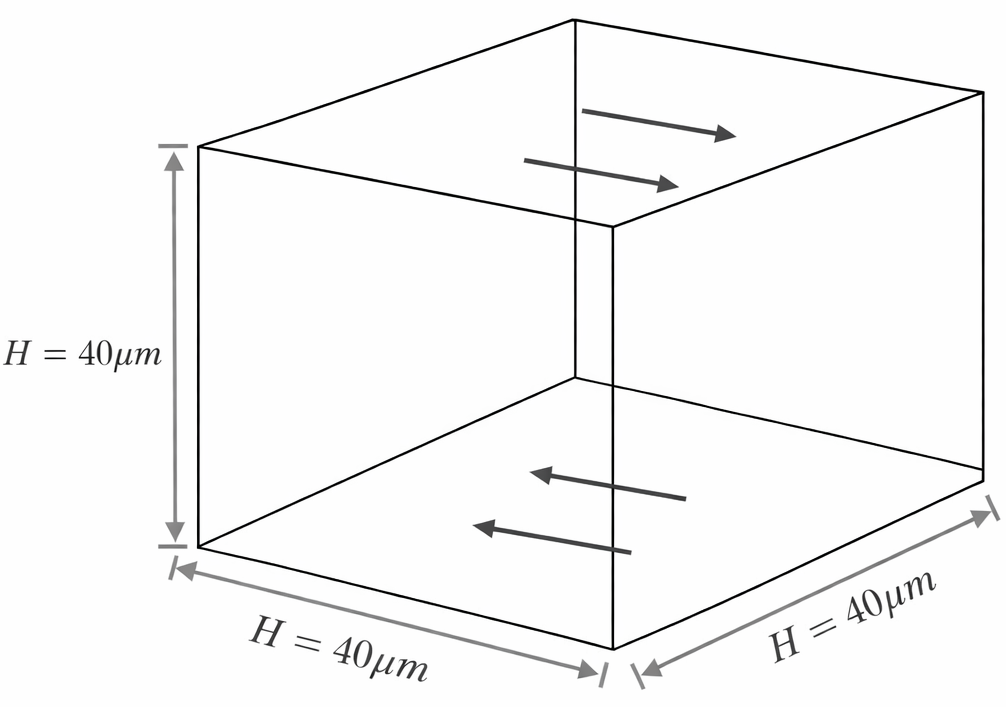}
        \caption{}
        \label{fig:3d_shear}
    \end{subfigure}
    \caption{Boundary conditions for the 3-d simulations. (a) Uniaxial compression used for calibration (b) Shearing with periodic boundary conditions on lateral faces used for simulating a shear band. Details of these boundary conditions on lateral faces are listed in Table.~\ref{tab:bc_cal_3d} and Table.~\ref{tab:bc_shear_3d}}
\end{figure}

\begin{table}[htbp]
\centering
\caption{Boundary conditions applied for the 3-d calibration.}
\label{tab:bc_cal_3d}
\begin{adjustbox}{max width=\textwidth}
\begin{tabular}{lll}
\hline
Boundary & Type & Condition \\
\hline
Top & Velocity and traction & $v_y = v_l$, $t_x = 0$, $t_z = 0$ \\
Bottom & Velocity and traction & $v_y = v_r$, $t_x = 0$, $t_z = 0$ \\
Front and Back & Traction-free & $\bft = \mathbf{0}$ \\
Left and Right & Traction-free & $\bft = \mathbf{0}$ \\
\hline
\multicolumn{3}{p{0.9\textwidth}}{\footnotesize $v_y$: $y$-component of velocity; $v_l$, $v_r$: prescribed velocities at the left and right boundaries from SHPB data; $t_x$, $t_z$: traction components in the $x$- and $z$-directions; $\bft$: traction vector.} \\
\hline
\end{tabular}
\end{adjustbox}
\end{table}

\begin{table}[htbp]
\centering
\caption{Boundary conditions applied to simulate a shear band in 3-d.}
\label{tab:bc_shear_3d}
\begin{adjustbox}{max width=\textwidth}
\begin{tabular}{lll}
\hline
Boundary & Type & Condition \\
\hline
Top & Prescribed velocity & $v_x = v_1 = AH/2$, uniform $v_y$ via MPC, $t_z = 0$ \\
Bottom & Prescribed velocity & $v_x = v_1 = -AH/2$, uniform $v_y$ via MPC, $t_z = 0$ \\
Left and Right & Periodic BC & Velocity periodicity in $x$ and $y$, $t_z = 0$ \\
Front and Back & Periodic BC & Velocity periodicity in $x$ and $y$, $v_z = 0$ \\
\hline
\multicolumn{3}{p{0.9\textwidth}}{\footnotesize $v_x$, $v_y$, $v_z$: components of the velocity vector; $A$: applied strain rate; $H$: height of the RVE; $t_z$: traction component in the $z$-direction; MPC: multi-point constraint enforcing uniform vertical displacement across the boundary. The loading is applied as a step in time, with the velocity equal to zero at $t = 0$ and held constant for all subsequent times.} \\
\hline
\end{tabular}
\end{adjustbox}
\end{table}

The CCP and MFDM approaches are studied next in 3-d simulations of adiabatic shear banding. The comparison is carried out along two lines: first, we examine the average stress-strain response obtained from both models and its sensitivity to mesh refinement; second, we analyze the spatial field distributions to investigate the nature of strain localization and the ability of each model to predict the experimentally observed finite width of adiabatic shear bands.
We then examine microstructure evolution through the GND density, and material strength. Without access to sophisticated tools for inferring localized 2-d field concentrations in 3-d regions of the simulated body, we  attempt to provide some rationalization for the potential formation of 2-d subgrain boundaries within the 3-d shear band region in the simulations.

\subsection*{Stress strain response of 3-d sample}

Before examining the localization patterns, we first verify mesh convergence of the stress-strain response under progressive mesh refinement, with mesh densities of 25, 50, and 100 elements per direction. As shown in Fig.~\ref{sfig:SS_MFDM}, the MFDM stress-strain curves remain nearly unchanged across these mesh densities, confirming mesh-independent behavior in 3-d. In contrast, the CCP formulation exhibits continued non-convergence with refinement, with the response changing as the mesh is refined, as shown in Fig.~\ref{sfig:SS_CCP}.

Figure~\ref{fig:3d_F_12_Comp} presents the shear component $F_{12}$ of the deformation gradient for both CCP and MFDM in the 3-d domain. Consistent with the 2-d results, CCP exhibits pronounced strain localization into narrow bands, while MFDM produces a more distributed deformation field throughout the domain. The 3-d simulations confirm that the mesh-independent behavior observed in 2-d extends to three dimensions, further demonstrating the capability of the MFDM framework for capturing finite shear band widths. 

The 3-d simulations also reproduce the size-dependent strengthening behavior discussed in the 2-d setting (Sec.~\ref{sec:size_2d}). Figure~\ref{sfig:Size_3d} shows results from a series of 3-d simulations in which the domain is self-similarly scaled from 10 to 160~$\mu m$, maintaining a constant grain-to-domain size ratio of approximately $1/10$. As in 2-d, CCP predicts an identical stress-strain response regardless of domain size, while MFDM captures a clear ``smaller is stronger'' trend arising from the GND density contribution to hardening. The size effect again diminishes with increasing domain size and saturates as the size of the grains increases. This is consistent with both the 2-d predictions and experimental observations \cite{Exp_Size_effect, Size_effect_2}.

\begin{figure}[htbp]
    \centering
    \begin{subfigure}{0.32\textwidth}  
        \centering
        \includegraphics[width=\textwidth]{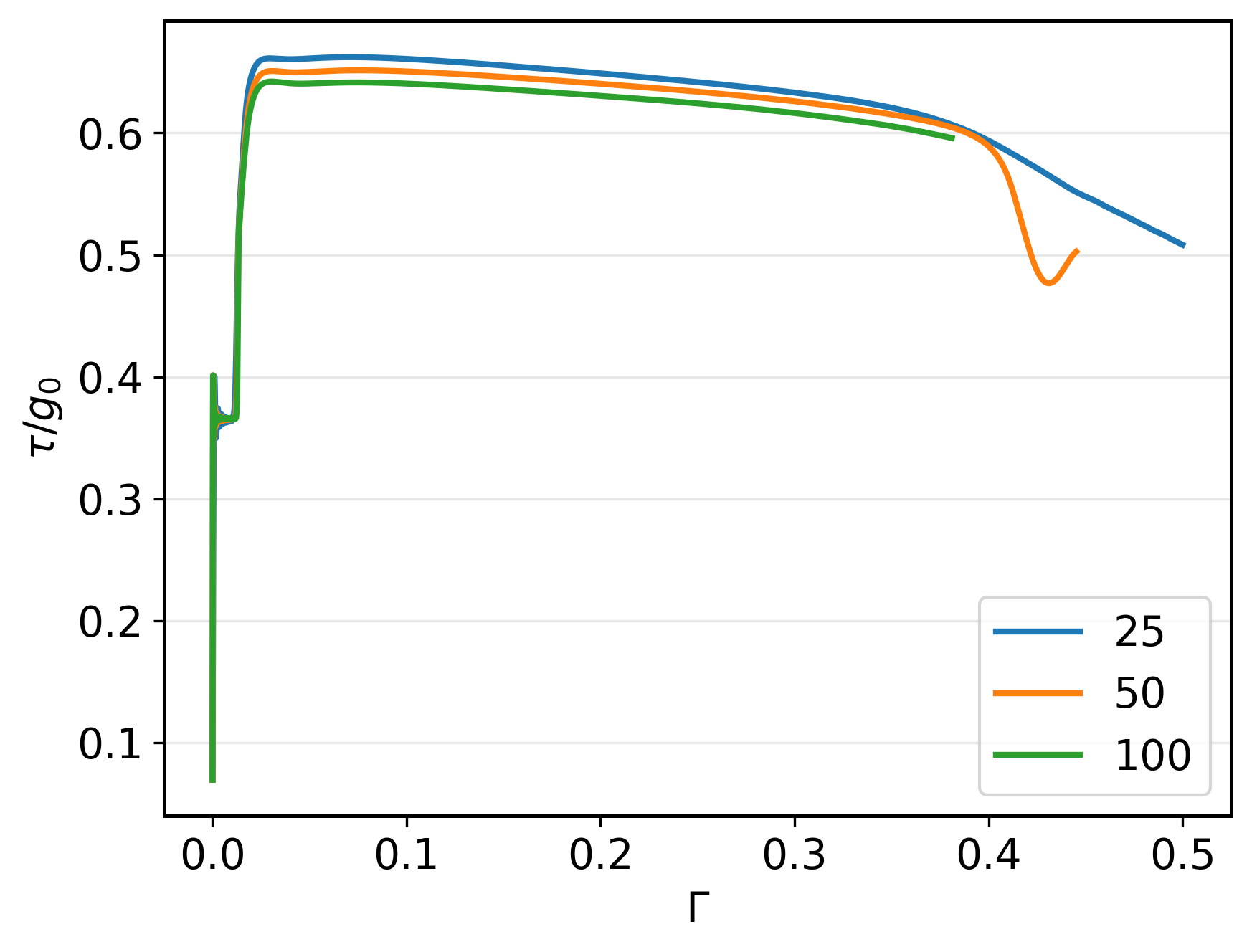}
        \caption{}
        \label{sfig:SS_CCP}
    \end{subfigure}
    \hfill
    \begin{subfigure}{0.32\textwidth}  
        \centering
        \includegraphics[width=\textwidth]{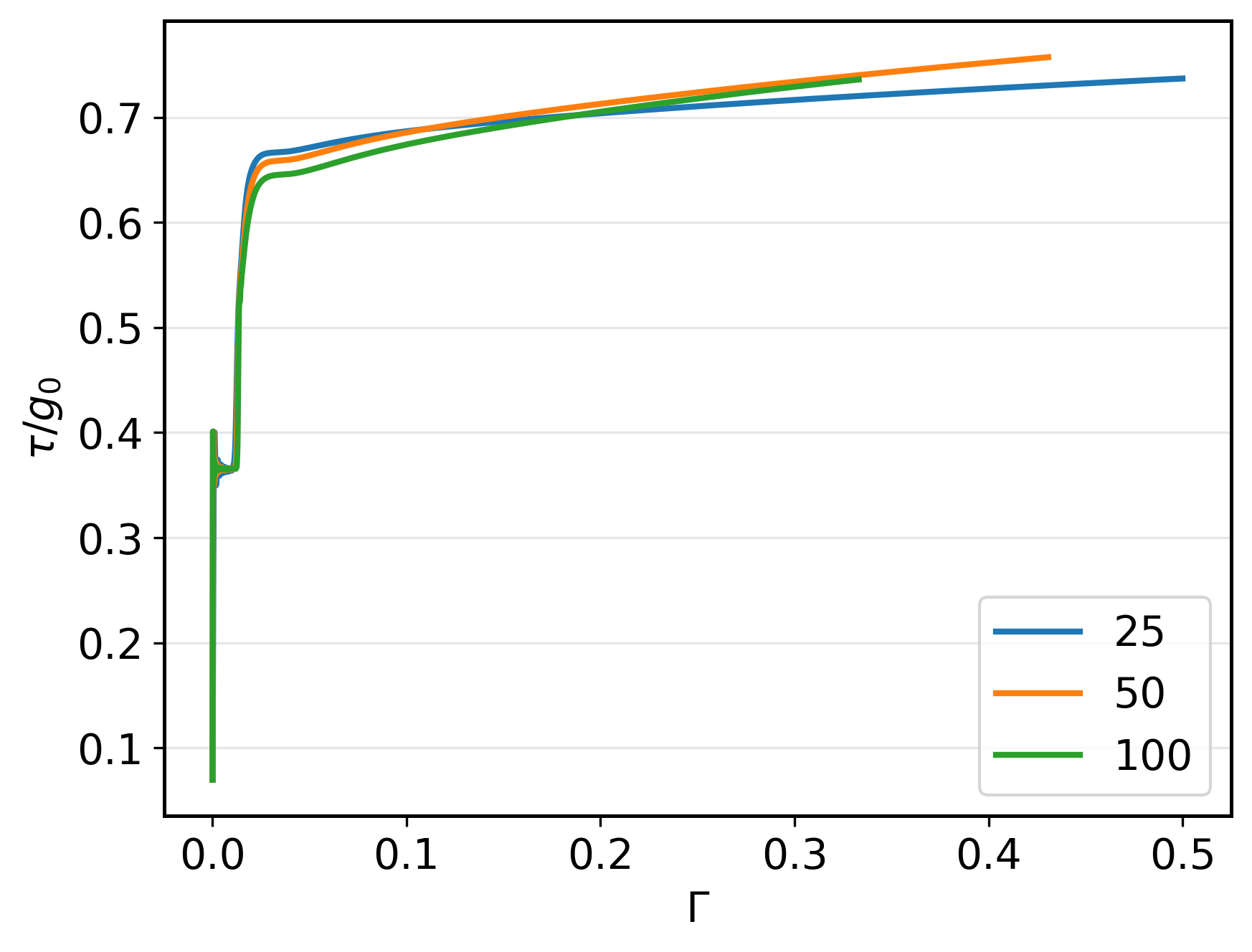}
        \caption{}
        \label{sfig:SS_MFDM}
    \end{subfigure}
    \hfill
    \begin{subfigure}{0.32\textwidth}  
        \centering
        \includegraphics[width=\textwidth]{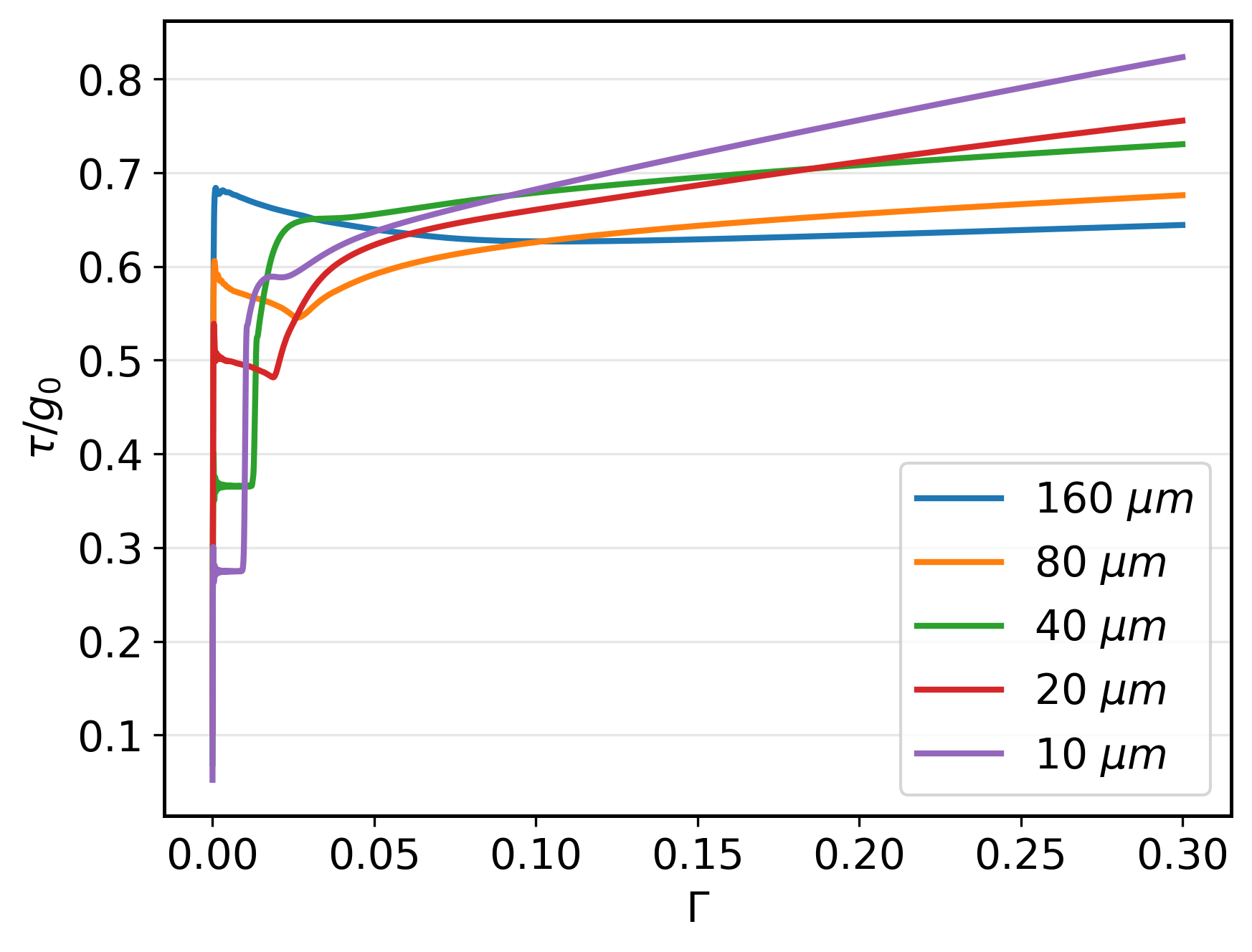}
        \caption{}
        \label{sfig:Size_3d}
    \end{subfigure}
    \caption{3-d simulation results: stress-strain response for (a) CCP and (b) MFDM on meshes of $25^3, 50^3, 100^3$ elements and (c) size-dependent strengthening predicted by MFDM for self-similarly scaled domains.}
    \label{fig:SS_CCP_MFDM}
\end{figure}

\subsection*{Microstructure evolution}

The evolution of GND density in the 3-d simulations, shown in Fig.~\ref{fig:3d_gnd}, follows trends qualitatively similar to those observed in 2-d. Starting from a dislocation-free initial state, GND density accumulates progressively at grain boundaries due to the intergranular plastic strain rate jump condition imposed across interfaces, while patterned structures emerge within grain interiors (Fig.~\ref{fig:Slice_GND}). Two notable distinctions arise in the 3-d case: first, the magnitude of the GND density is significantly higher due to the possible activation of all 48 slip systems; second, the subgrain structures that form exhibit a fully three-dimensional character.

The material strength field, shown in Fig.~\ref{fig:3d_strength}, is initially uniform and develops spatial heterogeneity as deformation proceeds, with grain boundaries hardening relative to grain interiors.

Fig.~\ref{fig:3d_Temp} shows the corresponding temperature evolution for the 3-d simulation. The overall progression is qualitatively similar to the 2-d case, with a uniform initial temperature field (Fig.~\ref{fig:T_1}) giving way to thermal bands at 0.15 (Fig.~\ref{fig:T_2}) and 0.30(Fig.~\ref{fig:T_3}) applied shear strain. Peak temperatures again reach approximately 410~K, and hot spots with temperatures exceeding the surrounding band are visible on the surface.

\begin{figure}
 \centering
    \begin{subfigure}{0.25\textwidth}  
        \centering
        \includegraphics[width=\textwidth]{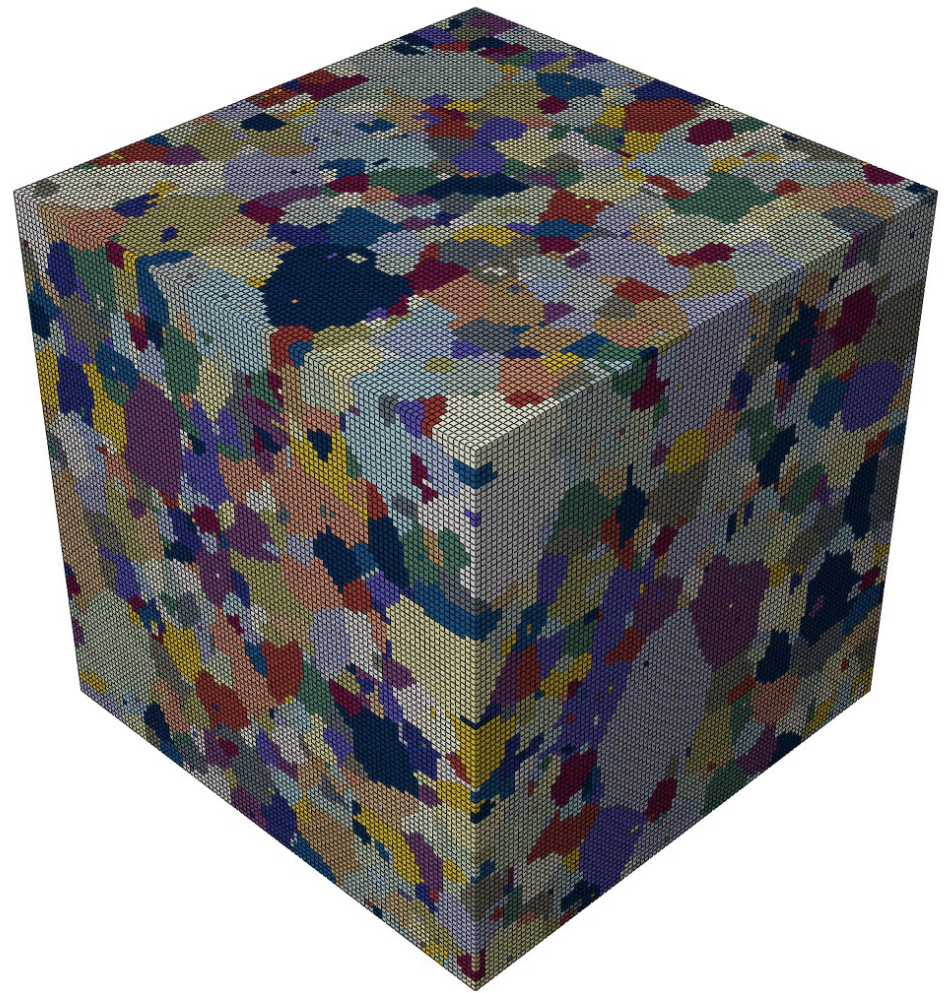}
        \caption{}
        \label{fig:grain_3d}
    \end{subfigure}
    \hfill
    \centering
    \begin{subfigure}{0.32\textwidth}  
        \centering
        \begin{overpic}[scale=0.12]{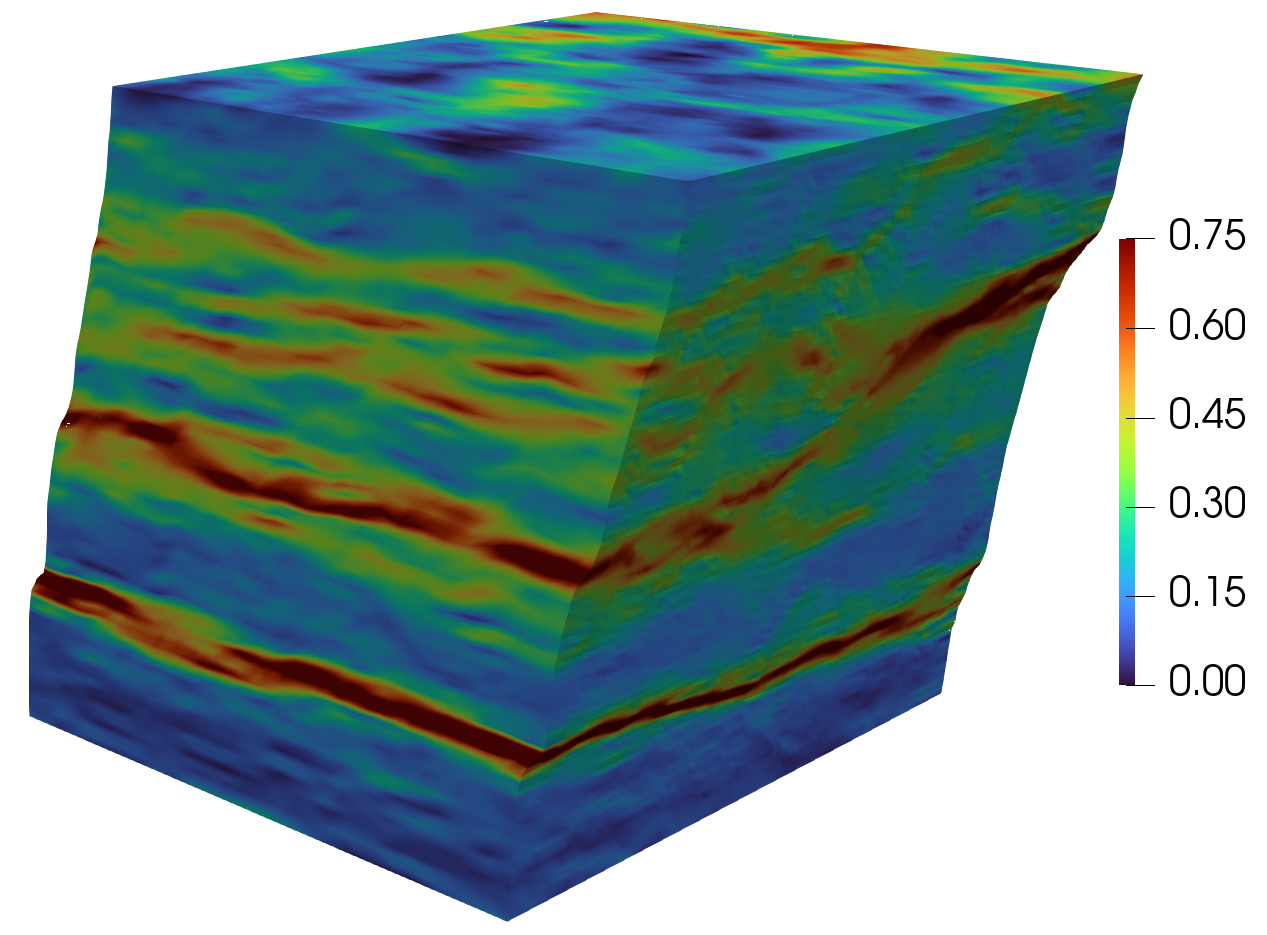}
            \put(90, 61){\fontsize{8}{5}\selectfont $F_{12}$}
        \end{overpic}
        \caption{}
        \label{fig:CCP_F_12}
    \end{subfigure}
    \hfill
    \begin{subfigure}{0.30\textwidth}  
        \centering
         \includegraphics[width=\textwidth]{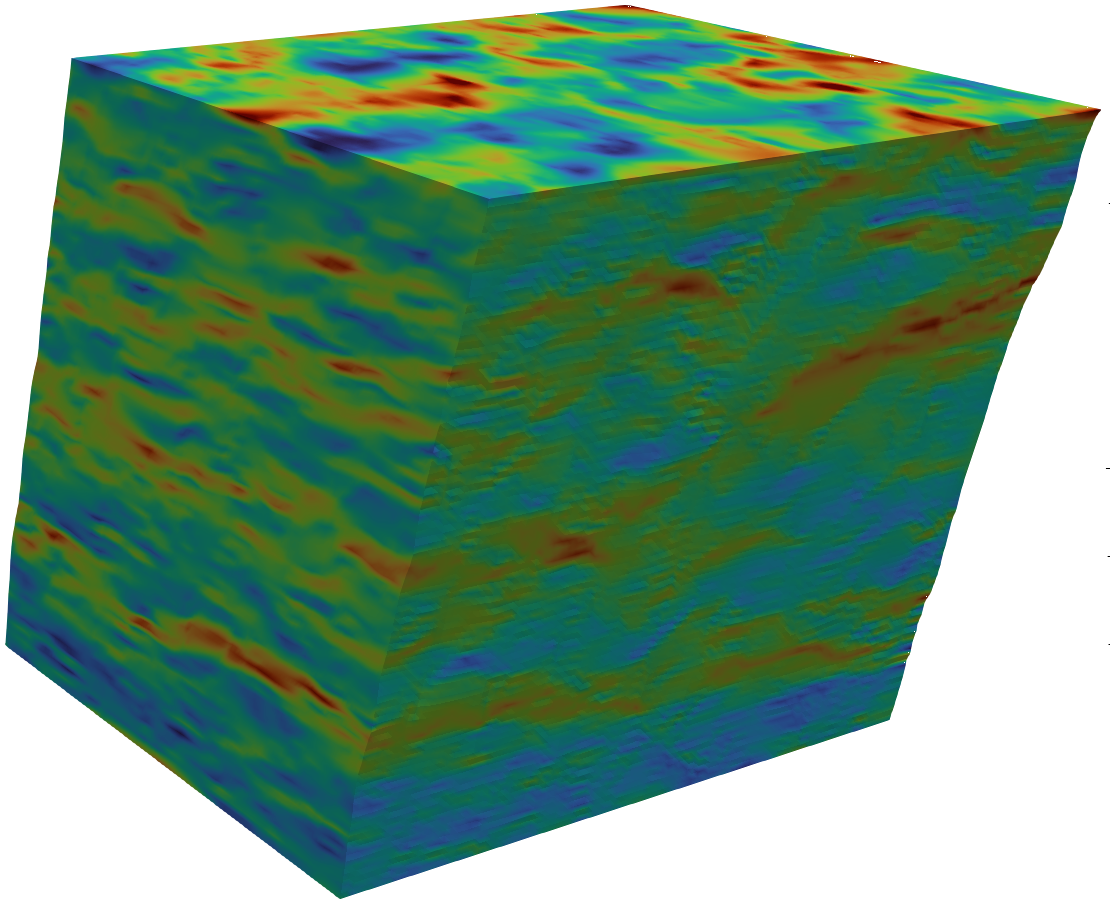}
        \caption{}
        \label{fig:MFDM_F_12}
    \end{subfigure}
    \caption{a) 3-d polycrystalline microstructure generated as \cite{SCHMELZER2025104318,Behnoudfar2026}, with grains colored by crystallographic orientation. Field plots of the shear component $F_{12}$ are shown at an applied strain of 0.3 for b) Classical crystal plasticity and c) MFDM.}
    \label{fig:3d_F_12_Comp}
\end{figure}

\begin{figure}[htbp]
    \centering
    \begin{subfigure}{0.32\textwidth}  
        \centering
        \begin{overpic}[scale=0.14]{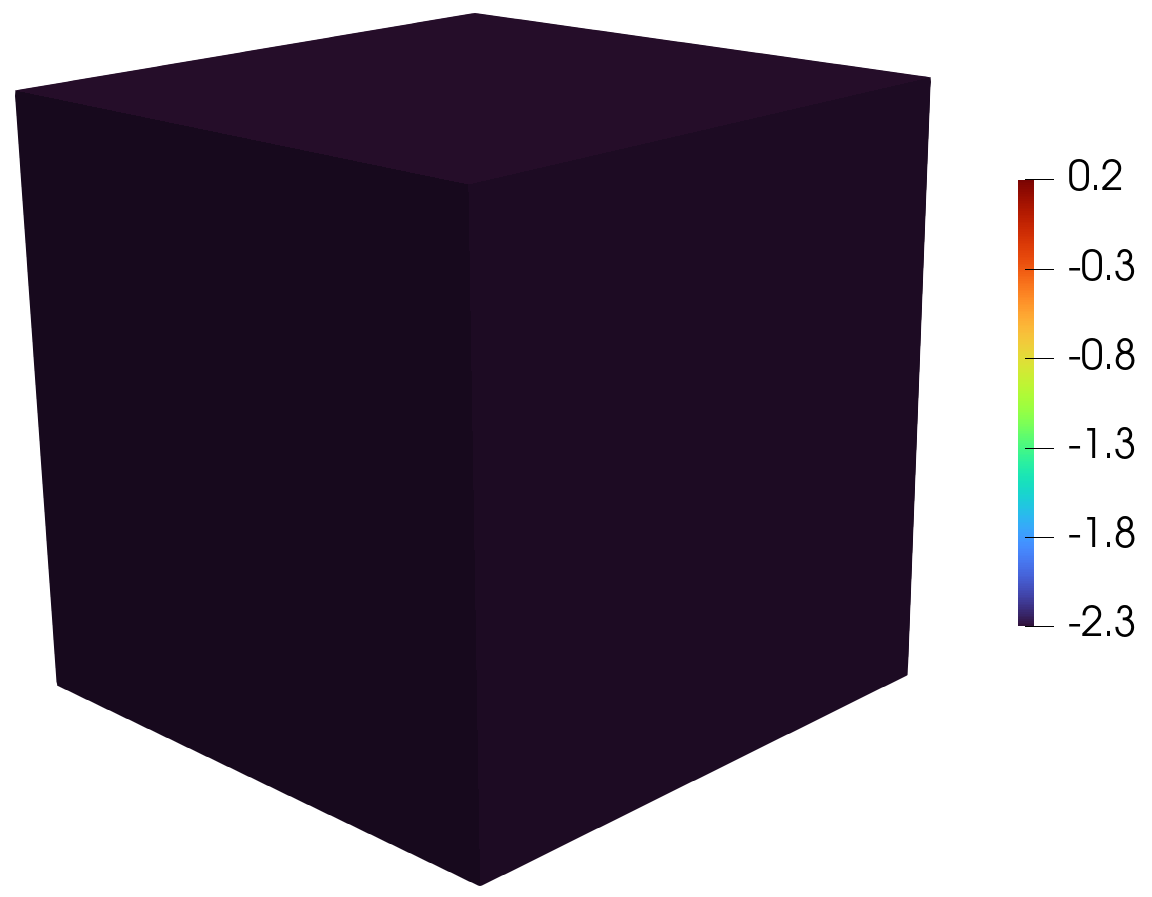}
            \put(82, 68){\fontsize{8}{5}\selectfont $\log(\rho_g/\bar{\rho}_s)$}
        \end{overpic}
        \caption{}
        \label{fig:gnd_0_3d}
    \end{subfigure}
    \hfill
    \begin{subfigure}{0.28\textwidth}  
        \centering
        \includegraphics[width=\textwidth]{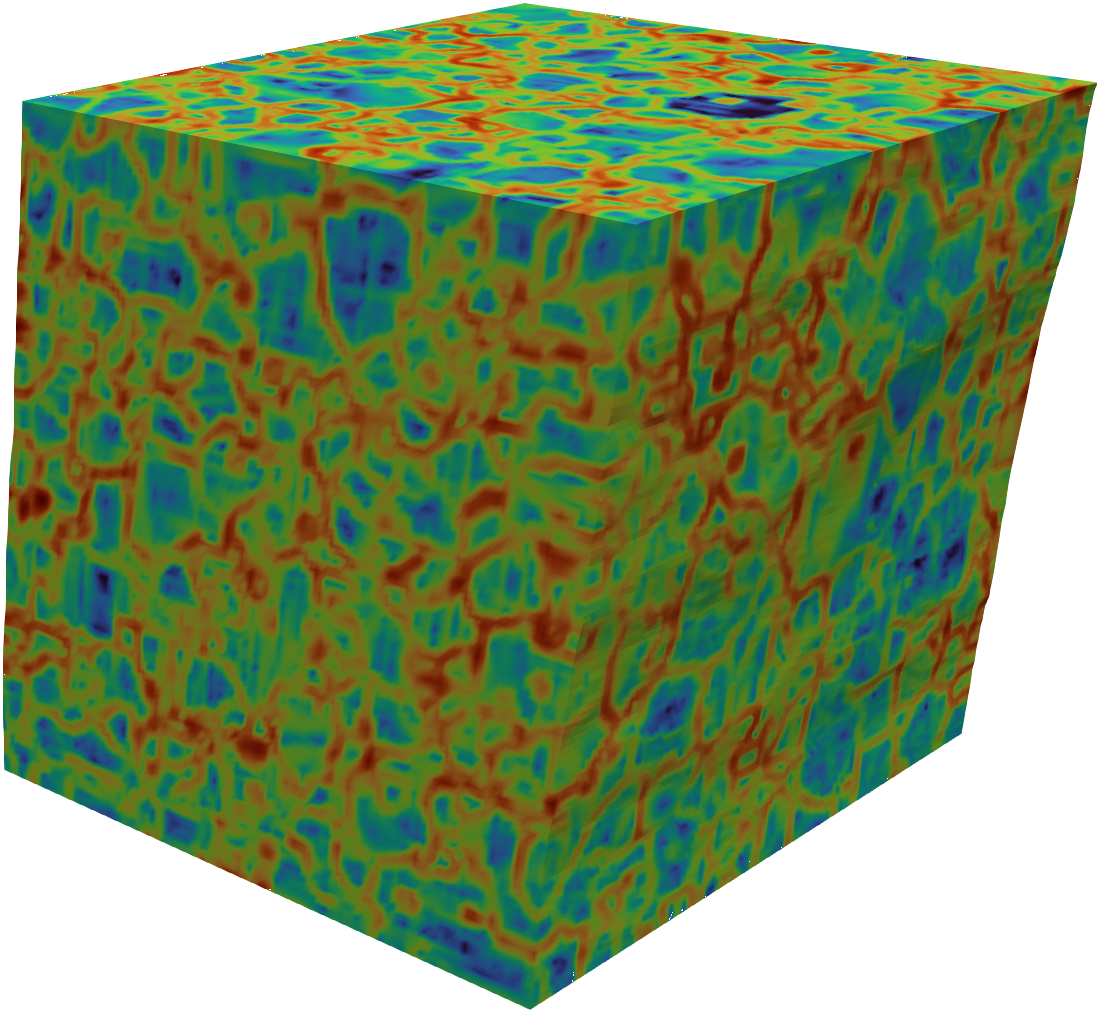}
        \caption{}
        \label{fig:gnd_1_3d}
    \end{subfigure}
    \hfill
    \begin{subfigure}{0.32\textwidth}  
        \centering
        \begin{overpic}[scale=0.12]{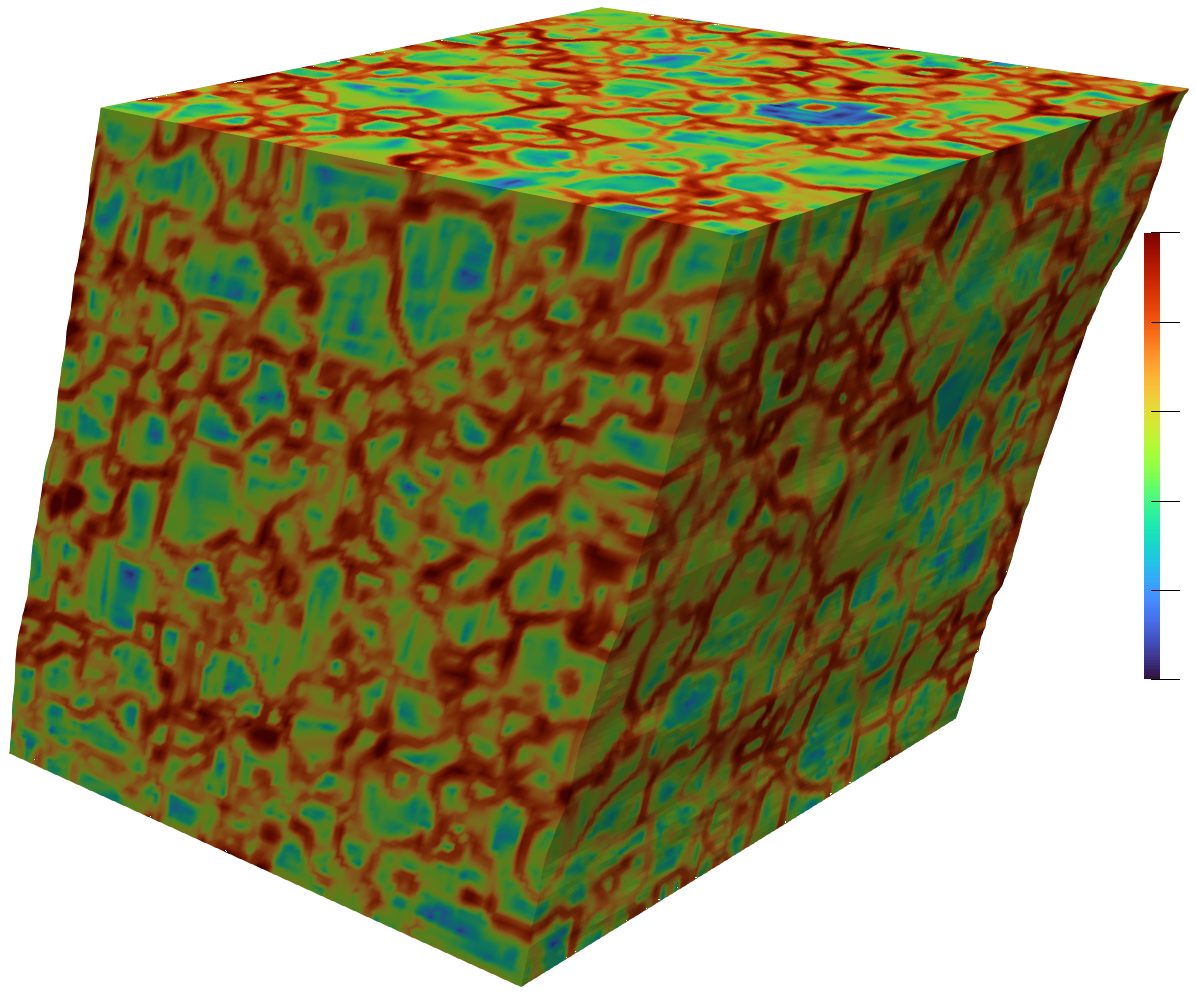}
            \put(95.5, 14){\textcolor{white}{\rule{3mm}{26mm}}}
        \end{overpic}
        \caption{}
        \label{fig:gnd_2_3d}
    \end{subfigure}
    \caption{Evolution of GND density plotted on a logarithmic scale ($\log(\rho_g/\bar{\rho}_s)$) at applied shear strains of a) 0.0 b) 0.15 and c) 0.30.}
    \label{fig:3d_gnd}
\end{figure}

\begin{figure}[htbp]
    \centering
    \begin{subfigure}{0.28\textwidth}  
        \centering
        \begin{overpic}[scale=0.13]{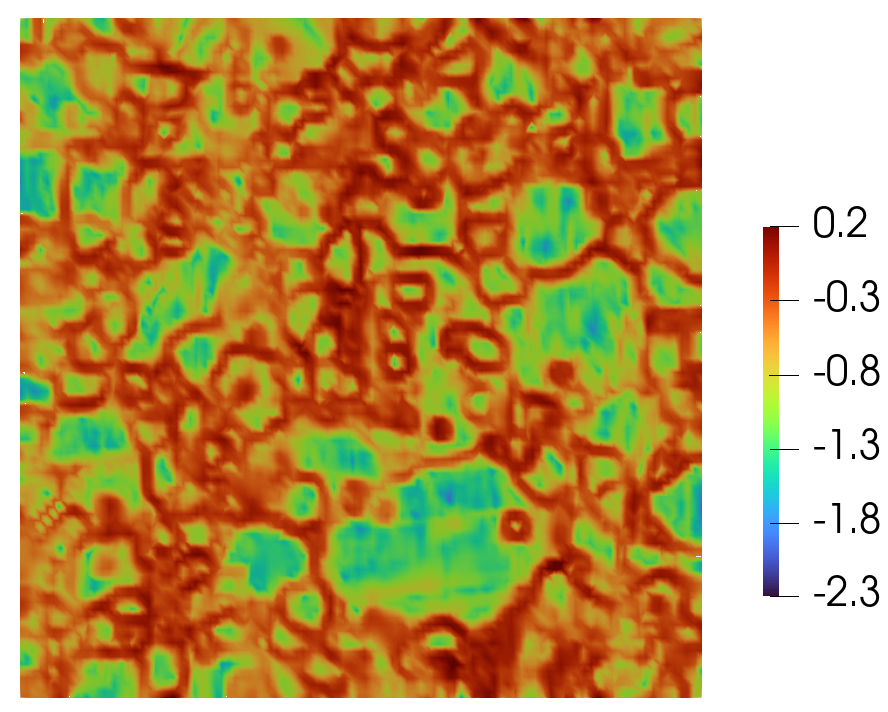}
            \put(85, 63){\fontsize{8}{5}\selectfont $\log(\rho_g/\bar{\rho}_s)$}
        \end{overpic}
        \caption{}
        \label{sfig:slice_x}
    \end{subfigure}
    \hfill
    \begin{subfigure}{0.28\textwidth}  
        \centering
        \includegraphics[scale=0.13]{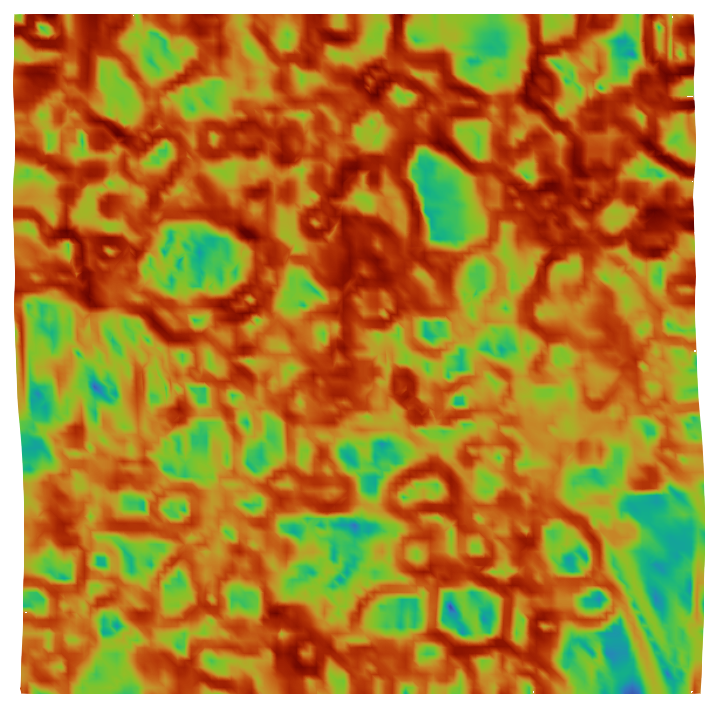}
        \caption{}
        \label{sfig:slice_y}
    \end{subfigure}
    \hfill
    \begin{subfigure}{0.28\textwidth}  
        \centering
        \includegraphics[scale=0.13]{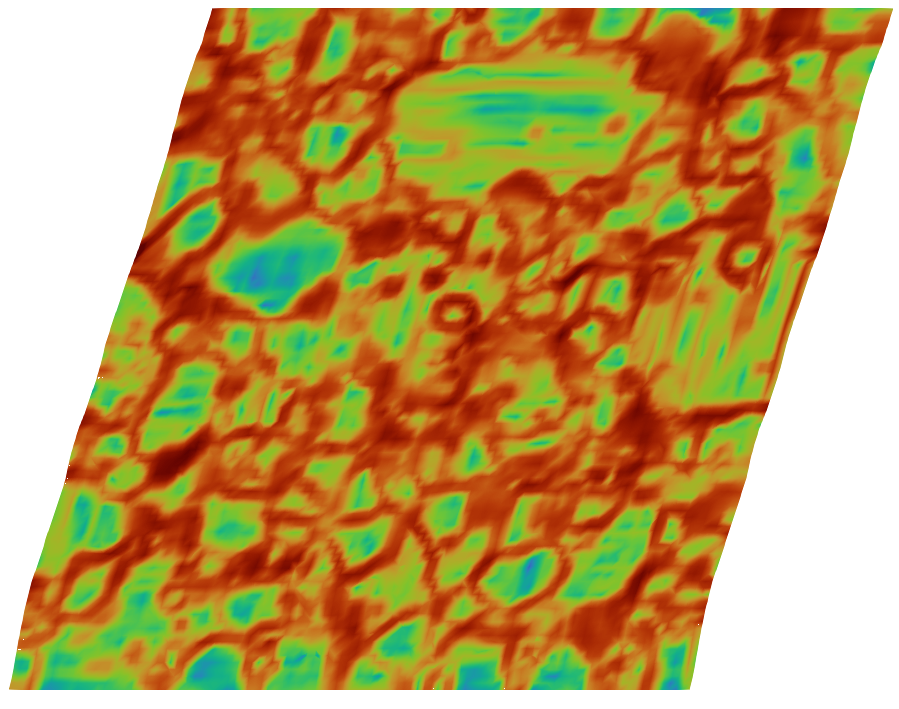}
        \caption{}
        \label{sfig:slice_z}
    \end{subfigure}
    \caption{GND density distribution on interior cross-sectional planes of the 3-d domain at applied shear strain of 0.30: (a) $x = 0$, (b) $y = 0$, and (c) $z = 0$ plane.}
    \label{fig:Slice_GND}
\end{figure}


\begin{figure}[htbp]
    \centering
    \begin{subfigure}{0.32\textwidth}  
        \centering
        \begin{overpic}[scale=0.11]{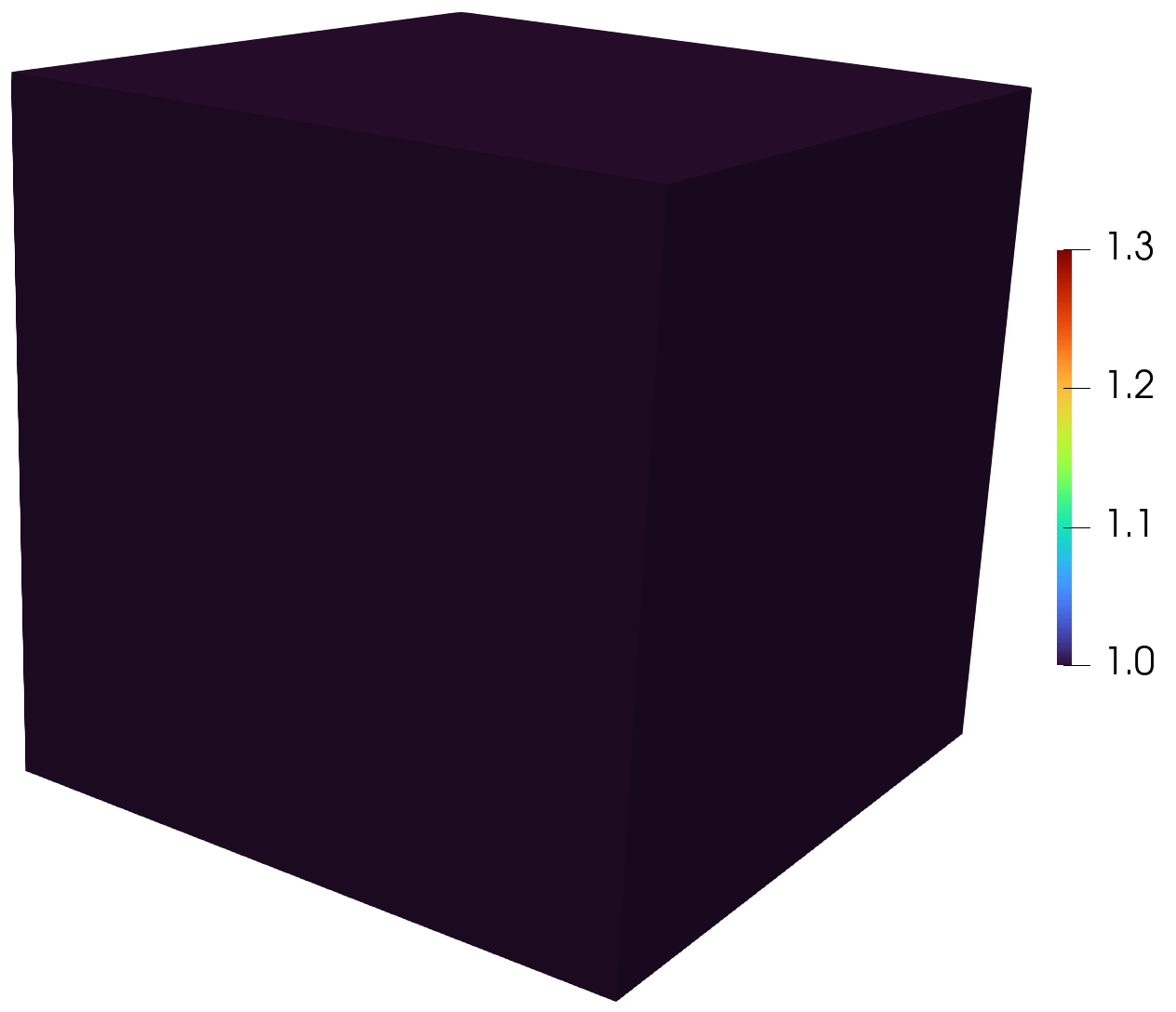}
            \put(90, 73){\fontsize{8}{5}\selectfont $g_{\theta}/g_0$}
        \end{overpic}
        \caption{}
        \label{fig:g_0_3d}
    \end{subfigure}
    \hfill
    \begin{subfigure}{0.32\textwidth}  
        \centering
        \includegraphics[width=\textwidth]{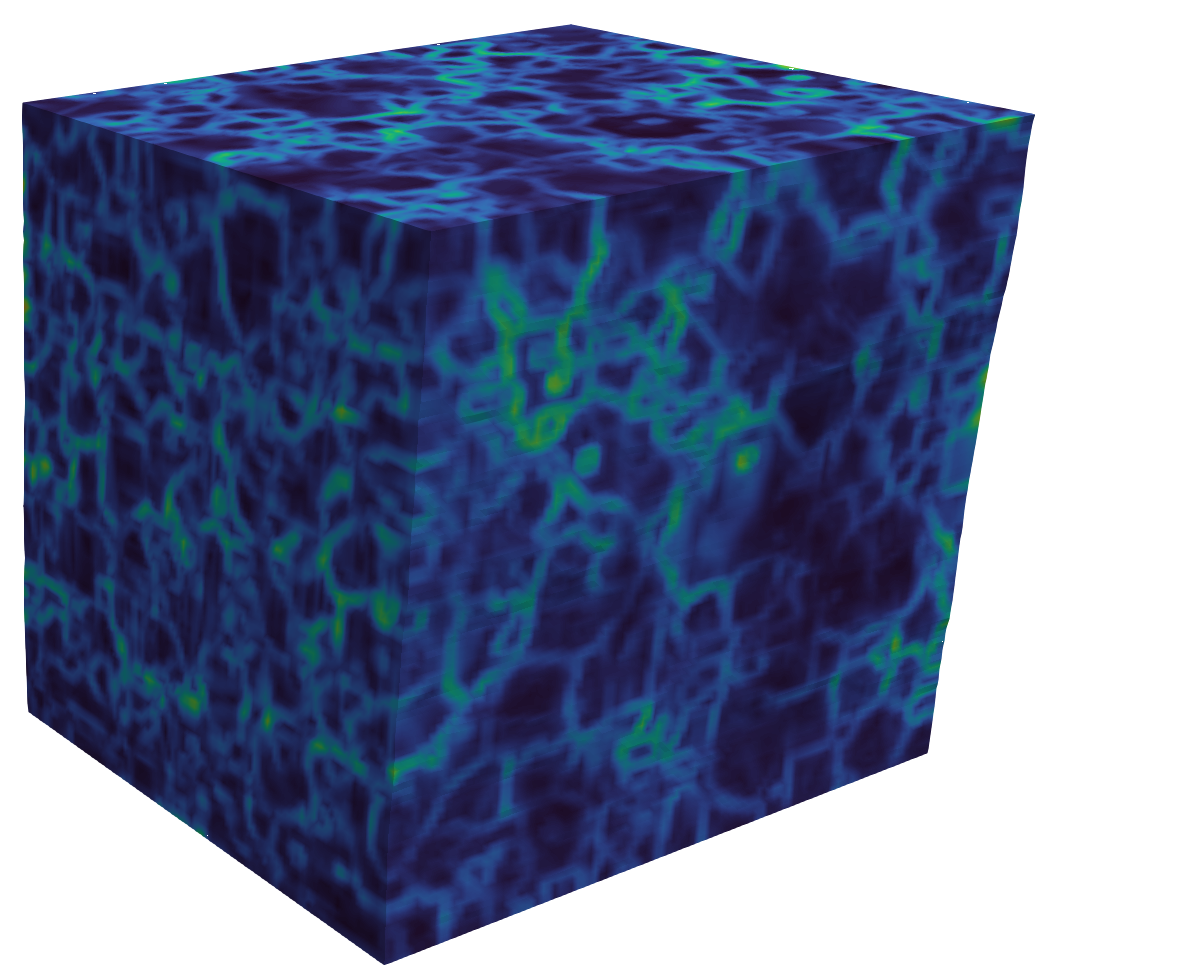}
        \caption{}
        \label{fig:g_1_3d}
    \end{subfigure}
    \hfill
    \begin{subfigure}{0.32\textwidth}  
        \centering
        \includegraphics[width=\textwidth]{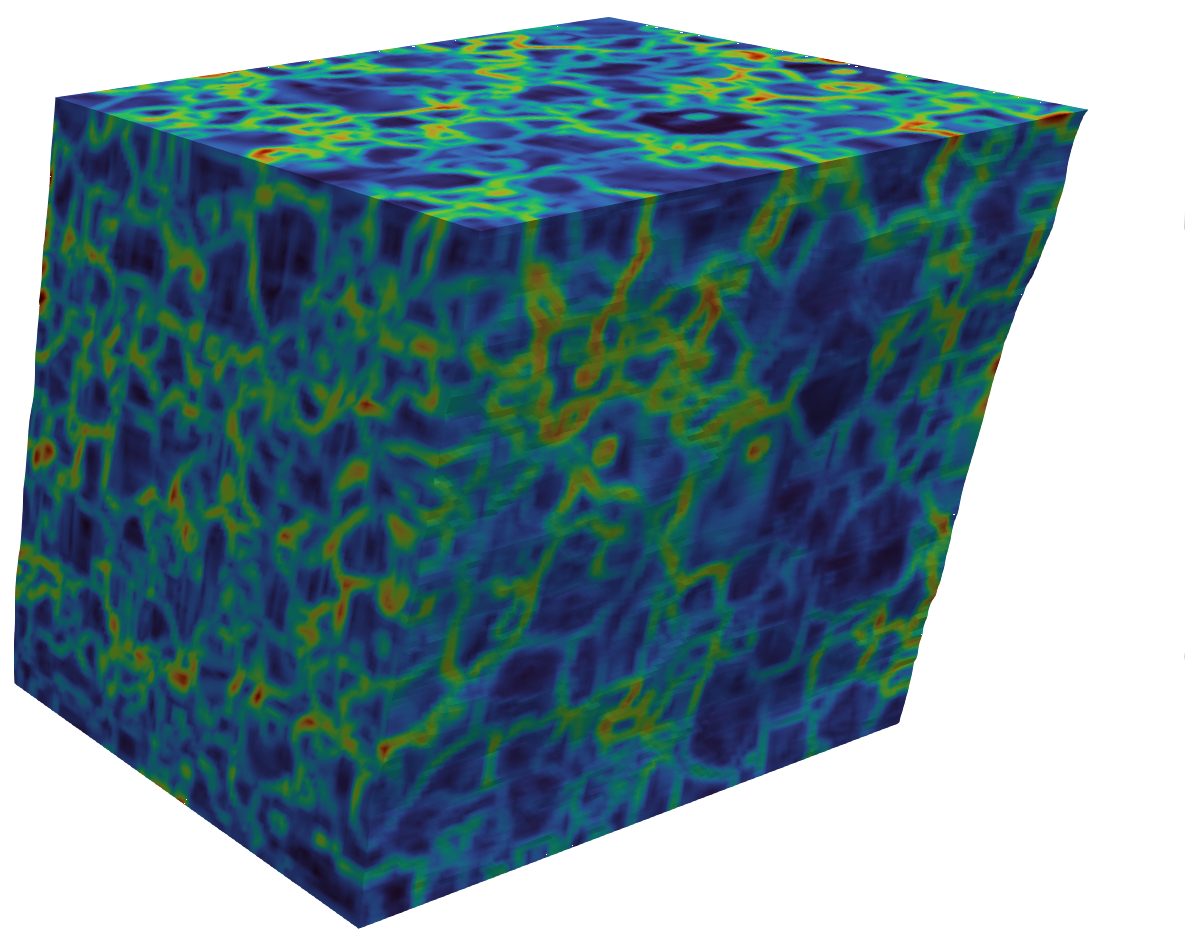}
        \caption{}
        \label{fig:g_2_3d}
    \end{subfigure}
    \caption{Evolution of normalized material strength ($g_\theta/g_0$) plotted at applied shear strains of a) 0.0 b) 0.15 and c) 0.30, where $g_0 = 0.40 \, GPa$ is the initial yield strength.}
    \label{fig:3d_strength}
\end{figure}

\begin{figure}[htbp]
    \centering
    \begin{subfigure}{0.32\textwidth}  
        \centering
        \begin{overpic}[scale=0.11]{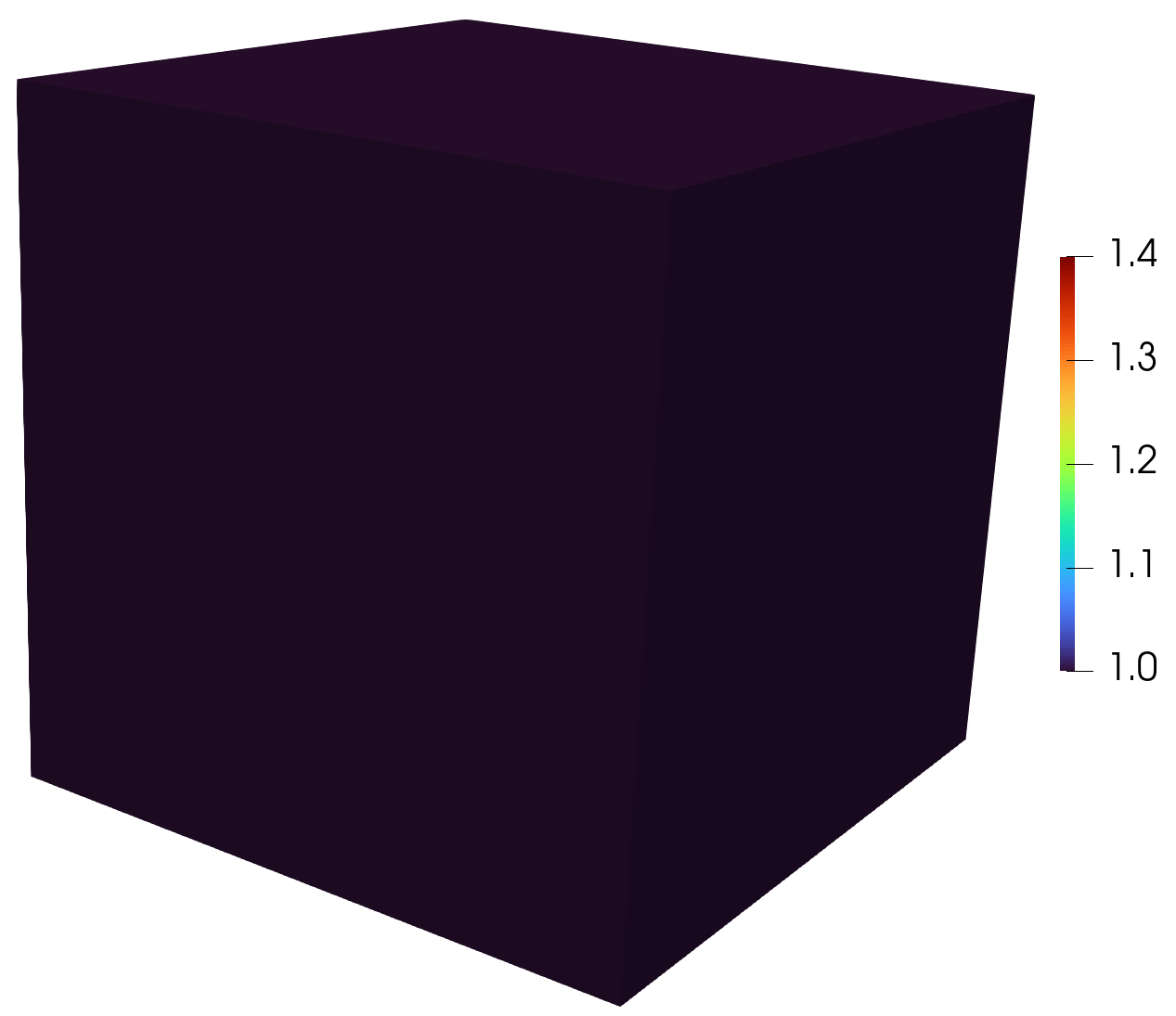}
            \put(90, 69){\fontsize{8}{5}\selectfont $\theta/\theta_0$}
        \end{overpic}
        \caption{}
        \label{fig:T_1}
    \end{subfigure}
    \hfill
    \begin{subfigure}{0.32\textwidth}  
        \centering
        \includegraphics[width=\textwidth]{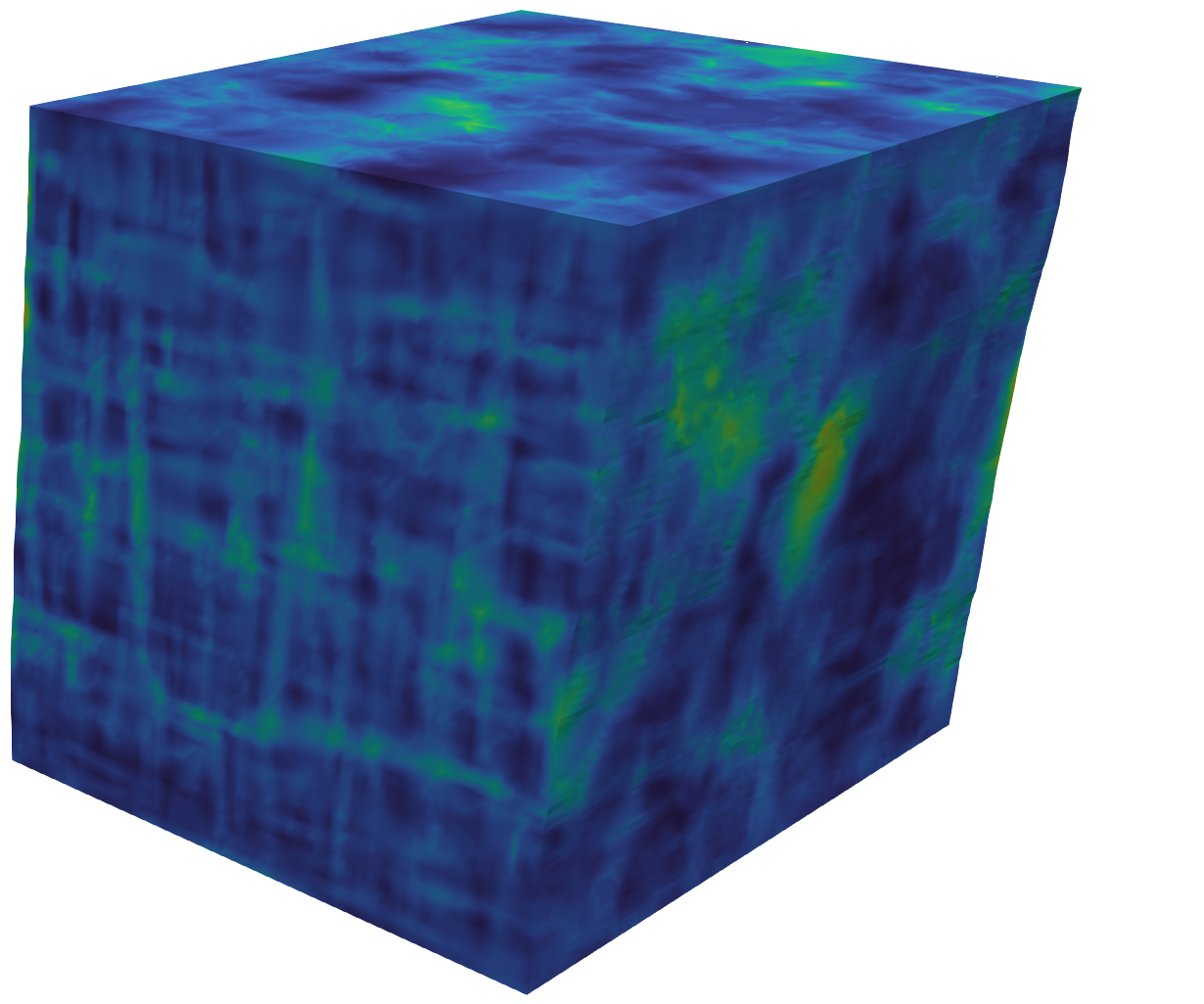}
        \caption{}
        \label{fig:T_2}
    \end{subfigure}
    \hfill
    \begin{subfigure}{0.32\textwidth}  
        \centering
        \includegraphics[width=\textwidth]{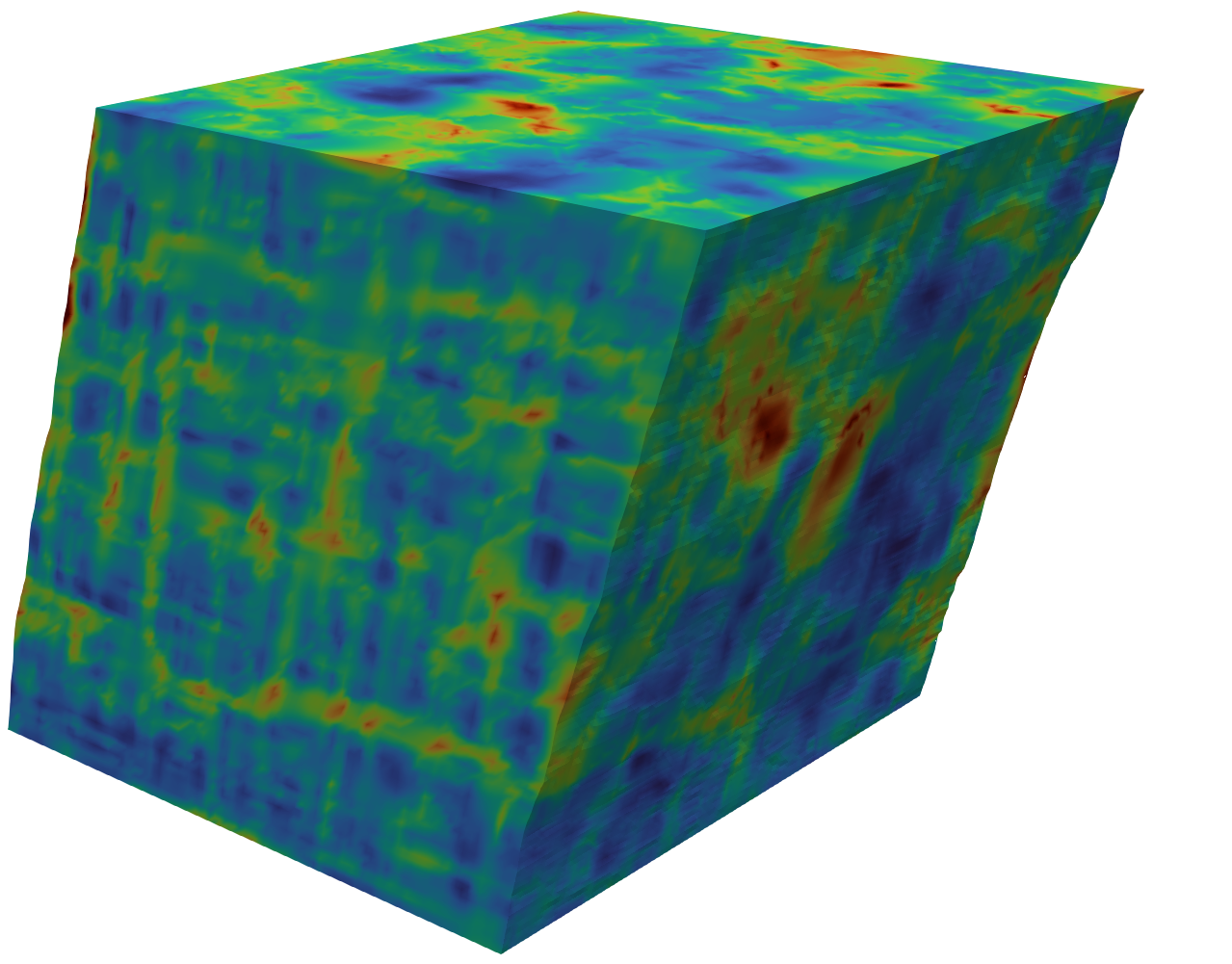}
        \caption{}
        \label{fig:T_3}
    \end{subfigure}
    \caption{Evolution of normalized temperature ($\theta/\theta_0$) plotted at applied shear strains of a) 0.0 b) 0.15 and c) 0.30, where $\theta_0 = 293 \, K$ is the reference temperature.}
    \label{fig:3d_Temp}
\end{figure}

\subsection*{Subgrain dislocation microstructure formation}

Analogous to the 2-d case, substructure formation is observed within individual grains in the 3-d polycrystal simulation as well. Unlike the planar substructure seen in 2-d, however, the substructure in 3-d is non-planar in nature, manifesting as 3-d subgrain formation within the grain interior. To illustrate this, we isolate a single grain from the polycrystal, which, as marked in Fig.~\ref{fig:3_axis_ang}, begins the simulation with a nominally uniform crystallographic orientation. Following deformation, the GND density distribution within this grain is presented in Fig.~\ref{fig:3_axis_gnd}, revealing the emergence of GND in the grain interior. This can be treated as a signature of subgrain formation. A zoomed view is provided in Fig.~\ref{fig:3_comb_sm} for additional clarity. Importantly, this substructure formation is not isolated to a single grain but is observed throughout the polycrystal domain, with the subgrain structure becoming more pronounced with increasing deformation.

\begin{figure}
 \centering
    \begin{subfigure}{0.4\textwidth}  
        \centering
        \includegraphics[width=\textwidth]{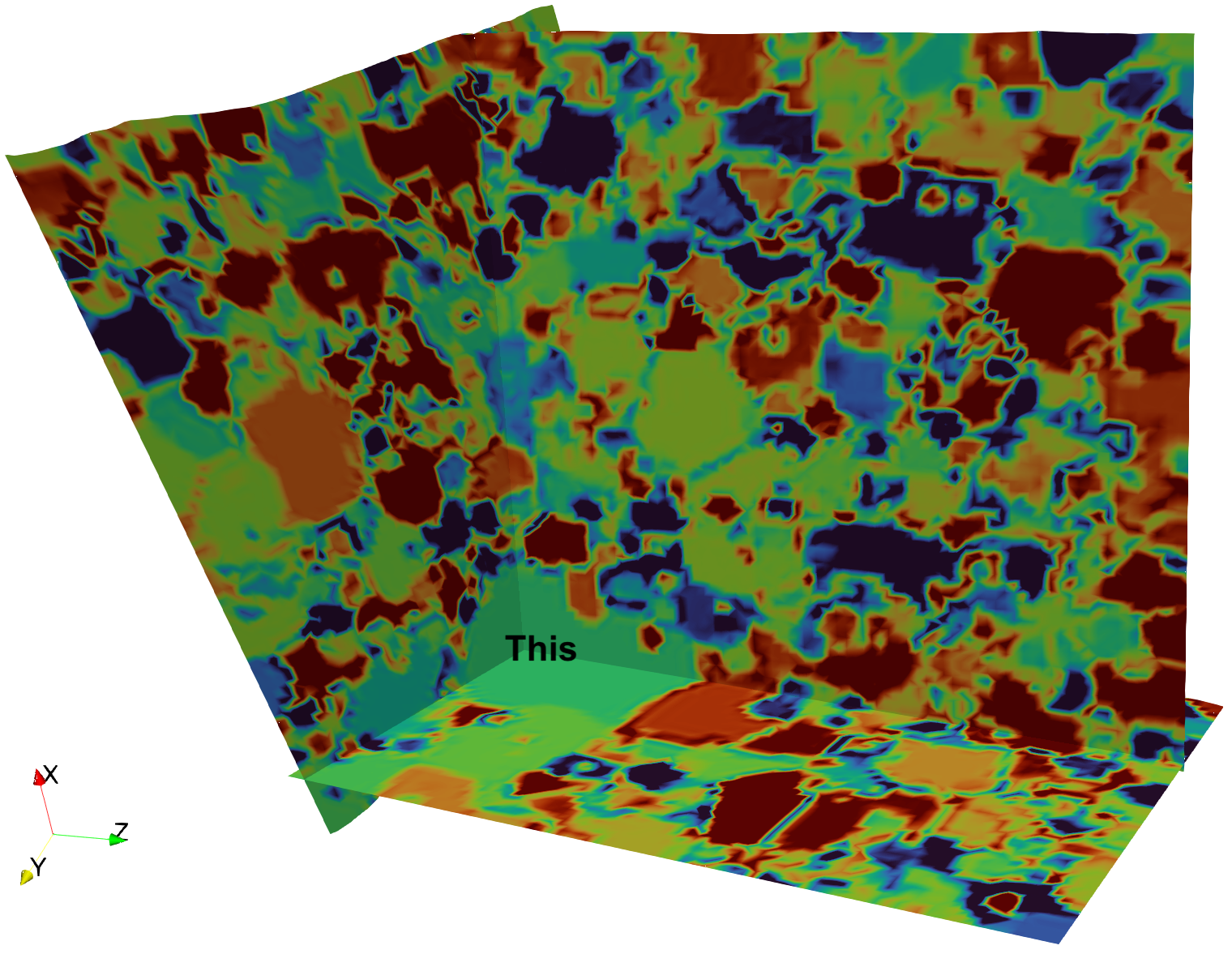}
        \caption{}
        \label{fig:3_axis_ang}
    \end{subfigure}
    \hfill
    \centering
    \begin{subfigure}{0.4\textwidth}  
        \centering
        \begin{overpic}[scale=0.11]{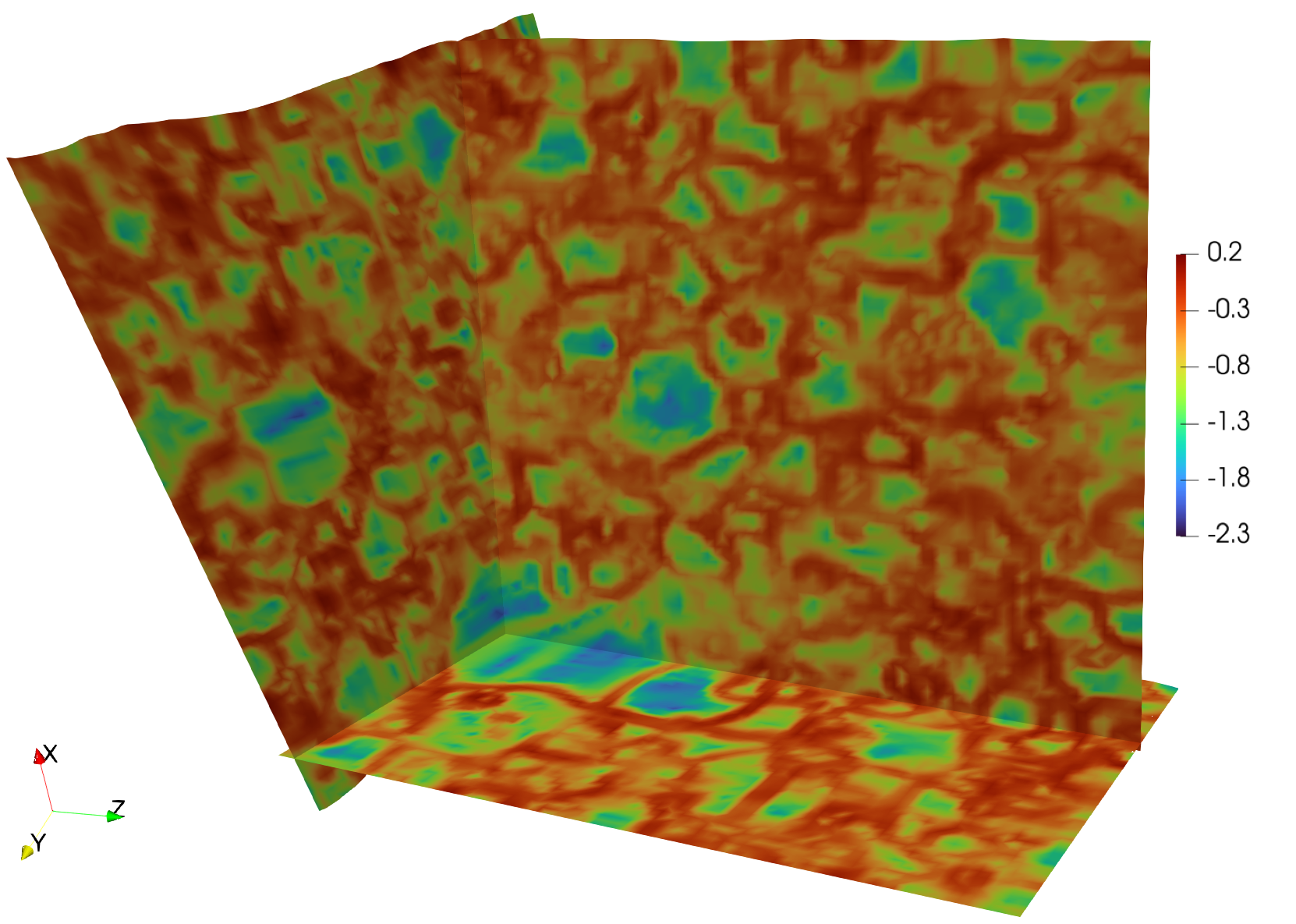}
            \put(90, 55){\fontsize{8}{5}\selectfont $\log(\rho_g/\bar{\rho}_s)$}
        \end{overpic}
        \caption{}
        \label{fig:3_axis_gnd}
    \end{subfigure}
    \caption{(a) Rotation angle field on three cross-sections of the 3-d polycrystal showing the initial grain structure. The grain selected for discussion in Fig.~\ref{fig:3_comb_sm} is marked. (b) GND density distribution within the isolated grain at an applied strain of 0.3, pointing to a potential non-planar subgrain substructure within the grain.}
    \label{fig:3axis_fig}
\end{figure}

\begin{figure}
 \centering
    \begin{subfigure}{0.4\textwidth}  
        \centering
        \includegraphics[width=\textwidth]{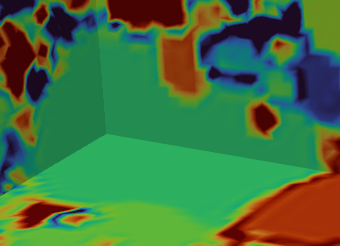}
        \caption{}
        \label{fig:3_axis_ang_sm}
    \end{subfigure}
    \hfill
    \centering
    \begin{subfigure}{0.4\textwidth}  
        \centering
        \includegraphics[width=\textwidth]{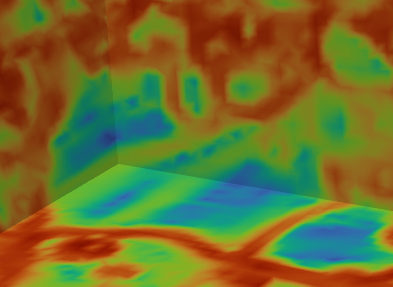}
        \caption{}
        \label{fig:3_axis_gnd_sm}
    \end{subfigure}
    \caption{Zoomed view of (a) the rotation angle field and (b) the GND density distribution within an isolated grain from the polycrystalline assembly marked in Fig.~\ref{fig:3axis_fig}, at an applied strain of 0.3.}
    \label{fig:3_comb_sm}
\end{figure}

\subsection{Non-localization under boundary and initial conditions corresponding to nominally homogeneous deformation}\label{sec:non_localized}

The experimental observations (Fig.~\ref{fig:th_combined}) corresponding to macroscopic top-hat experiments reveal that localized bands of deformation fall within a size range of $10 \mu m - 40 \mu m$ under dynamic loading conditions. We extend MFDM simulations on the selected RVE (Fig.~\ref{fig:grain_strct}) of size $80~\mu m^2$ to significantly higher strains compared to the discussion in Sec.~\ref{subsec:ASB_Comp} to investigate whether further localization or stress softening emerges beyond the initially observed response. A convergence study is performed using differently meshed domains, and the results remain converged to a fair degree considering the large overall strains involved, as shown in Fig.~\ref{fig:ss_MFDM_fur}. The local shear strain reaches values as high as 3.00 at an applied/average nominal engineering shear strain of 1.50. Even at these higher strains, no further localization develops, as shown in Fig.~\ref{fig:F12_MFDM_400}, nor does any softening appear in the stress-strain response. Contrast in $F_{12}$ is visible across the domain and is likely influenced by the presence of a relatively large grain (homogeneous material) in the microstructure. The temperature, GND density, and material strength profiles at 1.50 applied shear strain are shown in Fig.~\ref{fig:150_SE}.

Under the controlled boundary conditions for nominally homogeneous deformation (for a homogeneous material under quasi-static loading) as used in this study, it is not clear whether softening observed in macroscopic experiments should be expected here in the absence of ductile damage mechanism in the model; note the reductions in material strength observed within the band (Fig.~\ref{fig:Se_150_g}), which can possibly serve as sites for ductile damage nucleation. For the domain size and material properties considered, our model, which does not incorporate any damage mechanism, predicts no softening.

We have been able to produce cases in which MFDM exhibits softening in the polycrystal stress-strain response by increasing the value of the Taylor--Quinney factor ($\kappa$) or reducing the value of the GND hardening parameter ($k_0$). For certain values of the $\kappa$ and/or $k_0$, softening is observed; however, mesh convergence of the solution was not observed for the meshes considered, although the performance was much improved in comparison to the corresponding classical plasticity simulations. This suggests that there may exist a range of the parameters $\kappa$ and $k_0$ where softening with mesh convergence may be observed; the discussion at the end of the following paragraph illustrates another facet of this issue, even without any material parameter adjustment. Such matters are left for future study.

We note here that the dynamic nature of the loading induces a higher stress at the center of the domain where the stress waves from the top and bottom boundaries meet, leading to spatially heterogeneous yielding. However, as is observed, this heterogeneity is not sufficient to induce a persistent shear band with progressive loading in MFDM, in contrast to CCP. This trend persists even for the $J_2$-MFDM model of plasticity as shown in Fig.~\ref{fig:F12_J2_evolve} where the polycrystalline aggregate is treated as an effectively homogeneous material (with no corresponding initialization of a grain microstructure). In this latter scenario, a distinct sequence of bands emerge in the shear component $F_{12}$ of the deformation gradient. At 0.15 applied shear strain (Fig.~\ref{fig:F12_j2_1}), $F_{12}$ concentrates into a narrow band at the center of the domain, which persists as the dominant feature. As loading continues to 0.60 applied strain (Fig.~\ref{fig:F12_j2_2}), this central band remains intact while slip-band-like structures begin to emerge on either side of it, spreading outward from the center. By 0.90 strain (Fig.~\ref{fig:F12_j2_3}), these structures homogenize over the domain, with contrast of high strain to average strain remaining approximately uniform with progress of loading. This trend towards greater homogeneity in deformation is accompanied with a gradual, mesh-converged, softening response as shown in Fig.~\ref{fig:ss_MFDM_J2}, in contrast to the corresponding non-convergence of the classical plasticity response for the same problem shown in  Fig.~\ref{fig:ss_CCP_J2}. The gradual softening in the $J_2$-MFDM simulation arises due to the spatially-extended thermal softening following from delocalized plastic straining and the lesser degree of hardening due to reduced GND accumulation  because of the absence of grain boundaries, as compared to the MFDM response accounting for the grain microstructure. It is worth noting that this relative decrease in GND accumulation in $J_2$-MFDM occurs even though the model equations contain the same GND hardening term and enforce a common jump condition at surfaces of discontinuity in the total plastic strain rate, $\bfalpha \times \bfV + \bfL^p$, were they to exist.

Taken together, these results attest to the strong stabilizing and homogenizing effects of the GND hardening in MFDM, and the kinematically fundamental nature of GND accumulation in the theory, rooted in a conservation law for Burgers vector content.

\begin{figure}[htbp]
    \centering
    \begin{subfigure}{0.33\textwidth}  
        \centering
\includegraphics[scale=0.4]{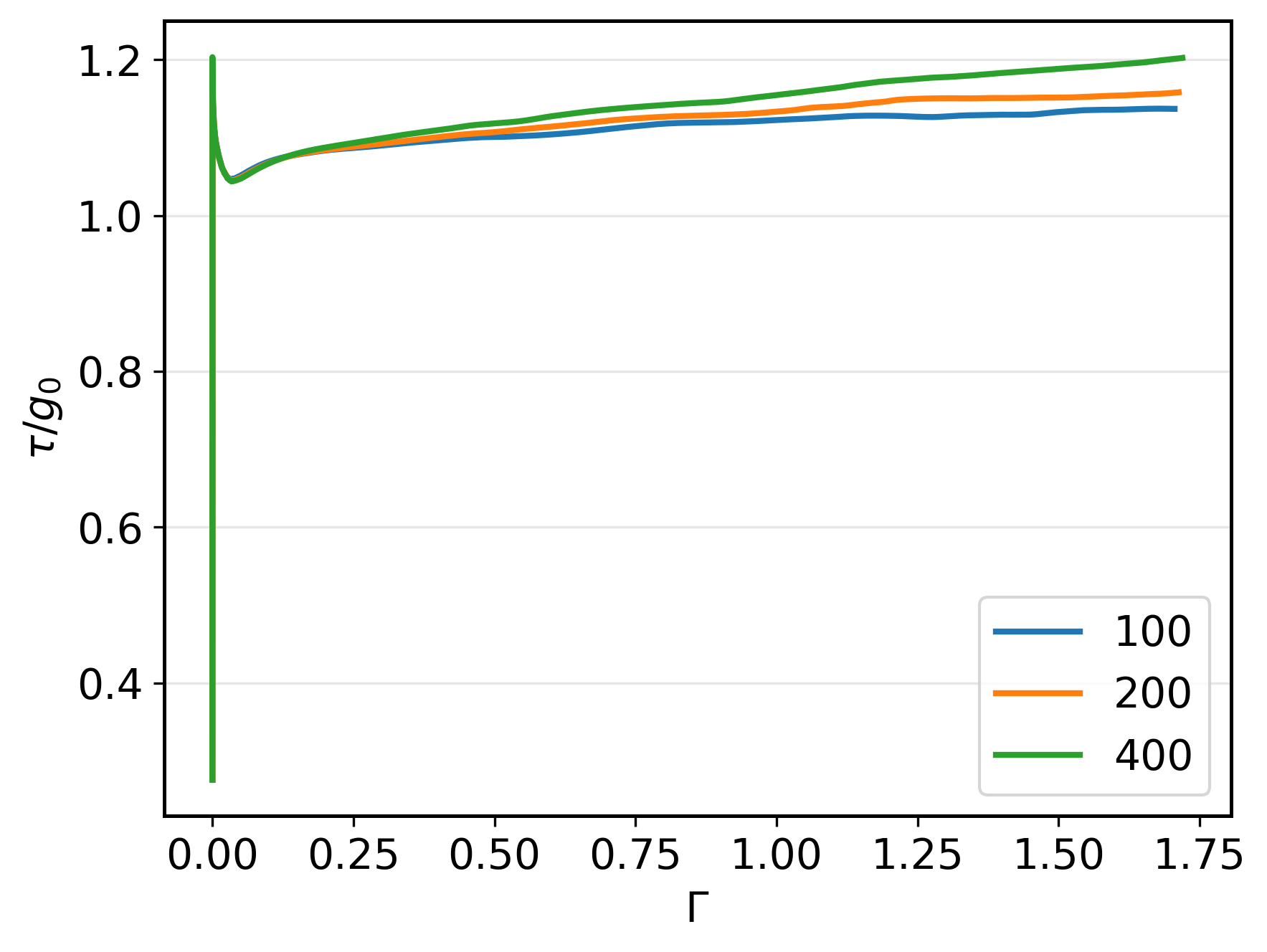}
        \caption{}
        \label{fig:ss_MFDM_fur}
    \end{subfigure}
    \hfill
    \begin{subfigure}{0.66\textwidth}  
        \centering
        \begin{overpic}[scale=0.16]{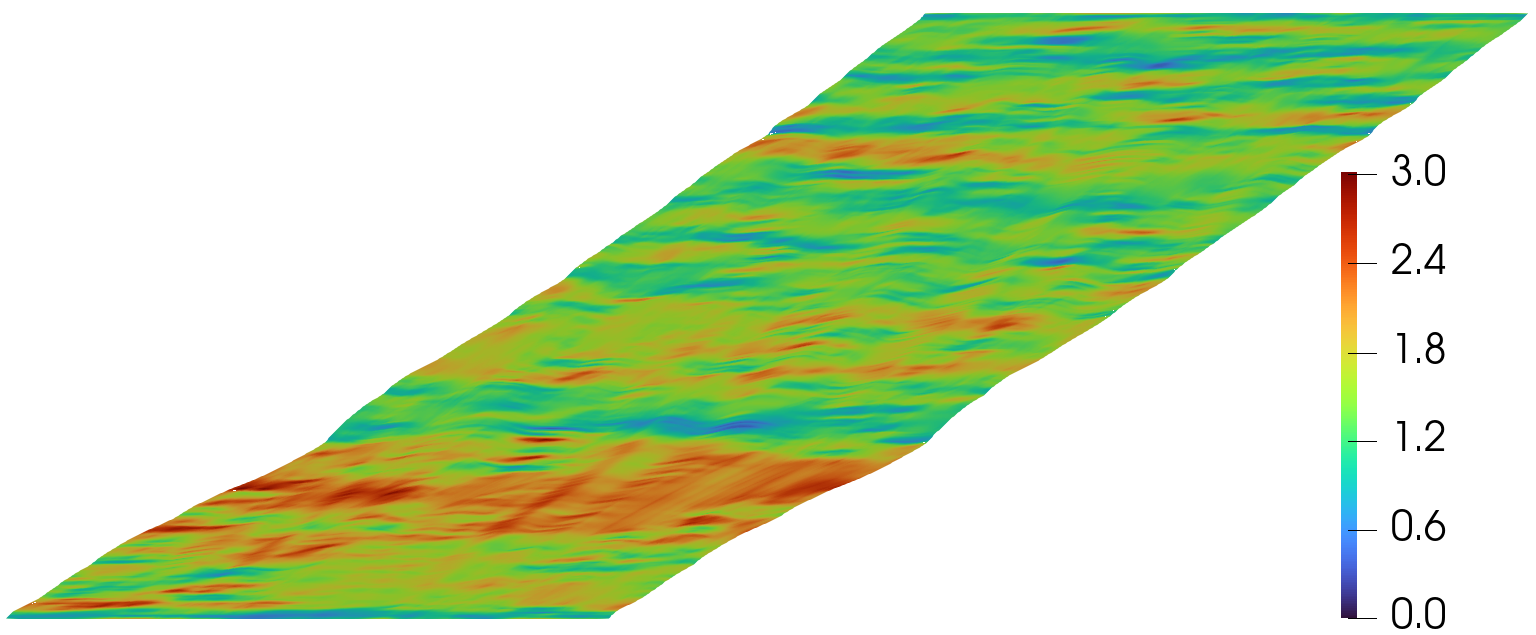}
            \put(96, 17){\rotatebox{0}{\fontsize{8}{5}\selectfont $F_{12}$}}
        \end{overpic}
        \caption{}
        \label{fig:F12_MFDM_400}
    \end{subfigure}
    \caption{Extended MFDM simulation results. (a) Normalized stress-strain response ($\tau/g_0$ vs.\ $\Gamma$) for three mesh resolutions ($100^2$, $200^2$, and $400^2$), demonstrating converged behavior up to 1.70 applied strain with no observed softening. (b) $F_{12}$ field at 1.50 strain, showing no further localization.}
    \label{fig:conv_stress_strain_2}
\end{figure}

\begin{figure}[htbp]
    \centering
    \begin{subfigure}{0.30\textwidth}  
        \centering
        \begin{overpic}[width=\textwidth]{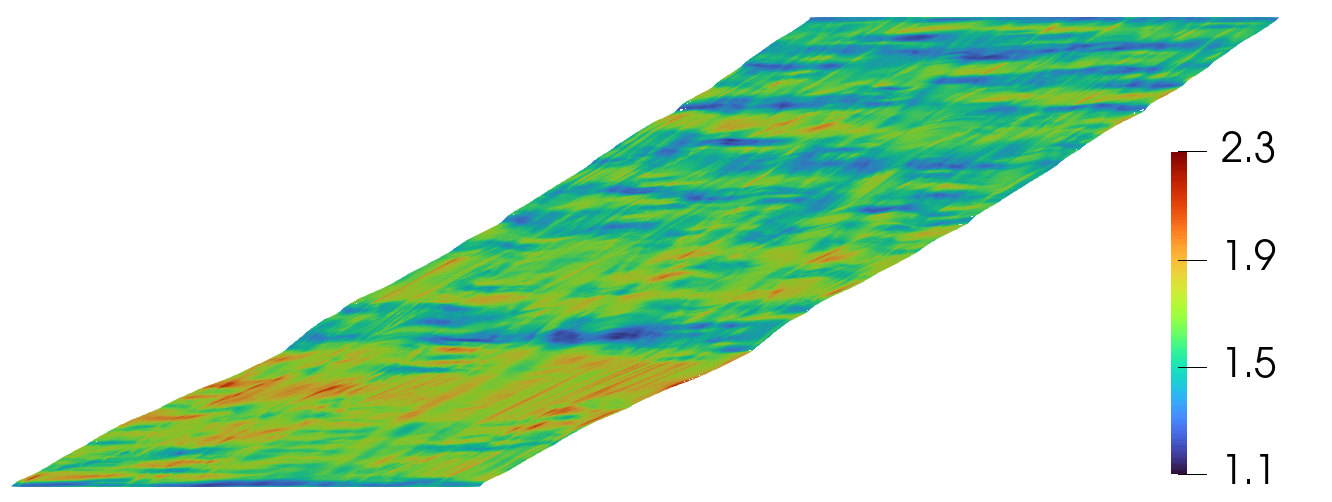}
            \put(100, 17){\rotatebox{0}{\fontsize{8}{5}\selectfont $\theta/\theta_0$}}
        \end{overpic}
        \label{fig:Se_150_temp}
    \end{subfigure}
    \hfill
    \begin{subfigure}{0.30\textwidth}  
        \centering
        \begin{overpic}[width=\textwidth]{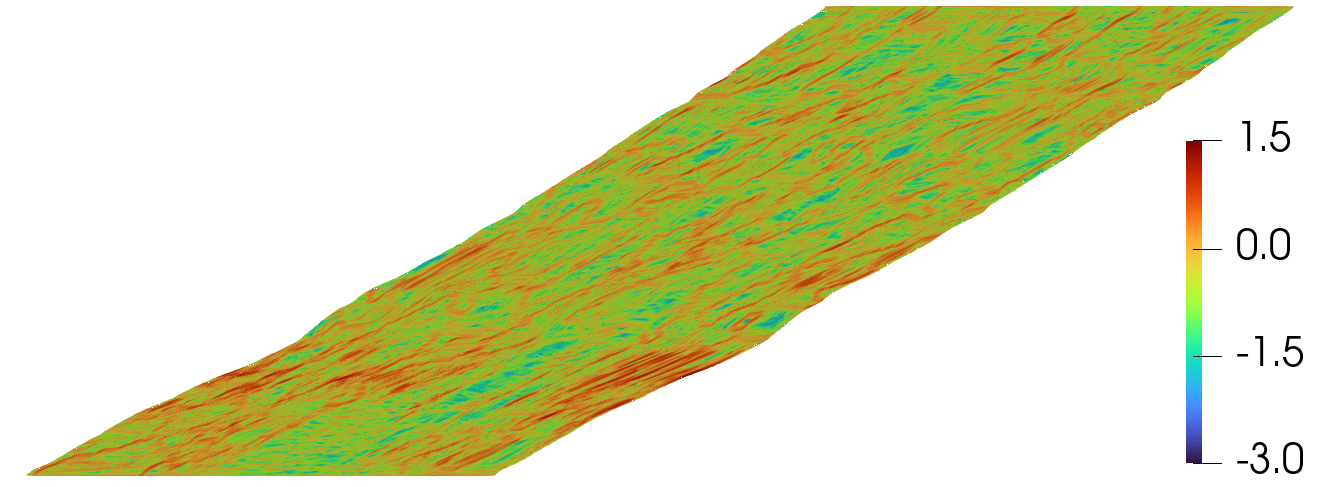}
            \put(100, 17){\rotatebox{0}{\fontsize{8}{5}\selectfont $\log(\rho_g/\bar{\rho}_s)$}}
        \end{overpic}
        \caption{}
        \label{fig:Se_150_GND}
    \end{subfigure}
    \hfill
    \begin{subfigure}{0.30\textwidth}  
        \centering
        \begin{overpic}[width=\textwidth]{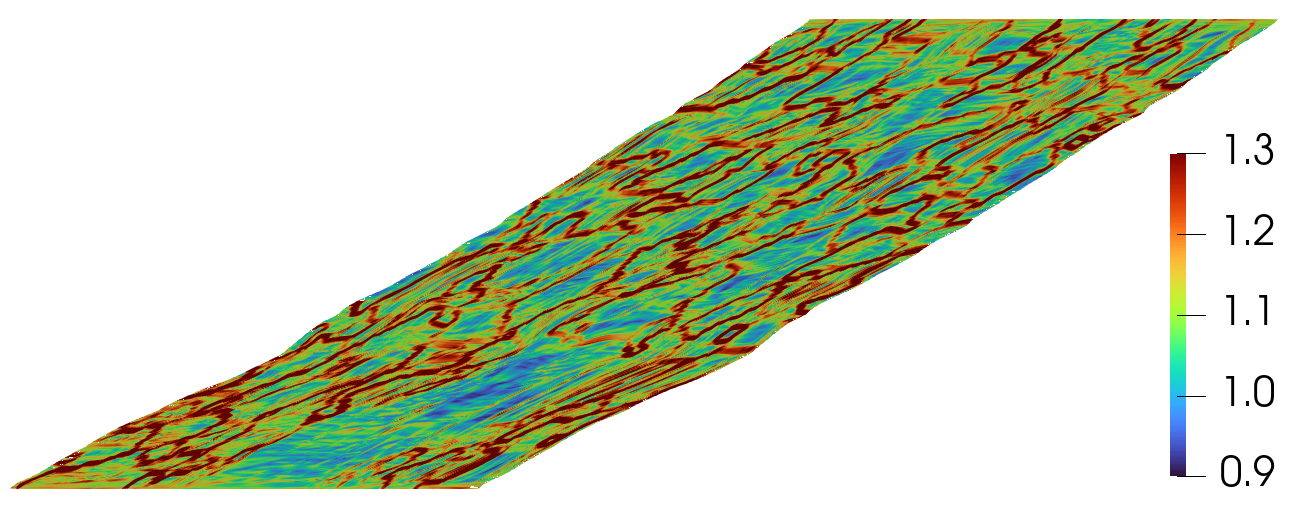}
            \put(100, 17){\rotatebox{0}{\fontsize{8}{5}\selectfont $g_{\theta}/g_0$}}
        \end{overpic}
        \caption{}
        \label{fig:Se_150_g}
    \end{subfigure}
    \caption{(a) Temperature, (b) GND density field, and (c) material strength (GPa) shown at 1.50 applied strain.}
    \label{fig:150_SE}
\end{figure}

\begin{figure}[htbp]
    \centering
    \begin{subfigure}{0.28\textwidth}  
        \centering
        \begin{overpic}[scale=0.10]{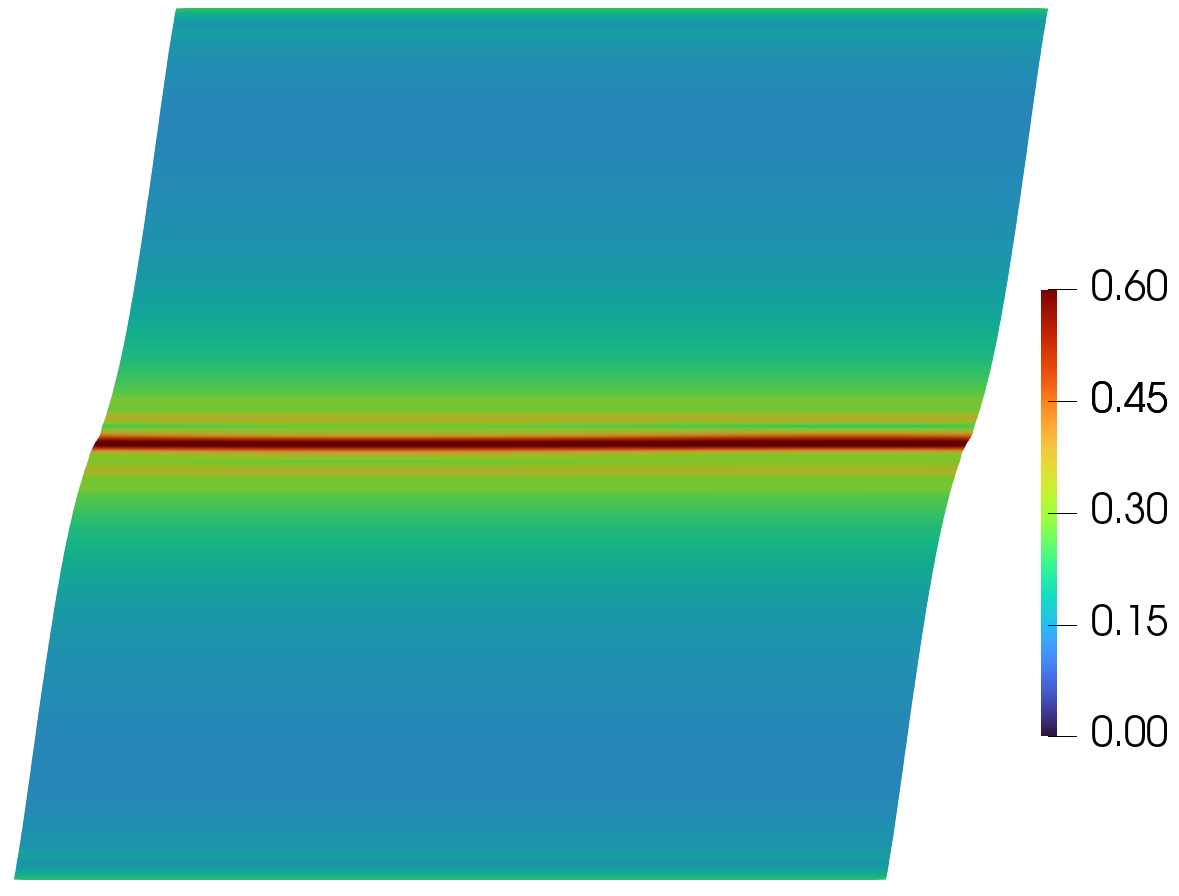}
            \put(102, 30){\rotatebox{0}{\fontsize{8}{5}\selectfont $F_{12}$}}
        \end{overpic}
        \caption{}
        \label{fig:F12_j2_1}
    \end{subfigure}
    \hfill
    \begin{subfigure}{0.28\textwidth}  
        \centering
        \begin{overpic}[scale=0.10]{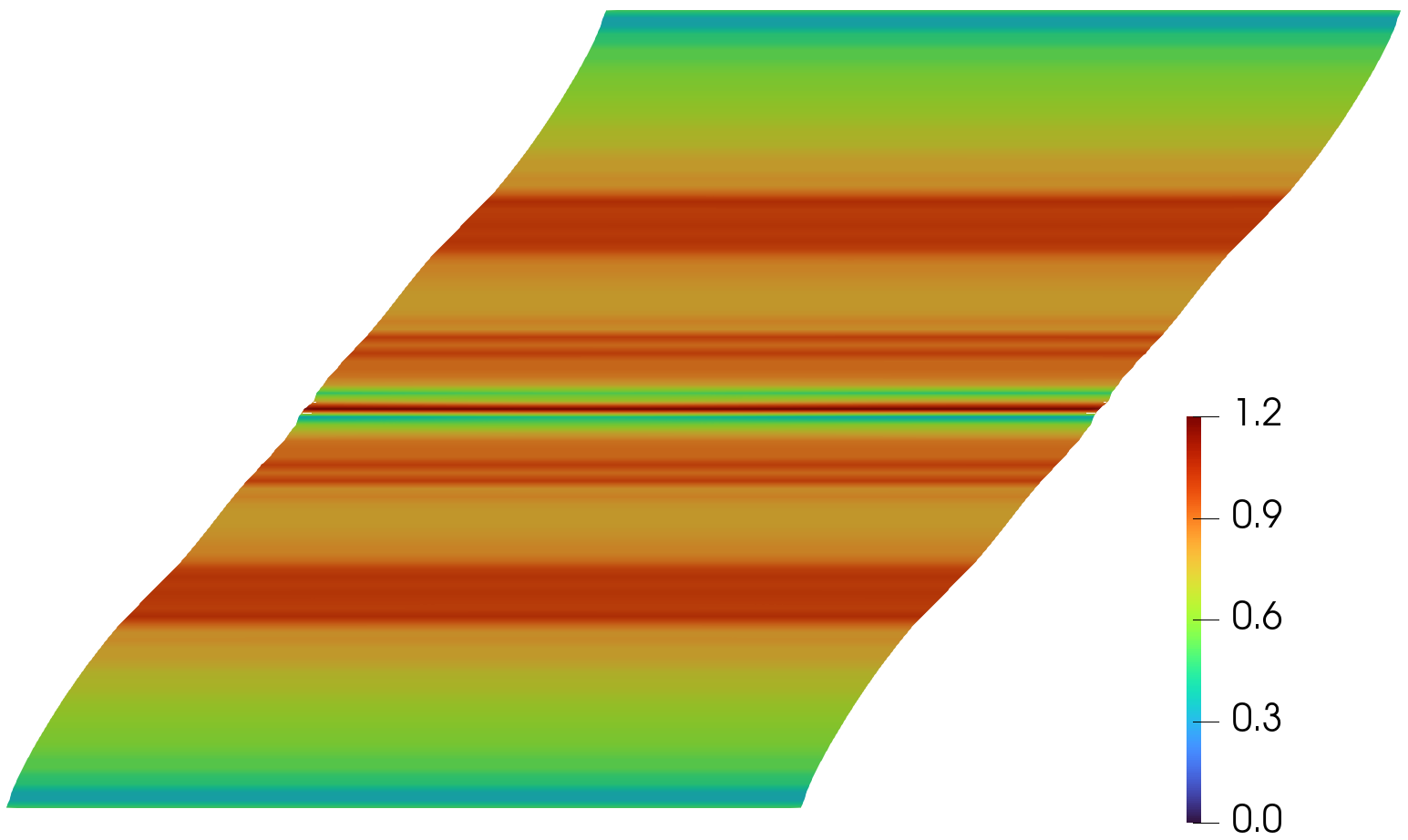}
            \put(92, 14){\rotatebox{0}{\fontsize{8}{5}\selectfont $F_{12}$}}
        \end{overpic}
        \caption{}
        \label{fig:F12_j2_2}
    \end{subfigure}
    \hfill
    \begin{subfigure}{0.28\textwidth}  
        \centering
        \begin{overpic}[scale=0.10]{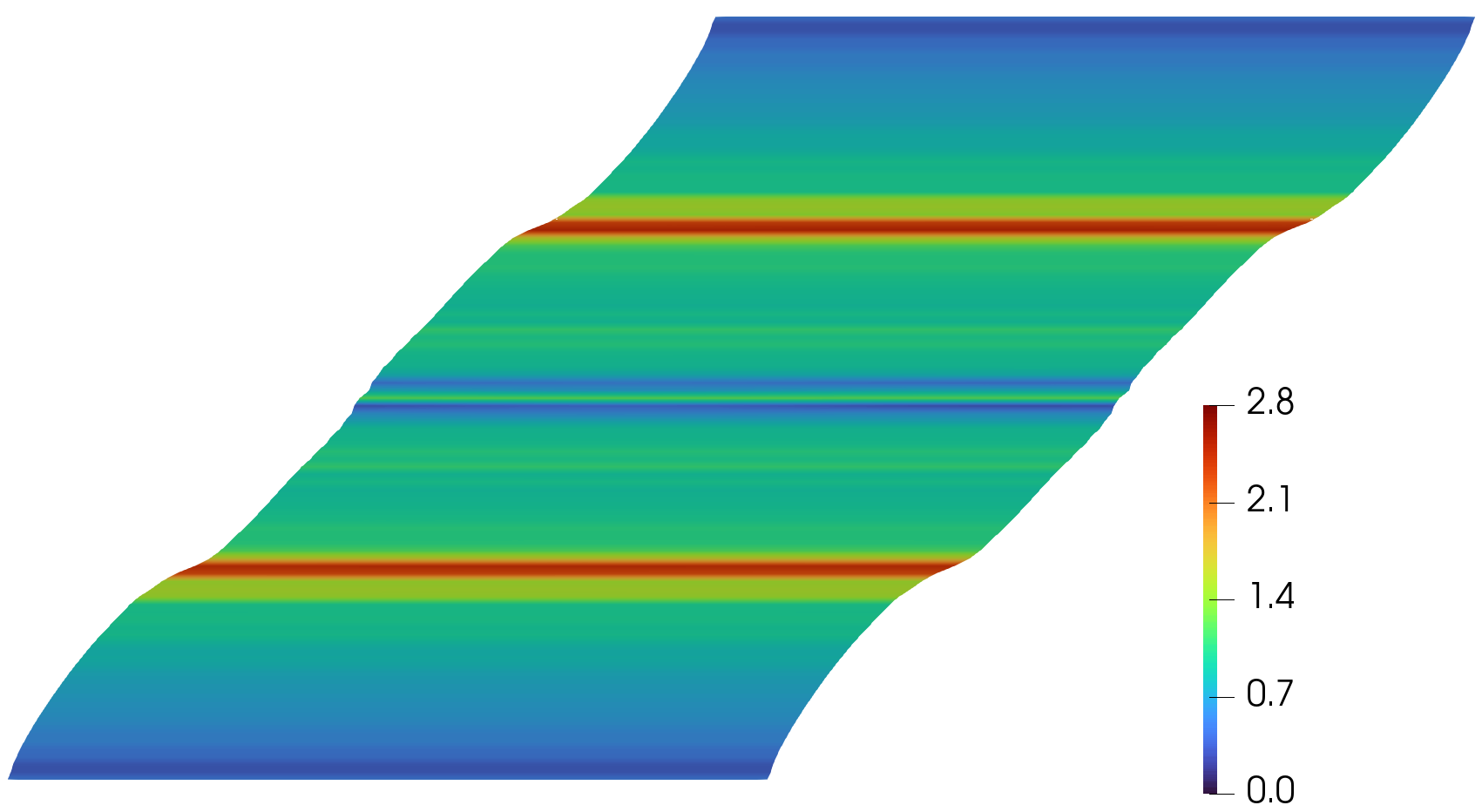}
            \put(90, 12){\rotatebox{0}{\fontsize{8}{5}\selectfont $F_{12}$}}
        \end{overpic}
        \caption{}
        \label{fig:F12_j2_3}
    \end{subfigure}
    \caption{Evolution of the shear component $F_{12}$ of the deformation gradient for $J_2$ model at (a) 0.15, (b) 0.60, and (c) 0.90 applied shear strain, showing the stages in which a central band forms and persists, before spreading outward into slip-band-like structures. The color bars for the sub-figures span different ranges.}
    \label{fig:F12_J2_evolve}
\end{figure}

\begin{figure}[htbp]
    \centering
    \begin{subfigure}{0.48\textwidth}  
        \centering
        \includegraphics[scale=0.5]{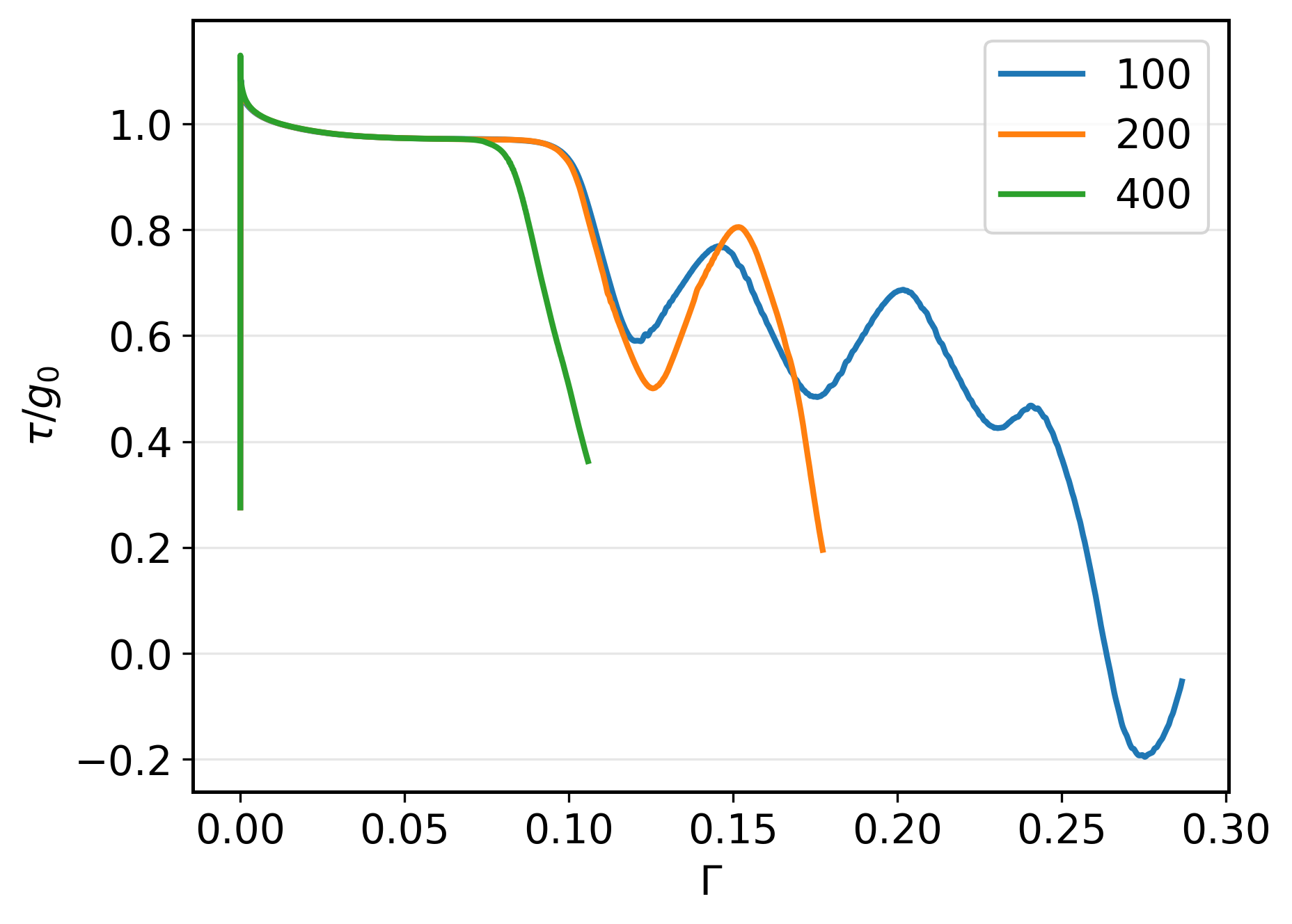}
        \caption{}
        \label{fig:ss_CCP_J2}
    \end{subfigure}
    \hfill
    \begin{subfigure}{0.48\textwidth}  
        \centering
        \includegraphics[scale=0.5]{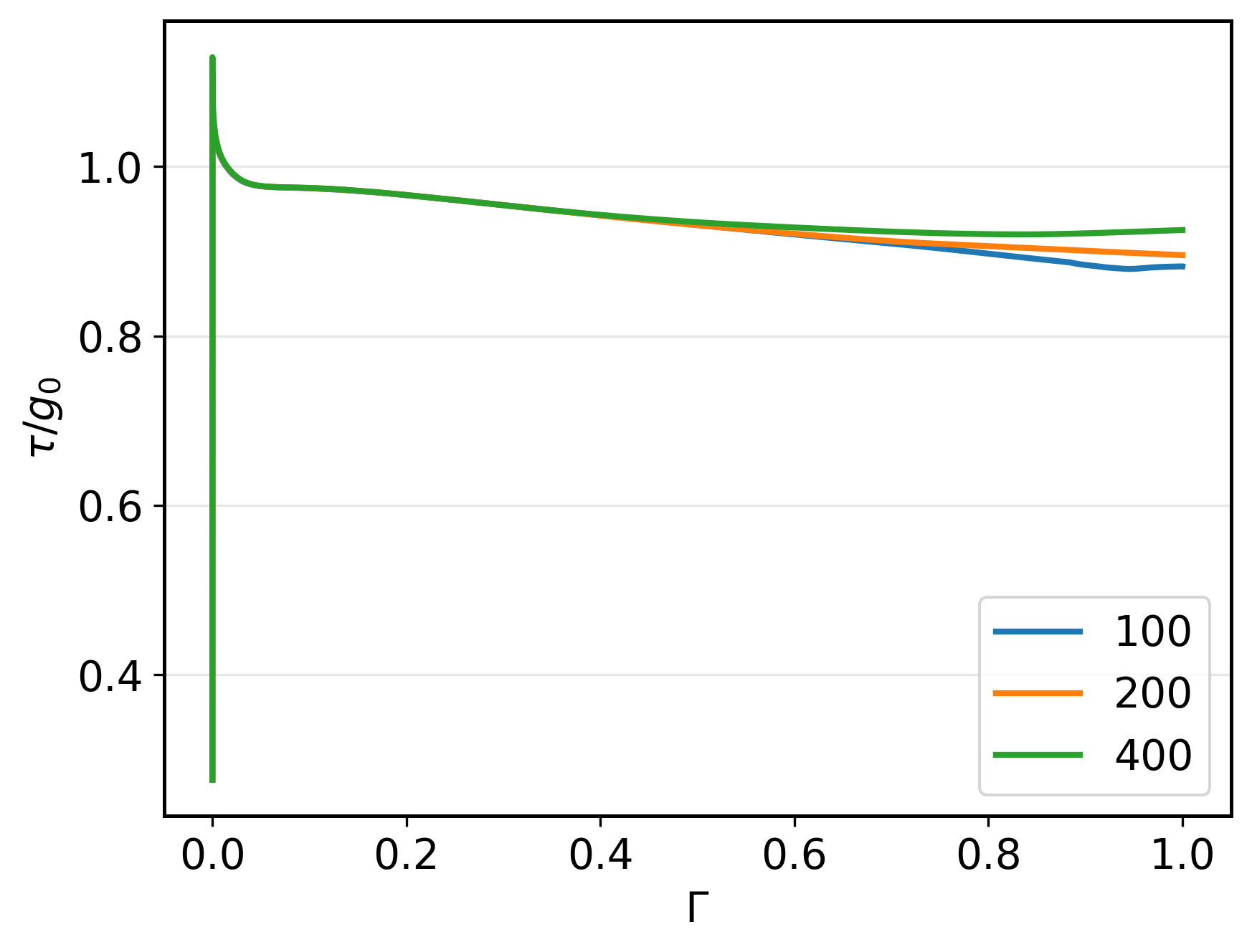}
        \caption{}
        \label{fig:ss_MFDM_J2}
    \end{subfigure}
    \caption{Stress-strain curves for (a) classical $J_2$ plasticity, and (b) $J_2$-MFDM, for three mesh resolutions of $100^2, 200^2, 400^2$ elements.}
    \label{fig:conv_stress_strain_J2}
\end{figure}

\subsection{Localization under boundary conditions corresponding to nominally homogeneous deformation with a spatial perturbation in the initial yield stress}

For the boundary conditions and initial conditions considered so far, MFDM does not produce further localization or softening. We now probe the response of MFDM under a different set of initial conditions and, in one case, also of geometry. 

A common approach in physical experiments and related modeling for studying localization is to utilize a perturbation in the initial conditions or in the geometry of the domain to induce heterogeneity \cite{Clifton1984,duffy1992measurement} (e.g., shear band studies in a torsional Kolsky bar experiment). Following this approach, we prescribe a reduced initial yield strength in a horizontal layer near the center of the domain, as shown in Fig.~\ref{sfig:lower_g_ini}.  All other aspects of the simulation, including the applied strain rate and the slip system configuration, remain as described in Sec.~\ref{sec:mat_mod_sim}. Simulations are performed using both MFDM and classical crystal plasticity (CCP). For a case shown in Fig~\ref{sfig:lower_g_F12_small}, the geometry and initial grain microstructure are self-similarly scaled to a domain size of $(40 \mu m)^2$.

We observe that the CCP results do not converge with mesh refinement, whereas MFDM yields a mesh-converged response, as shown in Fig.~\ref{sfig:lower_g_conv}.

More importantly, for this type of spatial perturbation in the initial yield stress, localization is observed in MFDM as shown in Figs.~\ref{fig:lower_g_2} (a,b,c), in contrast to the considerations of Sec.~\ref{sec:non_localized}. 

We explore some features of this localization w.r.t.~the length scales present in the problem. Dimensional analysis suggests that the normalized shear band width $W_b/b$ can depend on the following ratios related to length-scales (along with other non-dimensional numbers of the model):
\begin{equation*} \label{eq:dim_analysis_w}
    \frac{W_b}{b} = f\!\left(\frac{b}{H},\; \frac{b}{S_g},\; \frac{W_p}{b}, \ldots\right),
\end{equation*}
where $H$ is the domain size, $S_g$ is the average grain size, $W_p$ is the perturbation layer width, and $b$ is the Burgers vector.

We first examine the effect of the ratio related to the perturbation width by varying $W_p/b$ while holding $b/H$ and $b/S_g$ fixed. In Figs.~\ref{fig:lower_g_2} (a) and (b), the predicted band is narrower than the imposed perturbation layer: a $W_p = 10\;\mu m$ perturbation in an $80\;\mu m$ domain produces a band of approximately $7.5\;\mu m$, while reducing the perturbation to $W_p = 6\;\mu m$ in the same domain yields a band of approximately $5\;\mu m$. 

We next vary $b/S_g$ (and correspondingly $b/H$, by self-similar scaling of grain microstructure) while holding $W_p/b$ fixed. The expectation that the increase in this raio with $b$ fixed should result in stronger and more stable response is borne out in the simulations. Comparing the results of Fig.~\ref{sfig:lower_g_F12} and Fig.~\ref{sfig:lower_g_F12_small} corresponding to the average grain size being halved with $W_p = 10\;\mu m$ kept fixed, the predicted band width in Fig.~\ref{sfig:lower_g_F12} is $7.5\;\mu m$ whereas in Fig.~\ref{sfig:lower_g_F12_small} the band width is $6.8\;\mu\text{m}$ respectively, i.e., essentially unchanged. However, we note that as $b/S_g$ increases for $W_p$ fixed, the strain contrast between the band and its surroundings is reduced, as visible in Fig.~\ref{sfig:lower_g_F12_small} compared to Fig.~\ref{fig:lower_g_2}(a,b).

MFDM does not produce any softening in the stress-strain response in these simulations as well; instead, a stable flow stress is maintained. The temperature in the domain does rise to values as high as approximately $650\;\text{K}$, as shown in Fig.~\ref{sfig:lower_g_temp}. However, a competition between GND hardening and thermal softening results in a stable flow response for the material properties and boundary conditions considered here.

\begin{figure}[htbp]
    \centering
    \begin{subfigure}{0.28\textwidth}  
        \centering
        \includegraphics[width=\textwidth]{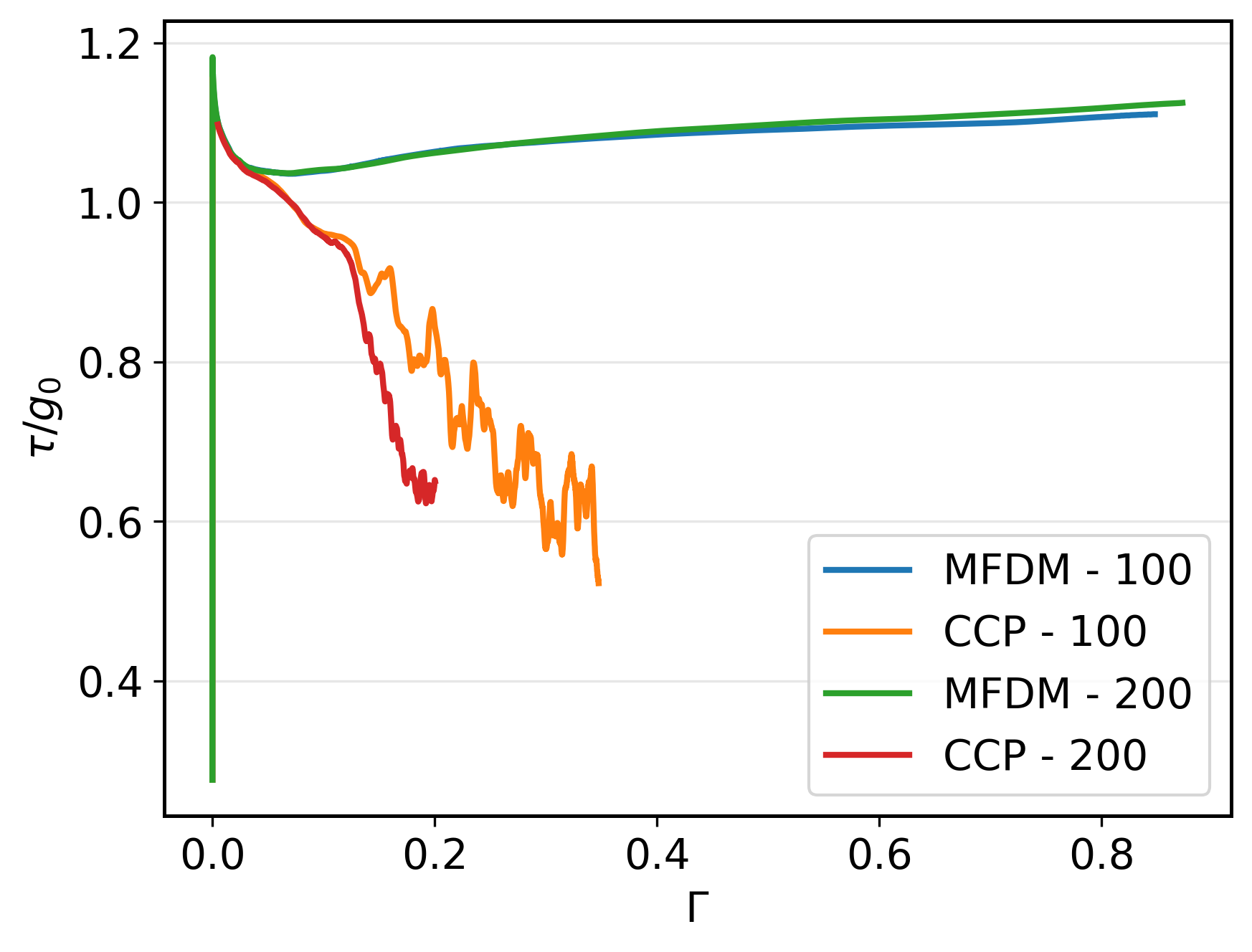}
        \caption{}
        \label{sfig:lower_g_conv}
    \end{subfigure}
    \hfill
    \begin{subfigure}{0.28\textwidth}  
        \centering
        \begin{overpic}[width=\textwidth]{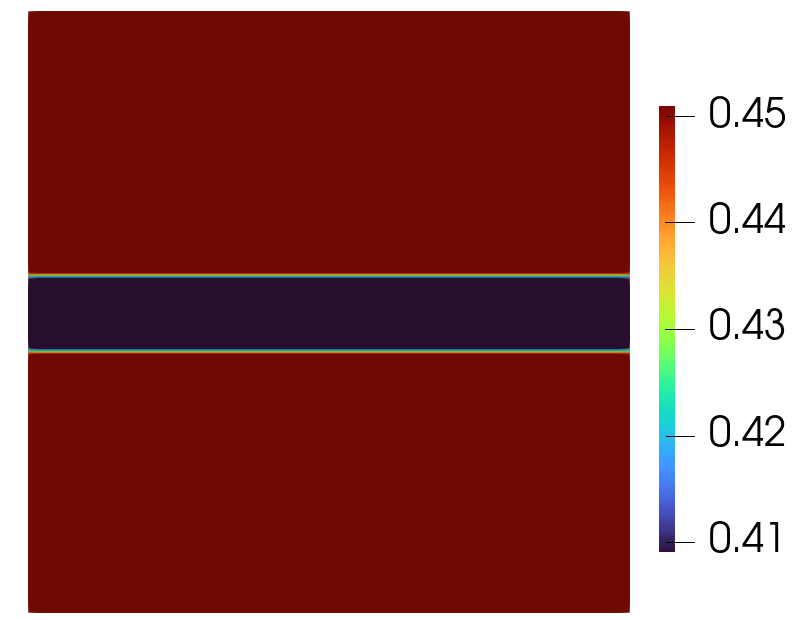}
            \put(84, 70){\rotatebox{0}{\fontsize{8}{5}\selectfont $g_{0}$}}
            \put(92, 70){\rotatebox{0}{\fontsize{8}{5}\selectfont $(GPa)$}}
        \end{overpic}
        \caption{}
        \label{sfig:lower_g_ini}
    \end{subfigure}
    \hfill
    \begin{subfigure}{0.40\textwidth}  
        \centering
        \begin{overpic}[width=\textwidth]{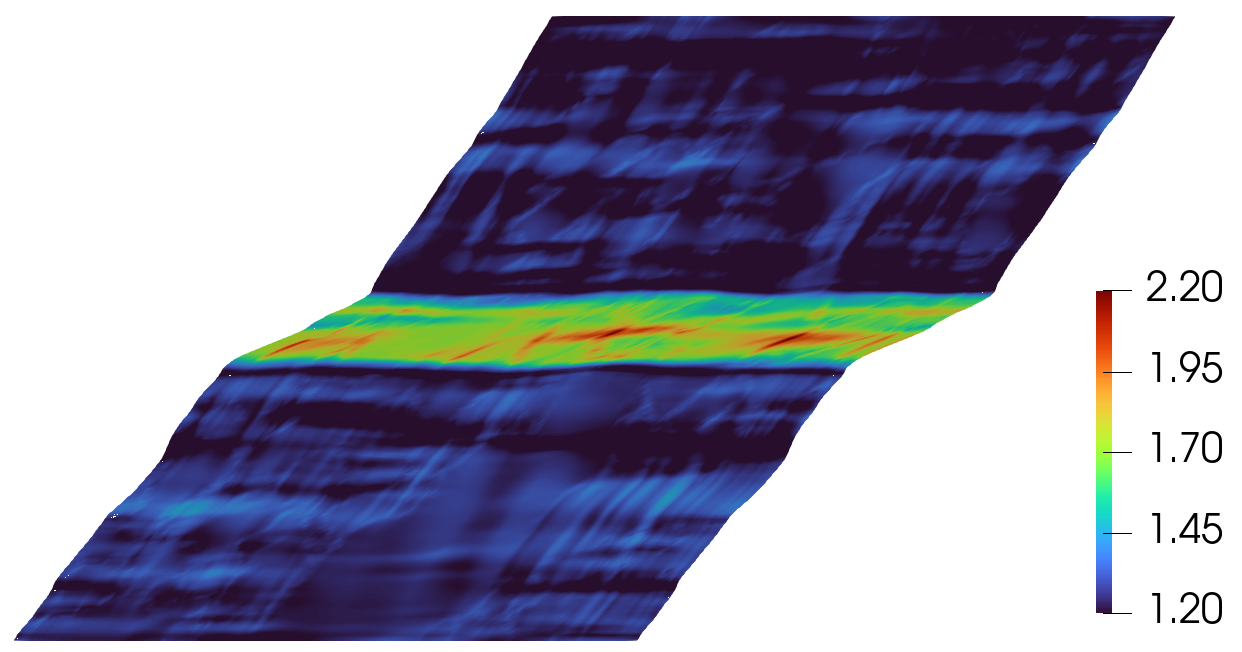}
            \put(88, 36){\rotatebox{0}{\fontsize{8}{5}\selectfont $\theta/\theta_0$}}
        \end{overpic}
        \caption{}
        \label{sfig:lower_g_temp}
    \end{subfigure}
    \caption{(a) Normalized stress-strain response ($\tau/g_0$ vs.\ $\Gamma$) for MFDM and classical crystal plasticity (CCP) at two mesh resolutions ($100^2$ and $200^2$); CCP results exhibit mesh dependence, whereas MFDM remains converged. (b) Initial yield strength ($g_0$) distribution showing the reduced-strength perturbation layer near the center of the domain. (c) Temperature field at an applied strain of 0.3, showing localization of deformation into a well-defined band within the MFDM framework.}
    \label{fig:lower_g_1}
\end{figure}

\begin{figure}[htbp]
    \centering
    \begin{subfigure}{0.32\textwidth}  
        \centering
        \begin{overpic}[width=\textwidth]{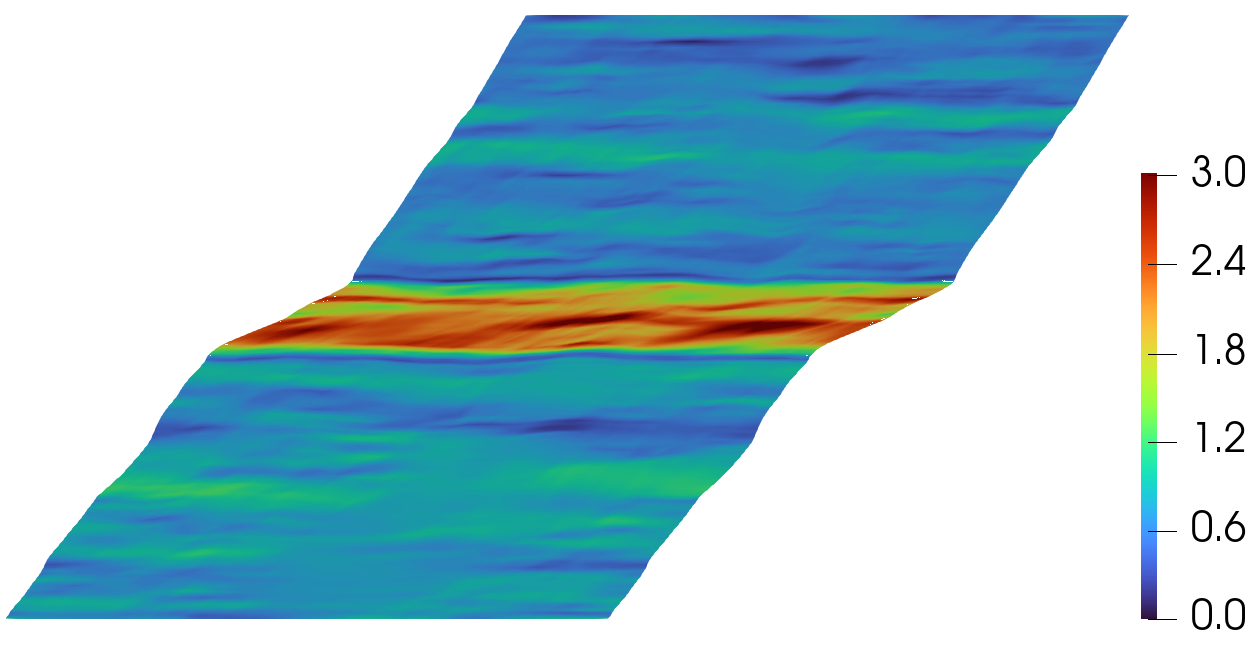}
            \put(90, 42){\rotatebox{0}{\fontsize{8}{5}\selectfont $F_{12}$}}
        \end{overpic}
        \caption{}
        \label{sfig:lower_g_F12}
    \end{subfigure}
    \hfill
    \begin{subfigure}{0.32\textwidth}  
        \centering
        \includegraphics[scale = 0.12]{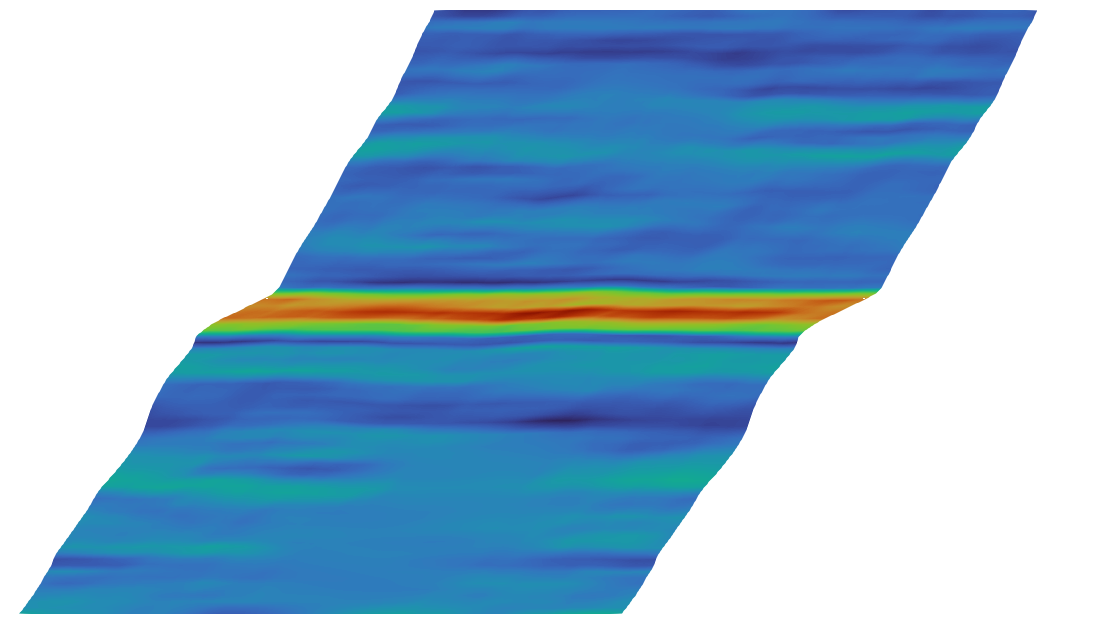}
        \caption{}
        \label{sfig:lower_g_F12_thin}
    \end{subfigure}
    \hfill
    \begin{subfigure}{0.32\textwidth}  
        \centering
        \includegraphics[scale = 0.12]{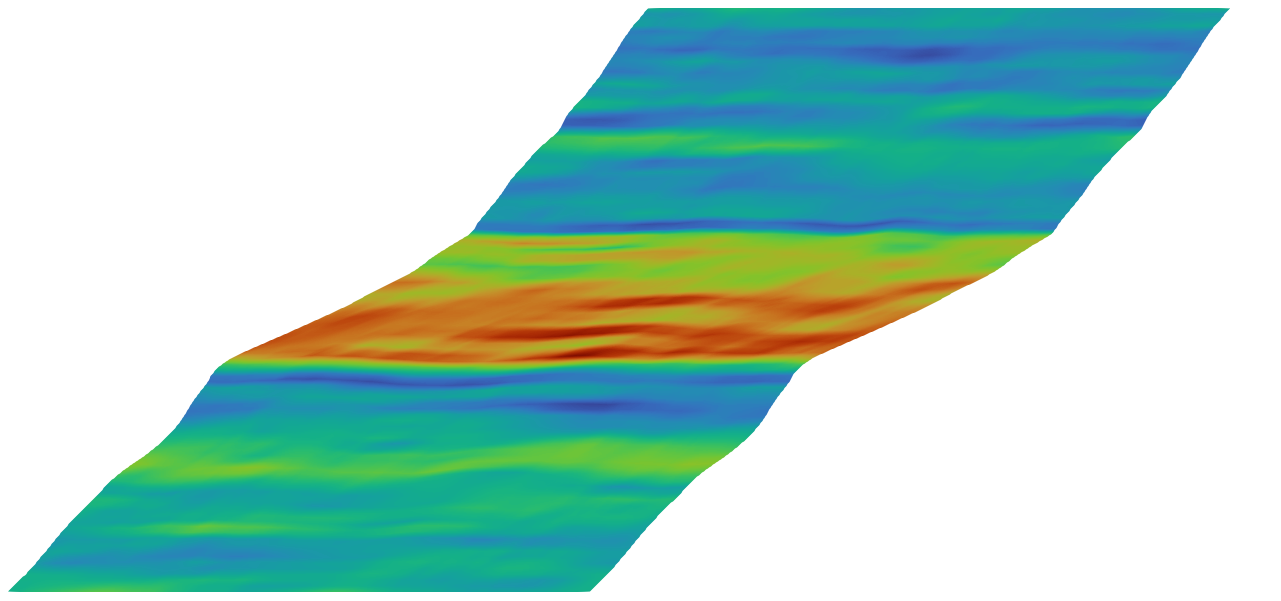}
        \caption{}
        \label{sfig:lower_g_F12_small}
    \end{subfigure}
    \caption{$F_{12}$ field for different perturbation layer widths and domain sizes. (a) Perturbation layer width of $10\;\mu m$ in an $80\;\mu m$ domain, resulting in a shear band width of approximately $7.5\;\mu m$. (b) Perturbation layer width of $6\;\mu m$ in an $80\;\mu m$ domain, resulting in a shear band width of approximately $5\;\mu m$. (c) Perturbation layer width of $10\;\mu m$ in a $40\;\mu m$ domain with a shear band width of $6.8\, \mu m.$ Fields in (a) and (b) are shown at an applied shear strain of $\sim 0.8$, while (c) is shown at an applied shear strain of $\sim 1.1$.}
    \label{fig:lower_g_2}
\end{figure}

\subsection{Structural softening and localization -- $J_2$-MFDM}

For the simulations presented in this Section, we employ the $J_2$-MFDM constitutive model (Sec.~\ref{sec:const}) within the MFDM framework. A polycrystalline microstructure is not initialized in the full domain, as sufficient information to do so is unavailable; however, the computational cost would remain comparable even if such information were available.

Here we demonstrate the ability of MFDM to capture localization into a finite-width band accompanied by softening in the structural response for a geometry and boundary conditions favoring heterogeneous deformation. The simulation is motivated by the idea of the macroscopic top-hat compression test for concentrating deformation. We use a tensile loading to avoid problems with contact, since our main concern here is to explore a point of principle about the model. We also consider a relatively larger domain than the RVEs previously selected (Sec.~\ref{sec:mat_mod_sim}).

The specimen geometry consists of two rectangular blocks connected through a narrow section, as shown in Fig.~\ref{fig:notch_schematic}. The left block has dimensions $200 \times 200\;\mu m^2$ and the upper-right block $200 \times 200\;\mu m^2$. The two blocks are offset vertically by $50\;\mu m$ and overlap horizontally over a $20\;\mu m$ wide strip, creating a narrow neck of $20 \times 50\;\mu m^2$ where localization is forced in order to accommodate the loading. The total specimen width is $380\;\mu m$ and the overall height is $350\;\mu m$.

The boundary conditions (shown in Fig.~\ref{fig:notch_schematic}) are chosen to promote localization through the connector region. On the bottom boundary of the left block the vertical velocity is constrained to vanish ($v_y = 0$) while leaving the horizontal direction free. On the right boundary of the upper-right block, the horizontal velocity is constrained to vanish ($v_x = 0$) while leaving the vertical direction free. The top boundary of the right block is subjected to a prescribed velocity in the $y$-direction with a rough applied strain rate of $4 \times 10^{3}~s^{-1}$, pulling the right block upward relative to the vertically constrained left block. All remaining boundaries are traction-free. This loading configuration produces a state of combined tension and shear in the narrow connector, promoting strain localization in the predetermined region.

\begin{figure}
    \centering
    \begin{tikzpicture}
        \node[anchor=south west, inner sep=0] (image) at (0,0) 
            {\includegraphics[scale=0.2]{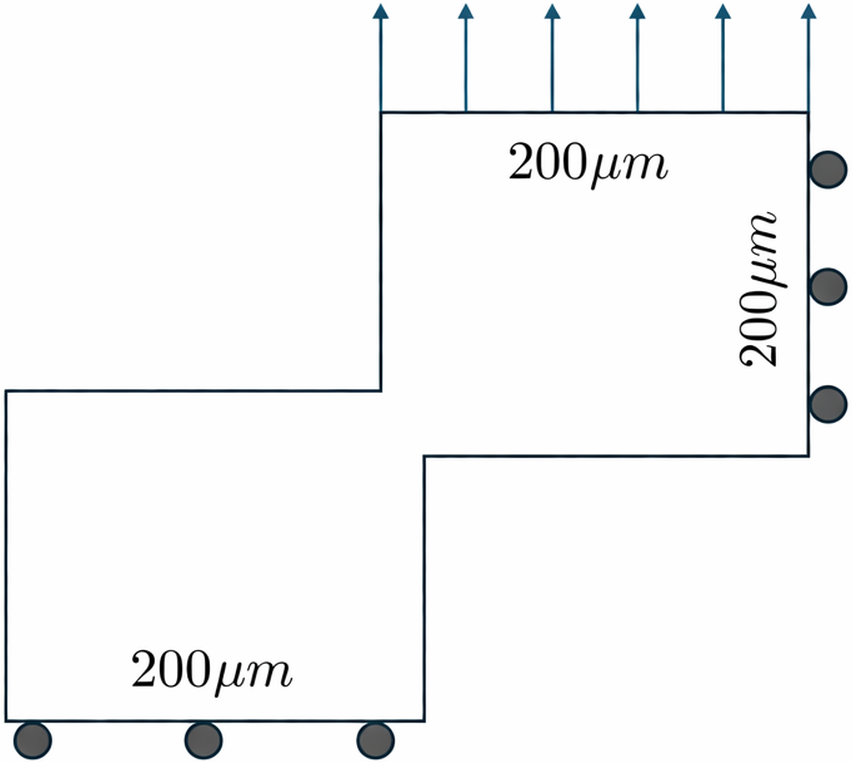}};
        \begin{scope}[x={(image.south east)}, y={(image.north west)}]
            \fill[white] (0.10, 0.08) rectangle (0.40, 0.18);
        \end{scope}
    \end{tikzpicture}
    \caption{Specimen geometry and boundary conditions. Rollers on the bottom boundary constrain $v_y = 0$; rollers on the right boundary constrain $v_x = 0$. The top boundary of the upper-right block is pulled upward with prescribed velocity $v_y = v(t)$. All other boundaries are traction-free.}
    \label{fig:notch_schematic}
\end{figure}

\begin{figure}
    \begin{subfigure}{0.40\textwidth}
    \centering
        \includegraphics[scale = 0.4]{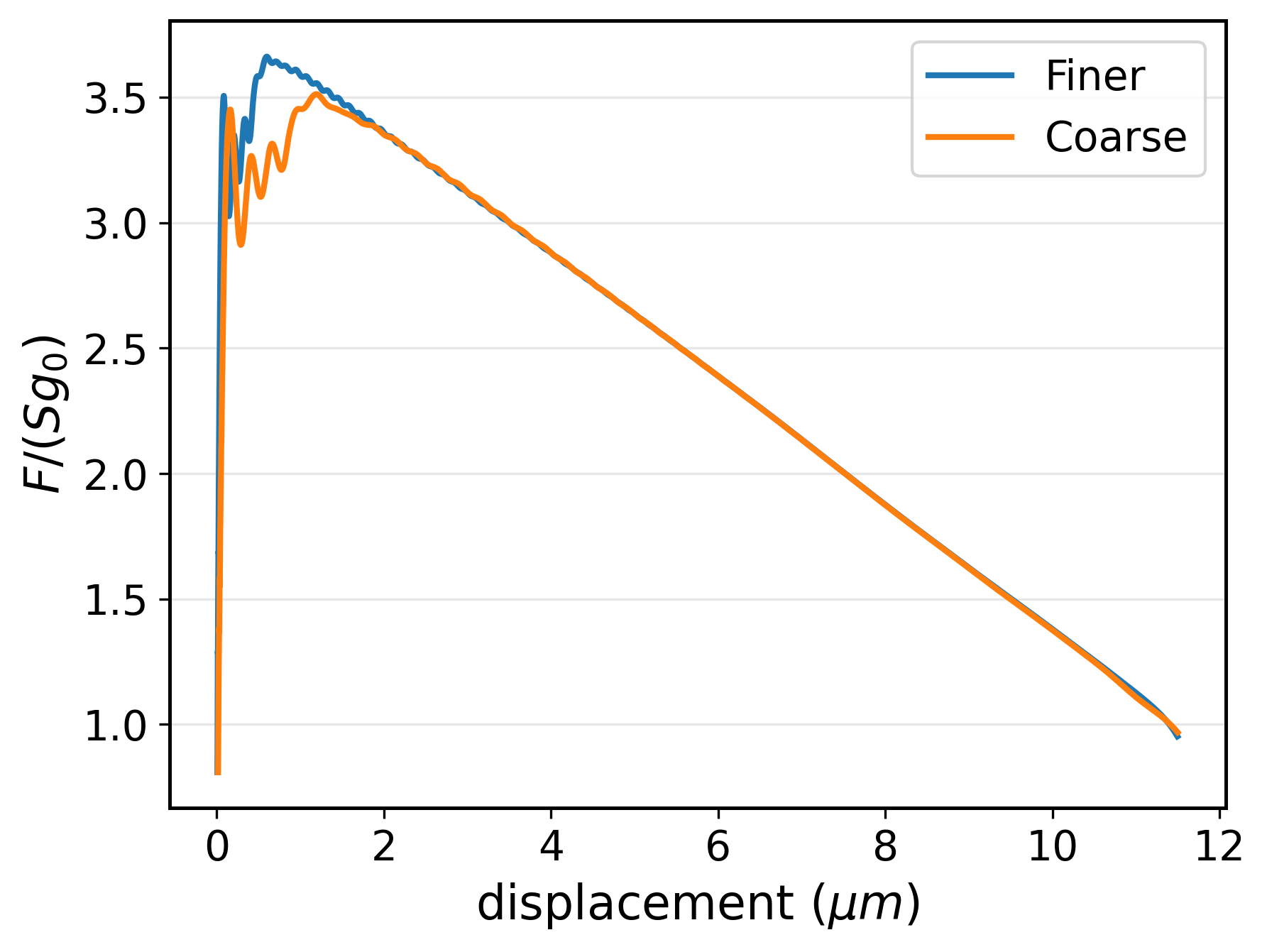}
        \caption{}
        \label{fig:ss_J2}
    \end{subfigure}
    \hfill
    \begin{subfigure}{0.40\textwidth}
    \centering
        \includegraphics[scale = 0.4]{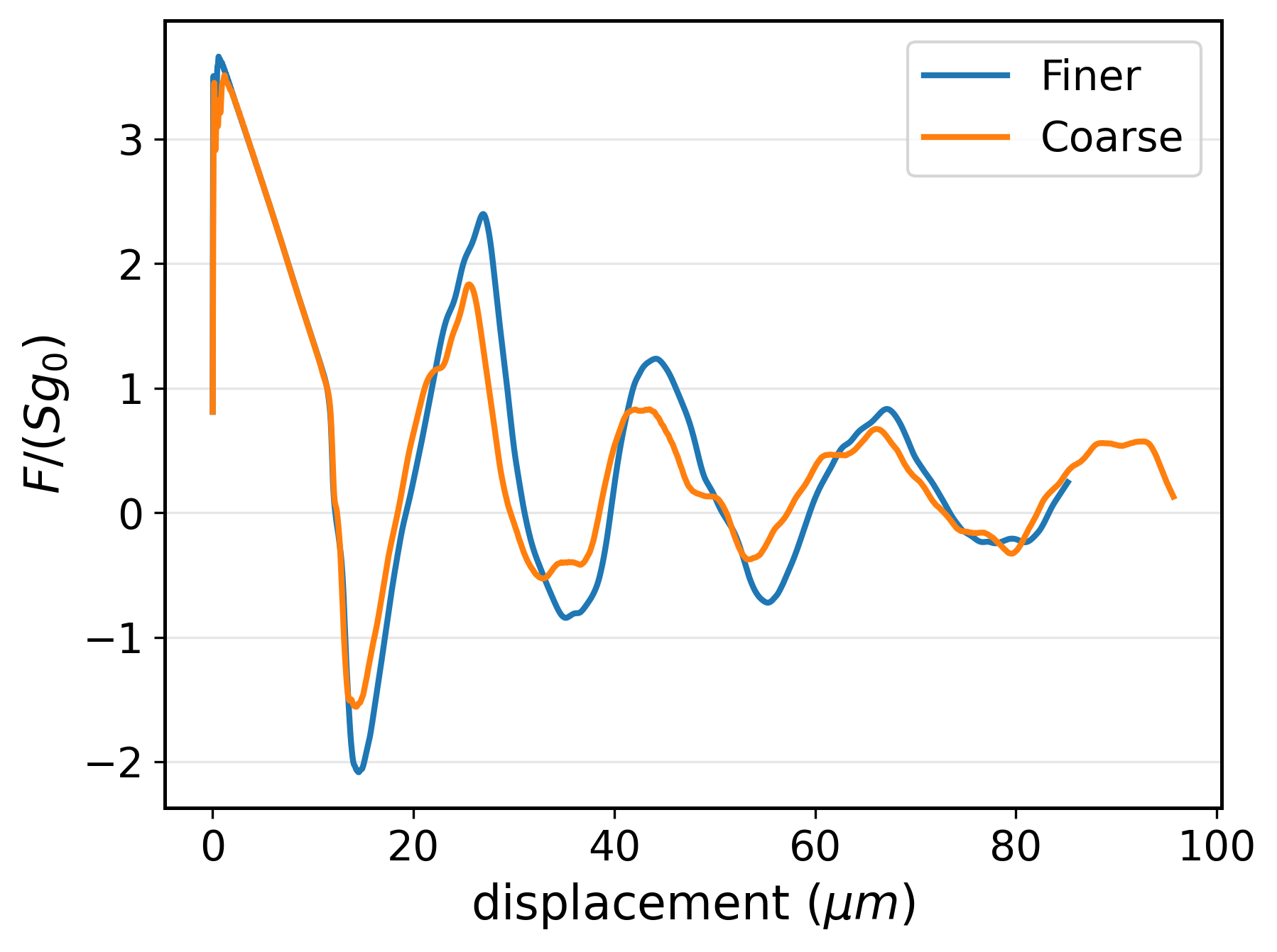}
        \caption{}
        \label{fig:ss_J2_extend}
    \end{subfigure}
    \caption{Force-displacement curve shown for (a) the softening regime and (b) beyond initial softening. The coarse mesh has approximately $50$k elements and the fine mesh approximately $200$k elements. Here, $F$ is the reaction force on the top-right boundary, and $S$ and $g_0$ are the cross-sectional area and initial yield strength, respectively. The oscillations arise from elastic wave reflections due to the small domain size and the specific boundary conditions employed.}
\end{figure}

The reaction force, normalized by the product of the cross-sectional area ($S$) and the initial yield strength ($g_0$), is plotted against the displacement of the top boundary of the right block in Fig.~\ref{fig:ss_J2}. A clear softening response is observed in the force-displacement curve. Furthermore, this softening behavior is shown to be converged across two levels of mesh refinement, with the coarser mesh containing approximately $50$k elements and the finer mesh approximately $200$k elements.

\subsubsection*{Mesh convergence and band width}

To assess whether MFDM produces mesh-convergent localization, we perform a series of simulations with progressively refined meshes while keeping all other parameters fixed. The deformation is characterized in terms of the norm of the logarithmic strain tensor $|\ln \bfV|$, where $\bfV$ is the left stretch tensor obtained from the left polar decomposition of the deformation gradient $\bfF = \bfV\bfR$, defined with respect to the initial configuration at $t = 0$.

Figure~\ref{fig:notch_lnv} shows the spatial distribution of $|\ln \bfV|$ in the deformed configuration for the two mesh refinements. In both cases, deformation localizes into a band within the narrow connector. The key observation is that both the width of the localized band and the magnitude of $|\ln \bfV|$ within the band remain approximately unchanged across mesh resolutions. This demonstrates that MFDM provides mesh-convergent predictions of localization, including the a finite width of the deformation band.

The converged band width is approximately $15\,\mu m$, governed by the interplay between the intrinsic length scale of MFDM (through the Burgers vector and the GND hardening coefficient $k_0$), the applied loading, and thermal softening. Thus, MFDM naturally produces a finite-width localized band when the domain is sufficiently large, and predicts softening in the structural response when the geometry and loading conditions permit it.

\begin{figure}[htbp]
    \centering
    \begin{subfigure}{0.48\textwidth}
        \centering
        \begin{overpic}[width=\textwidth]{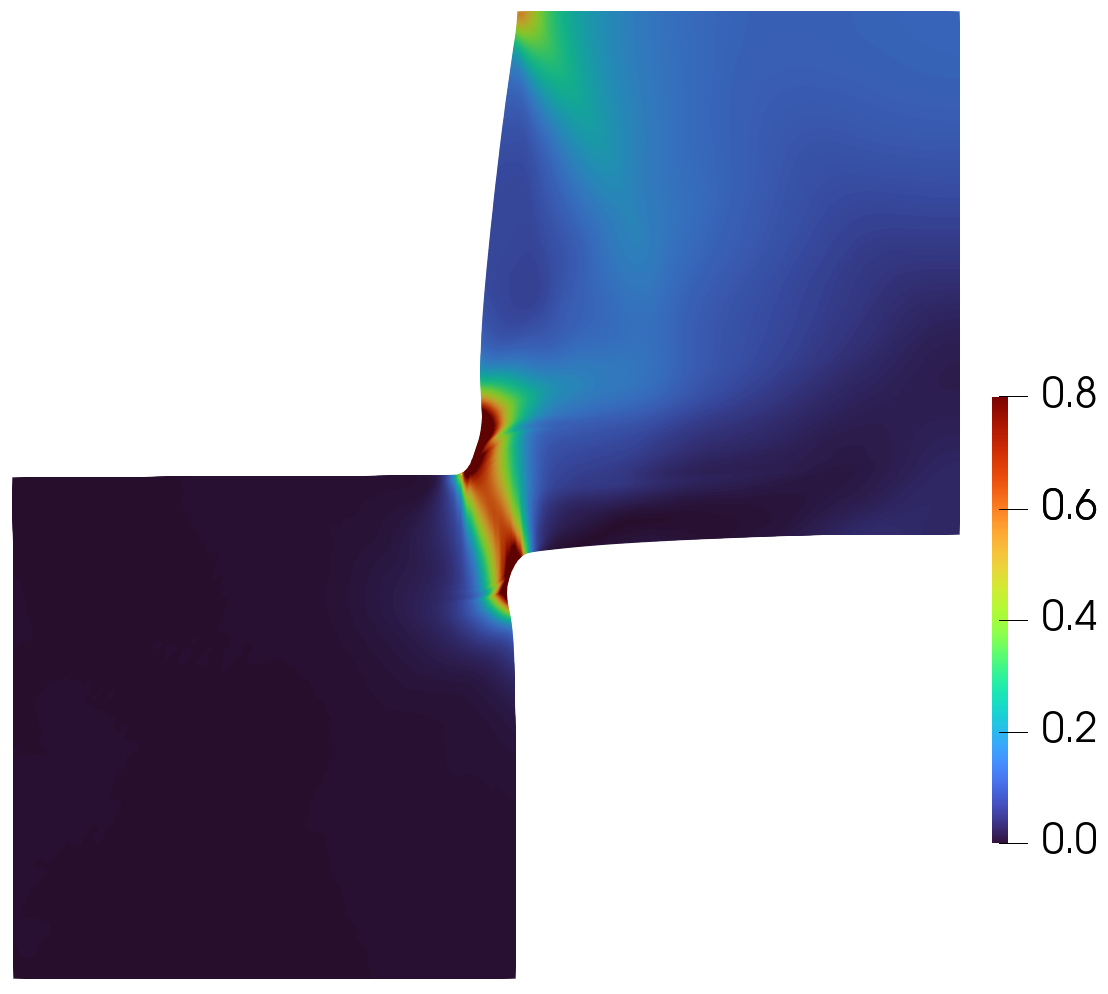}
            \put(90, 59){\rotatebox{0}{\fontsize{8}{5}\selectfont $|\ln \bfV|$}}
        \end{overpic}
        \caption{Coarse mesh ($\sim 50$k elements)}
        \label{fig:NS_1}
    \end{subfigure}
    \hfill
    \begin{subfigure}{0.48\textwidth}
        \centering
        \includegraphics[width=\textwidth]{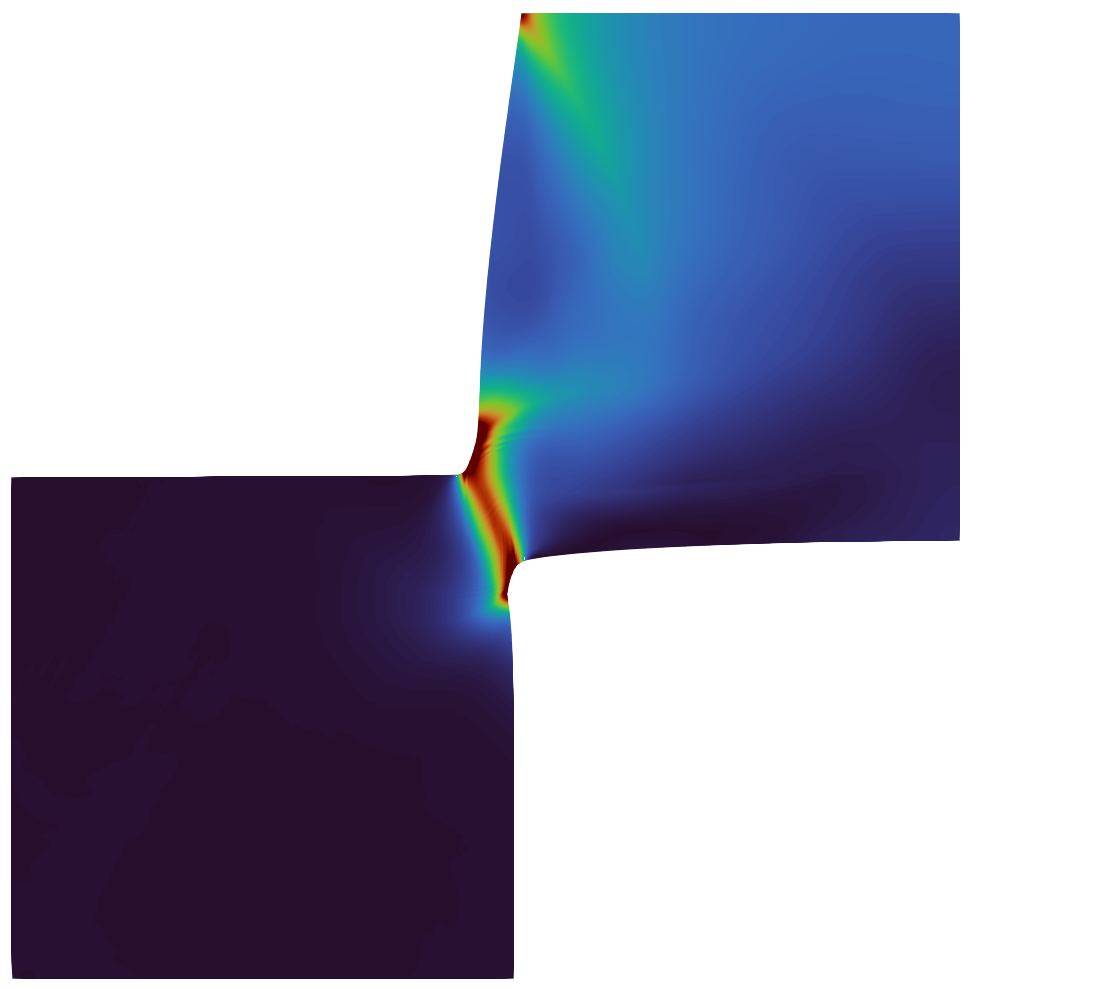}
        \caption{Fine mesh ($\sim 200$k elements)}
        \label{fig:NS_2}
    \end{subfigure}
    \caption{Spatial distribution of $|\ln \bfV|$ in the deformed configuration for two successively refined meshes, with the element size halved between refinements. The width (approx.~$\sim 15 \mu m$) and intensity of the localized band remain essentially unchanged, demonstrating mesh convergence of the localized solution.}
    \label{fig:notch_lnv}
\end{figure}

\section{Conclusion}

Adiabatic shear bands involve complex microstructural changes that are difficult to resolve experimentally at the mesoscopic length scale. Informed by Split Hopkinson Pressure Bar experiments for dynamic shear banding, we studied the phenomenon through theory and simulation. A dimensional analysis was employed to make our mesoscale simulations computationally tractable, especially the large-scale 3-d simulations presented in this work.

Finite width of adiabatic shear bands, consistent with experimental observations, were obtained as part of our results. In the absence of heat conduction, the effect of thermal softening was balanced by GND hardening in our simulations, suggesting a mechanism for the finite widths of the bands.  Dislocation substructure formation was also observed, with similarity to experimental observations reported in the literature. A comparison between classical crystal plasticity (CCP) and MFDM demonstrated the well-known mesh-dependent strain localization in CCP into increasingly narrow bands upon refinement. MFDM yielded mesh-convergent fields and stress-strain response across all mesh resolutions tested, in both 2-d and 3-d.

Size-dependent strengthening was observed for grain sizes ranging from $1$ to $20 \, \mu m$, with the effect saturating as the size was increased. Low-angle subgrain boundaries formed within grain interiors, with misorientations consistent with Incidental Dislocation Boundaries observed in deformed metals. These boundaries correlated spatially with GND density patterns, confirming that dislocation accumulation drives microstructural refinement during shear band formation.

The response of the MFDM theory/model was probed under different combinations of boundary conditions, initial conditions, and geometry to assess when localization and softening emerge. Under boundary conditions corresponding to nominally homogeneous deformation (for a homogeneous material deforming quasi-statically), MFDM predicts no further localization or softening even when the simulation was extended to shear strains as high as 300\% (locally), as the competition between GND hardening and thermal softening sustained a stable flow response in the absence of a damage mechanism. A boundary condition-induced stress heterogeneity was not found to be sufficient to form a persistent shear band in MFDM. 

When a spatial perturbation in the initial yield stress was introduced, MFDM yielded a mesh-converged localized band whereas classical crystal plasticity did not. Within the material parameter range probed, mesh convergent strain localization with softening was not obtained in MFDM with thermal softening (in the absence of other ductile damage mechanisms in the model).

For a top-hat-inspired geometry simulated with $J_2$-MFDM, the model captured finite-width localization of approximately $15~\mu m$ together with a clear, mesh-convergent softening response in the force--displacement curve. These results demonstrate that MFDM produces finite-width bands and structural softening when the geometry and loading conditions permit heterogeneous deformation, while remaining stable under nominally homogeneous loading conditions, without recourse to ad-hoc/physically dubious regularization mechanisms.

These results provide a mechanistic picture of microstructure evolution within adiabatic shear bands, from GND accumulation and strength heterogeneity to subgrain boundary formation, and establish a framework for further investigation of shear localization in polycrystalline metals.

With regard to advancing theory and simulation tools for the study of plasticity, our concern here has mainly been the study of plasticity problems on smaller scales where (R)MFDM has been validated against experiments relating to size effects and patterning. The approach has been quite successful in addressing a broad array of problems in modern plasticity theory, including size-dependent strengthening and the ``smaller is stronger'' trend observed in micron and sub-micron scale experiments (thin films, micropillar compression), orientation-dependent mechanical response in confined thin films, spontaneous dislocation patterning and microstructure emergence without introducing non-convexity in the energy, kink band formation in Cu-Nb nanometallic laminates enabled by natural jump/interface conditions, slip transmission constraints at grain boundaries, directionality of the sharp yield point and L\"{u}ders-type behavior in strain-aged steels, intermittent plasticity associated with dislocation transport, grain size effects on work hardening in polycrystals, and particle size and interlacing effects on the mechanical response of metal-matrix composites \cite{Arora_acharya,Arora_acharya_2,Arora_zhang_acharya,a_arora,a_arora_2,roy2006size,Ach_07,PRAD10,MBA10,FAB11,PAR11,PDA11,DAS16,FBE09,TVC07,TBFB10,RWF11,TVFB08,DTBF15,VBF09,DBTL,BTL20,GBG20}. This study is the first application of the MFDM framework to the modeling of shear banding phenomena. 

Together with the above list of validations, this work establishes MFDM as a framework that passes critical experimental tests of theory posed in the research community for modeling mesoscale plasticity and associated length-scale dependent phenomena. At the level of constitutive specification, it does so with a minimalistic adjustment beyond a basic, and well-accepted, model of classical plasticity. 

However, it is to be recognized, as we do, that the mechanisms of strengthening and softening reflected in the constitutive statements we have utilized (Table \ref{tab:constitutive_relation_g}, Eqns.~\eqref{eqn:g_theta}, \eqref{eq:Beta_form} and \eqref{eq:ES_evolve}) are at best rudimentary vis-a-vis capturing the averaged mesoscale behavior of dislocation interactions and energy dissipation, and the softening of elastic moduli at high (and low) temperatures. In the absence of serious, ground-up theoretical understanding of such (very challenging) coarse-graining issues as they pertain to strongly nonlinear, time-dependent thermomechanical plastic response, we have let simplicity guide our constitutive choices for engineering purposes and qualitative understanding, with obvious room for improvement from the research community interested in such matters (but hopefully not through the use of more and more complicated ad-hoc kinematic and empirical constitutive assumptions).

Finally, our framework for 3-d, finite deformation, dynamic simulations of statistically representative RVE with a state-of-the-art model for mesoscale response of polycrystalline aggregates sets the stage for computationally efficient, large-scale statistical data collection of the detailed (thermo)mechanical response of metallic materials to inform coarser scale models, e.g.~a principled way of including a grain size effect in the macroscopic response of polycrystalline materials valid for a wide range of grain sizes. 

\section*{Acknowledgment}
This work was supported by the Army Research Laboratory funded Center for Extreme Events in Structurally Evolving Materials (contract W911NF2320073). Generous grants of computer time from the NSF ACCESS program through the Pittsburgh Supercomputing Center are acknowledged.

\bibliographystyle{alphaurl}\bibliography{main.bib}
\end{document}

%% file: temp-definitions.tex
\usepackage{amsmath}
\usepackage{amsbsy}
\usepackage{amssymb}
\usepackage{amscd}
\usepackage{amsfonts}

\newcommand{\bfa}{{\mathbold a}}
\newcommand{\bfb}{{\mathbold b}}
\newcommand{\bfc}{{\mathbold c}}
\newcommand{\bfd}{{\mathbold d}}
\newcommand{\bfe}{{\mathbold e}}
\newcommand{\bff}{{\mathbold f}}

\newcommand{\bfm}{{\mathbold m}}
\newcommand{\bfn}{{\mathbold n}}

\newcommand{\bft}{{\mathbold t}}

\newcommand{\bfv}{{\mathbold v}}

\newcommand{\bfx}{{\mathbold x}}

\newcommand{\bfz}{{\mathbold z}}

\newcommand{\bfA}{{\mathbold A}}
\newcommand{\bfB}{{\mathbold B}}

\newcommand{\bfE}{{\mathbold E}}
\newcommand{\bfF}{{\mathbold F}}

\newcommand{\bfL}{{\mathbold L}}

\newcommand{\bfR}{{\mathbold R}}

\newcommand{\bfT}{{\mathbold T}}

\newcommand{\bfV}{{\mathbold V}}
\newcommand{\bfW}{{\mathbold W}}
\newcommand{\bfX}{{\mathbold X}}

\newcommand{\beq}{\begin{equation}}
\newcommand{\eeq}{\end{equation}}
\newcommand{\beqs}{\begin{eqnarray}}
\newcommand{\eeqs}{\end{eqnarray}}
\newcommand{\beql}{\begin{equation} \label}

\newcommand{\bfsigma}{\mathbold{\sigma}}

\newcommand{\bfalpha}{\mathbold{\alpha}}

\newcommand{\bfzero}{\mathbf{0}}

\newcommand{\grad}{\mathop{\rm grad}\nolimits}
\newcommand{\divergence}{\mathop{\rm div}\nolimits}
\newcommand{\curl}{\mathop{\rm curl}\nolimits}